\documentclass[twocolumn]{aastex62}

\usepackage{hyperref}
\usepackage{amsmath}
\usepackage{graphicx}
\usepackage{graphbox}
\usepackage{tikz}
\usepackage{float}
\usepackage{multirow}
\usepackage{color}

\usepackage[colorinlistoftodos]{todonotes}

\accepted{in The Astrophysical Journal}

\begin{document}
	\title{An exploration of model degeneracies with a unified phase curve retrieval analysis: The light and dark sides of WASP-43\,b}
	
	\correspondingauthor{Q. Changeat}
	\email{quentin.changeat.18@ucl.ac.uk}
	\author[0000-0001-6516-4493]{Q. Changeat}
	\affil{Department of Physics and Astronomy \\
		University College London \\
		Gower Street,WC1E 6BT London, United Kingdom}
	\author[0000-0003-2241-5330]{A.F. Al-Refaie}
	\affil{Department of Physics and Astronomy \\
		University College London \\
		Gower Street,WC1E 6BT London, United Kingdom}
	\author[0000-0002-5494-3237]{B. Edwards}
	\affil{Department of Physics and Astronomy \\
		University College London \\
		Gower Street,WC1E 6BT London, United Kingdom}
	\author[0000-0002-4205-5267]{I.P. Waldmann}
	\affil{Department of Physics and Astronomy \\
		University College London \\
		Gower Street,WC1E 6BT London, United Kingdom}
	\author[0000-0001-6058-6654]{G. Tinetti}
	\affil{Department of Physics and Astronomy \\
		University College London \\
		Gower Street,WC1E 6BT London, United Kingdom}
	

\begin{abstract}

The analysis of exoplanetary atmospheres often relies upon the observation of transit or eclipse events. While very powerful, these snapshots provide mainly 1-dimensional information on the planet structure and do not easily allow precise latitude-longitude characterisations. The phase curve technique, which consists of measuring the planet emission throughout its entire orbit, can break this limitation and provide useful 2-dimensional thermal and chemical constraints on the atmosphere. As of today however, computing performances have limited our ability to perform unified retrieval studies on the full set of observed spectra from phase curve observations at the same time. Here, we present a new phase curve model that enables fast, unified retrieval capabilities. We apply our technique to the combined phase curve data from the Hubble and Spitzer space telescopes of the hot-Jupiter WASP-43\,b. We tested different scenarios and discussed the dependence of our solution to different assumptions in the model. Our more comprehensive approach suggests that multiple interpretation of this dataset are possible but our more complex model is consistent with the presence of thermal inversions and a metal rich atmosphere, contrasting with previous data analyses, although this likely depends on the Spitzer data reduction. The detailed constraints extracted here demonstrate the importance of developing and understanding advanced phase curve techniques, which we believe will unlock access to a richer picture of exoplanet atmospheres. \\\vspace{5mm}
\end{abstract}

\section{Introduction}

In the field of transiting exoplanetary atmospheres, observational studies are dominated by two main techniques: transit and eclipse spectroscopy. The first consists of analysing the changes in the wavelength dependent transit depth when a planet passes in front of its host star while the second relies on observing the changes in the observed flux when the planet passes behind the star. These techniques are complementary, probing two distinct regions of the planet and being sensitive to different physical processes. Transmission spectra are most sensitive to the planetary radius, the cloud structure and the atmospheric chemical species. Spectra from eclipse observations are sensitive to the combination of thermal changes with altitude and chemical abundances, making these observations particularly useful to study exoplanet thermal structures \citep{Seager_2010,Tinetti_2013,Madhusudhan_2019_review, Changeat_2021_K9}. Most current retrieval codes extract information from such spectra using 1-dimensional descriptions \citep{rodger_retrievals,irwin2008, Madhu_retrieval_method,chimera, Waldmann_taurex1,Waldmann_taurex2,Gandhi_retrieval,Mollire_petitrad, al-refaie_taurex3, Zhang_platon, Edwards_2020, min2020arcis}. These models have been benchmarked, proving their consistency \citep{Barstow_2020_compar} when the same assumptions are taken.
In these models, the different processes can be modeled using free or self-consistent approaches. Self-consistent approaches attempt to reduce the number of free variables by modeling the physical phenomena from physico-chemical principles. These can include for example solving the full radiative-convective equilibrium equations, including dis/equilibrium chemistry, 2-stream temperature profile or micro-physical clouds. On the opposite, a free approach does not assume much about the physical state of the considered system and uses parametric descriptions. As of today, there is no consensus on what should be adopted when analysing real spectra and different assumptions can lead to different interpretations. 

By design transit and eclipse techniques offer the projection of a 3-dimensional atmosphere to a 1-dimensional spectrum with a wavelength dependence, from which it is difficult to extract the geometrical repartition of chemical and thermal properties \citep{feng_2016,Line_2016, caldas_3deffects, Drummond_2020,MacDonald_2020_cold,Pluriel_2020,skaf_2020_ares, taylor_2020, feng2020_2d}. To overcome these limitations and characterise the longitudinal structure of exoplanets, photometric and spectral phase curves have been used \citep[e.g.][]{Esteves_2013,Placek_2017,sing_2018observational,Parmentier_2018,deming_2020highlights,barstow_2020outstanding}. The technique consists in following the combined light (reflected and emitted) from the planet and star along the entire planet orbit, thus capturing the planet signal as a function of its phase. This technique is challenging, requiring a particularly high temporal stability, whereas current instruments limits its application to planets with short orbital periods. Only a handful of the known exoplanets have been observed in phase curves with the Hubble Space Telescope (HST), Spitzer, Kepler and TESS \citep{Parmentier_2018, bell2020comprehensive}. Amongst those observations, HST provided particularly good constraints between 1.1$\mu$m and 1.7$\mu$m for WASP-43\,b \citep{stevenson_w43_1,stevenson_w43_2}, WASP-103\,b \citep{Kreidberg_2018} and WASP-18\,b \citep{Arcangeli_2019_W18}. Spitzer obtained additional phase curve data at 3.6$\mu$m and 4.5$\mu$m for many planets, allowing to inform on circulation processes. Spitzer phase-curve observations, for example, includes 55\,Cancri\,e \citep{Demory_2016}, HD\,209458\,b \citep{Zellem_hd209_phase}, HD\,189733\,b \citep{Knutson_2012}, WASP-43\,b \citep{stevenson_w43_2}, WASP-33\,b \citep{Zhang_2018}, HD\,149026\,b \citep{Zhang_2018},  WASP-12\,b \citep{Bell_2019} and KELT-1\,b \citep{Beatty_2019}. Of these targets, WASP-43\,b possesses the most complete set of observations, with high quality spectra from both HST and Spitzer. The data was first analysed in \cite{stevenson_w43_1,stevenson_w43_2}, providing a unique insight into the properties of this world. Their spectral retrieval analysis was performed for each of the individual spectra obtained at each phase: in these analyses no correlation between the different observed phases was considered. More recently, \cite{Irwin_w43b_phase} used optimal estimation techniques to investigate
the combined information content of the spectra at different phases. It was the first successful attempt to analyse the spectral phase curve of an exoplanet with a unified model. However, due to the computing cost of their model, concessions had to be made in the sampling technique. Their paper highlighted the limitations induced when using optimal estimation technique in exoplanet atmospheric studies, where observations have a low signal-to-noise and the prior knowledge on the solution is unknown. Other studies highlighted the importance of using phase curve data to break the degeneracies coming from the 3-dimensional aspect of exoplanets \citep{feng_2016, taylor_2020,feng2020_2d}. In particular, \cite{feng2020_2d} performed the first combined retrieval of the WASP-43\,b phase curve using full exploration of the parameter space with the MultiNest sampler \citep{Feroz_multinest,buchner_pymultinest}. 

In this paper, we propose an update of the forward model used in \cite{changeat_2020_phase1} to compute the phase dependent emission of tidally locked planets. Our semi-analytical computation of atmospheric spectra enables Bayesian retrieval capabilities with full exploration of the parameter space. We applied the technique on the available WASP-43\,b phase curve data from HST and Spitzer, showing the potential of phase curve techniques in extracting detailed 3-dimensional atmospheric properties of exoplanet atmospheres and breaking degeneracies in transmission. We explore various model assumptions to investigate the information content of phase curve data and highlight model dependant behaviours linked to these complex datasets.

\section{Model Description}\label{sec:model}

\renewcommand{\floatpagefraction}{.9}%

The analysis of the WASP-43\,b phase curve spectra is performed using the Bayesian retrieval framework TauREx-3 \citep{al-refaie_taurex3, Waldmann_taurex1, Waldmann_taurex2}. Taking advantage of the new `plugin system' \citep{al-refaie_taurex3.1}, we developed a dedicated Phase Curve model enabling unified phase-curve retrieval analysis. We updated the geometry presented in \cite{changeat_2020_phase1} and built a phase dependant model that is more adapted to calculate the emission of tidally locked planets. The planet is considered as a single entity, physically separated in 3 regions of similar properties (hot-spot, day-side and night-side). In tidally locked planets, the hot-spot region corresponds to the region of highest temperatures (the sub-stellar point). The day-side represents the remaining part of the day-side atmosphere, facing the star. The night-side refers to the part in the atmosphere which does not receive direct stellar radiations. This separation is relevant for the study of irradiated tidally locked planets that are showing asymetric emission and day-night contrasts in their observed phase curves, such as WASP-43\,b \citep{stevenson_w43_1,stevenson_w43_2}. These features are potentially due to a directional redistribution of the atmosphere creating offsets in the observed brightness temperature (hot-spot) and a cooler night-side. In the 3 regions of our model, we consider that the chemistry, the temperature and the cloud properties can be considered constant with latitude and longitude.  

For each phase angle, the emission contribution of each region is calculated semi-analytically via the computation of contribution coefficients: C$^h_i$, C$^d_i$ and C$^n_i$ for respectively the hot-spot, the day and the night region. For clarity, the details about the calculation of these coefficients are shown in Appendix \ref{apx:phase_model}. By analysing all the spectra together, the model ensures that the redundancy of the information content between the different spectra is taken into account. We also updated the transit model to account for the predicted strong differences in the day and night-side atmospheres in these types of planets \citep{caldas_3deffects, Pluriel_2020}. The mathematical description of the new transmission model is detailed in Appendix \ref{apx:trans_model}.

In the complete model (phase emissions + transmission), the calculation of the coefficients C$^h_i$, C$^d_i$ and C$^n_i$ is trivial for a computer and takes negligible time. In comparison to a standard emission or transmission model, we observe that this phase curve forward model is about 5 times slower. It can be understood by the fact that the complete model includes 3 emission models (1 per region) and two transmission models (day and night-sides) for which the quantities (temperature, chemistry, clouds) and the optical depth must be computed (see \cite{changeat_2020_phase1} for a more detailed discussion on the performances). We also highlight the fact that the most recent version of TauREx 3.1 \citep{al-refaie_taurex3.1} introduced the new plugin system, allowing the user to benefit from any other TauREx module without requiring a particular adaptation. As the phase curve model is based on the standard TauREx emission and transmission models, which have GPU accelerated plugins, our phase curve model is automatically compatible with GPU architectures.

\section{Retrieval setup} \label{sec:retrieval}

The hot-Jupiter WASP-43\,b was first reported in 2011 \citep{Hellier_2011} and, while its radius is similar to that of Jupiter, it is twice as massive \citep{Hellier_2011,bonomo_2017}. It was immediately recognised as an extraordinary laboratory for atmospheric studies thanks to its very short orbit (0.8 days) and the data's particularly high signal to noise ratio. The entire phase has been observed with both Hubble and Spitzer, providing one of the most complete datasets to date. The complete phase curve was first analysed in \cite{stevenson_w43_1,stevenson_w43_2}, which unveiled variations in the chemical abundances of H$_2$O, CO$_2$ and CH$_4$. They also revealed a particularly low emission from the planet night-side, suggesting inefficient heat redistribution and/or a large cloud cover. We take WASP-43\,b as an example for our phase-curve investigation. We start by exploring the behaviour of our technique on mock-up spectra for this planet (Section \ref{sec:results_mock}). Then, we analyse the phase curve data from the HST and Spitzer phase-curve observations using a large range of retrieval scenarios (Section \ref{sec:result}). 

\subsection{Observations}\label{sec:w43_observations}

Real observations of WASP-43\,b were used to test and illustrate the phase curve model presented in this paper. They were obtained from \cite{stevenson_w43_1} and \cite{stevenson_w43_2} with no modifications. These consist of 15 already reduced spectra of the phase curve of WASP-43\,b: 0.0625, 0.125, 0.1875, 0.25, 0.3125, 0.375, 0.4375, 0.5, 0.5625, 0.625, 0.6875, 0.75, 0.8125, 0.875, 0.9375. The data was originally obtained by the Hubble Space Telescope in November 2013 (Program GO-13467) during 3 full orbital phases, 3 transits and 2 eclipses with the WFC3 G141 grism (wavelength coverage from 1.1 $\mu$m to 1.6 $\mu$m). These consisted of spatially scanned images corresponding to 13-14 HST orbits for the phases and 4 HST orbits for the transit and eclipses. Complementary phase curve observations were obtained by the Spitzer Space Telescope, using the 3.6 $\mu$m (2 visits, including 1 discarded) and 4.5 $\mu$m (1 visit) photometric channels (Programs 10169 \& 11001, PI: Kevin Stevenson). In addition to this, we also include the transmission spectrum from \cite{Kreidberg_2014} (in our model, this corresponds to phase 0.0) as this provides additional constraints on the day-night limb and allows us to extract the planetary radius with greater accuracy.

In emission, a large wavelength coverage is usually required to ensure that a sufficient pressure range of the atmosphere is probed, thus constraining the temperature structure (see Appendix \ref{apx:tp_investigations}). Normally, this is done by combining the observations from HST and Spitzer to obtain spectra spanning 1.1$\mu$m to 4.5$\mu$m. Combination of instruments however can bring additional difficulties as nothing guarantees the compatibility of the observations. Sources of discrepancies can come from different instrument systematics, the use of different orbital elements, different reduction pipelines, stellar or planet temporal variations \citep{Yip_lightcurve, Yip_W96, Changeat_k11}. As a consequence, independent studies of HST WFC3 data often lead to similar spectral shapes but different absolute depths \citep{Changeat_k11}. Studies of the WASP-43\,b Spitzer data \citep{stevenson_w43_2,Mendon_a_2018,morello_w43_phase,May_2020,bell2020comprehensive} have also proved that independent reduction pipelines can obtain different results for the same Spitzer dataset. In this paper, we use the Spitzer data from \cite{stevenson_w43_2} as-is and do not investigate the potential implications of these effects. We however present a complementary retrieval in the Discussion section with the HST only data, which highlights how the Spitzer points affect our solution. To date, this WASP-43\,b dataset is one of the most complete. Our integrated phase-curve framework accumulates the information at all phases to extract extremely precise constraints on the atmospheric properties of this planet.

\subsection{Opacity sources}\label{sec:opacities}

For this study, we assumed that the planet was mainly composed of hydrogen and helium with a ratio He/H$_2$ = 0.17. We considered collision induced absorption of the H$_2$-H$_2$ \citep{abel_h2-h2,fletcher_h2-h2} and H$_2$-He \citep{abel_h2-he} pairs and included opacities induced by Rayleigh scattering \citep{cox_allen_rayleigh} and clouds. Since clouds are most constrained by the transmission spectrum, we model them with a fully opaque cloud layer above a given pressure, restricted to the day and night-side regions. For the chemistry, we use the molecular line-lists from the Exomol project \citep{Tennyson_exomol, Tennyson_2020_exomol, Chubb_2021_exomol}, HITEMP \citep{rothman} and HITRAN \citep{gordon}. While many molecules are considered in the chemical equilibrium scheme, we only include molecular opacities for H$_2$O \citep{barton_h2o,polyansky_h2o}, CH$_4$ \citep{hill_xsec,exomol_ch4}, CO \citep{li_co_2015}, CO$_2$ \citep{rothman_hitremp_2010}, C$_2$H$_2$ \citep{2016_WILZEWSKI_C2H2}, C$_2$H$_4$ \citep{2018_Mant_C2H4}, NH$_3$ \citep{Yurchenko_2011_NH3} and HCN \citep{Barber_2013_HCN,Harris_2006_HCN}.

\subsection{Mock retrieval configuration}

Section \ref{sec:results_mock} aims to validate our method by presenting a retrieval on mock-up data where the true solution is known in advance. This approach allows to check that our model is correctly implemented and will ensure that the results presented on real data, later on, are not artefacts of our method. This example is performed using Ariel simulated spectra, which also gives us the opportunity to investigate the Ariel capabilities to perform phase-curves observations. The planet is simulated using the stellar and planet parameters from the literature \citep{bonomo_2017}. We first create the forward model at high resolution using our phase curve model with three distinct regions: hot-spot, day-side and night-side. Each region is composed of 100 layers spaced in log-pressures. We assume homogeneous temperature profiles and a chemical compositions for the three regions inspired from previous works \citep{changeat_2020_phase1, stevenson_w43_1, stevenson_w43_2}. For simplicity, the chemistry is modeled constant with altitude and only H$_2$O, CH$_4$ and CO are included. In this example, the H$_2$O abundance is set at 6$\times$10$^{-3}$ for the hot-spot and 1$\times$10$^{-4}$ for the day and night-sides. For CH$_4$, the atmosphere contains 1$\times$10$^{-7}$, 1$\times$10$^{-5}$ and 1$\times$10$^{-4}$ for respectively the hot-spot, the day-side and the night-side. Finally, the CO abundance is set to 1$\times$10$^{-3}$ on the hot-spot and 1$\times$10$^{-4}$ for the other regions. For this test, the hot spot parameters were fixed to an hot-spot offset of 12.2 degrees, consistent with \cite{stevenson_w43_1}, and a hot-spot size of 40 degrees, matching findings from \cite{Kataria_2015_W43GCM,Irwin_w43b_phase}. This leads to 15 spectra from phase 0.0625 to phase 0.9375. We then bin down those spectra to Ariel observations assuming Tier 2 resolution and the observation of 4 complete planet revolutions. The noise was simulated using the Ariel Radiometric Model (ArielRad) from \cite{mugnai_Arielrad}. We do not use random noise instances of our simulated observations for the retrievals, as we aim to study the retrieval biases arising from our retrieval technique \citep{feng_2016,Changeat_2019_2l, Mai_2019,changeat2020_alfnoor}. Finally, we perform our phase-curve retrieval test leaving free the planetary radius, the temperature profiles, the constant chemical abundances and the hot-spot parameters. The parameter space is explored uniformly with large bounds (see Table \ref{tab:priors}). The results of this exercise are described in Section \ref{sec:results_mock}.

\subsection{Detailed retrieval configurations} \label{sec:details_retrieval}

Section \ref{sec:result} presents the results for the retrievals performed on the real data. Here, we consider the reduced observations described in Section \ref{sec:w43_observations}. We also use the star parameters from \cite{bonomo_2017} and fix the planet mass \citep{changeat2019_impact} to its radial velocity measurement \citep{bonomo_2017}. Since we always include the transmission spectrum in our retrievals, which is very sensitive to the planetary radius, we let this parameter as free. We note that complementary tests without including the transmission spectrum did not affect the results presented here. We performed an analysis of each spectrum individually (see Appendix \ref{apx:append_indiv_spectra}) and carried on with unified analyses using 3 classes of models of increasing complexity: \\ \\

- {\bf 2-Faces free}: We considered a simplified geometry with no hot-spot.  We remove the influence from the hot-spot region by coupling all parameters (temperature, chemistry, clouds) between the hot-spot and the day-side regions. This allows us to check the consistency of our model and compare with previous results from the literature. This setup uses a similar geometry to \cite{feng2020_2d}. We considered free constant with altitude abundances for the molecules H$_2$O, CH$_4$, CO, CO$_2$ and NH$_3$. In the main scenario, these were coupled between all three regions.  The free temperature profiles and cloud properties were left independent between the day and night regions.  For the clouds, we use a simplistic fully opaque grey cloud. Two complementary runs were also used to explore the effect of \cite{Guillot_TP_model} T-p profile and de-coupled chemistry (different chemistry between the night and day-sides). \\

- {\bf 2-Faces equilibrium}: For the second scenario, we used the same geometry but increased the complexity of our model by considering equilibrium chemistry for the day and night-side regions. We used the scheme ACE from \cite{Agundez_2dchemical_HD209_HD189}. It calculates thermo-chemical equilibrium abundances for H, He, C, O and N bearing species, which is expected for hot-Jupiters between 1000 K and 2000 K. Only a fraction of the calculated molecules possess cross sections so we limit the actively absorbing species to the ones described in Section \ref{sec:opacities}. The chemistry was de-coupled between the day and night regions (different chemical profiles) but the free parameters (metallicity and C/O ratio) are shared between all regions. \\

- {\bf Full}: Finally, we performed simulations on the full model by introducing the hot-spot region described in Section \ref{sec:model}. The definition of this region requires the additional hot-spot offset ($\Delta$) and hot-spot size ($\alpha$) parameters. In the first place, we attempted to retrieve these parameters but encountered inconsistencies with previous values from the literature (see posterior distribution in Appendix \ref{apx:free_hs_run}). Since these degrees of freedom were already included in the reduction steps leading to the spectra \citep{stevenson_w43_1,stevenson_w43_2}, we decided to fix the hot-spot shift to the value in \cite{stevenson_w43_1}: -12.2 degrees. For the hot-spot size, we tested the values 30, 40 and 50 degrees, consistent with standard predictions from recent models for this planet \citep{Kataria_2015_W43GCM,Irwin_w43b_phase} but only present the model with 40 degrees in the result section. The complementary retrievals are presented in the discussion section and explore the impact of hot-spot size, hot-spot clouds and the addition of the Spitzer data. We caution the fact that these apparent inconsistencies might be linked to incompatibilities of the spectra following the reduction process (this was not highlighted by our individual analysis in Appendix \ref{apx:append_indiv_spectra}), issues with our model assumptions (different planet geometry), or the lack of information content in the considered spectra. \\

In all scenarios, except when stated otherwise, the parameterisation of the temperature profiles was done using the N-point profile from TauREx-3 \citep{al-refaie_taurex3}. This heuristic profile interpolates linearly between N freely moving temperature-pressure points. The profile is then smoothed over 10 layers to avoid inflexion points. For the hot-spot and day regions, we retrieve 7 temperature points at fixed pressures (10$^6$, 10$^5$,10$^4$, 10$^3$, 10$^2$, 1 and 0.01 Pa). Since the information content decreases on the night-side due to the lower emission, we found that retrieving 5 points (10$^6$, 10$^5$,10$^3$, 10 and 0.1 Pa) was more suitable. These choices are investigated in Appendix \ref{apx:tp_investigations}, where various free temperature structures are explored. This level of complexity was chosen to maximise the Bayesian evidence in the retrievals of the WASP-43\,b data from HST+Spitzer with the new phase curve model. We highlight, however, that other profiles lead to equivalent Bayesian evidences in the HST only case.

The exploration of the parameter space is performed with uniform priors using the nested sampling algorithm MultiNest \citep{Feroz_multinest,buchner_pymultinest} with 500 live points and a log likelihood tolerance of 0.5. These choices ensure an optimal free sampling of the parameter space. All the free parameters considered in this retrieval and their uniform priors are described in Table \ref{tab:priors}. The results of our investigation on the real data from WASP-43\,b observations are presented in Section \ref{sec:result}.

\onecolumngrid

\begin{figure}[H]
\centering
    \includegraphics[width = 0.8\textwidth]{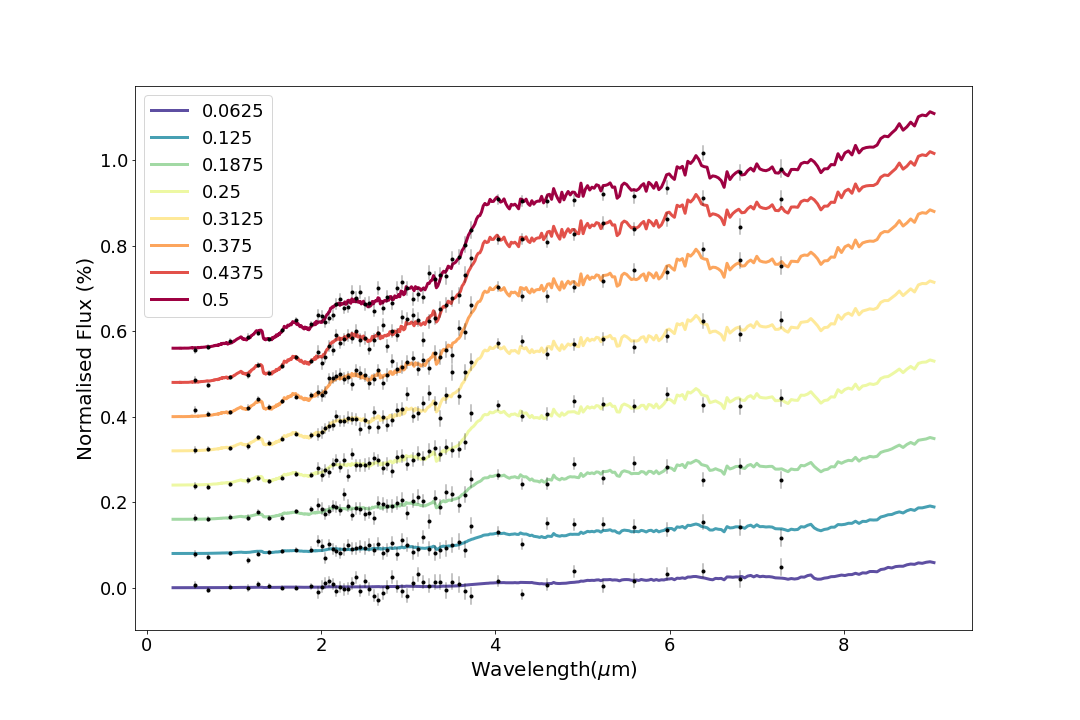}
    \includegraphics[width = 0.8\textwidth]{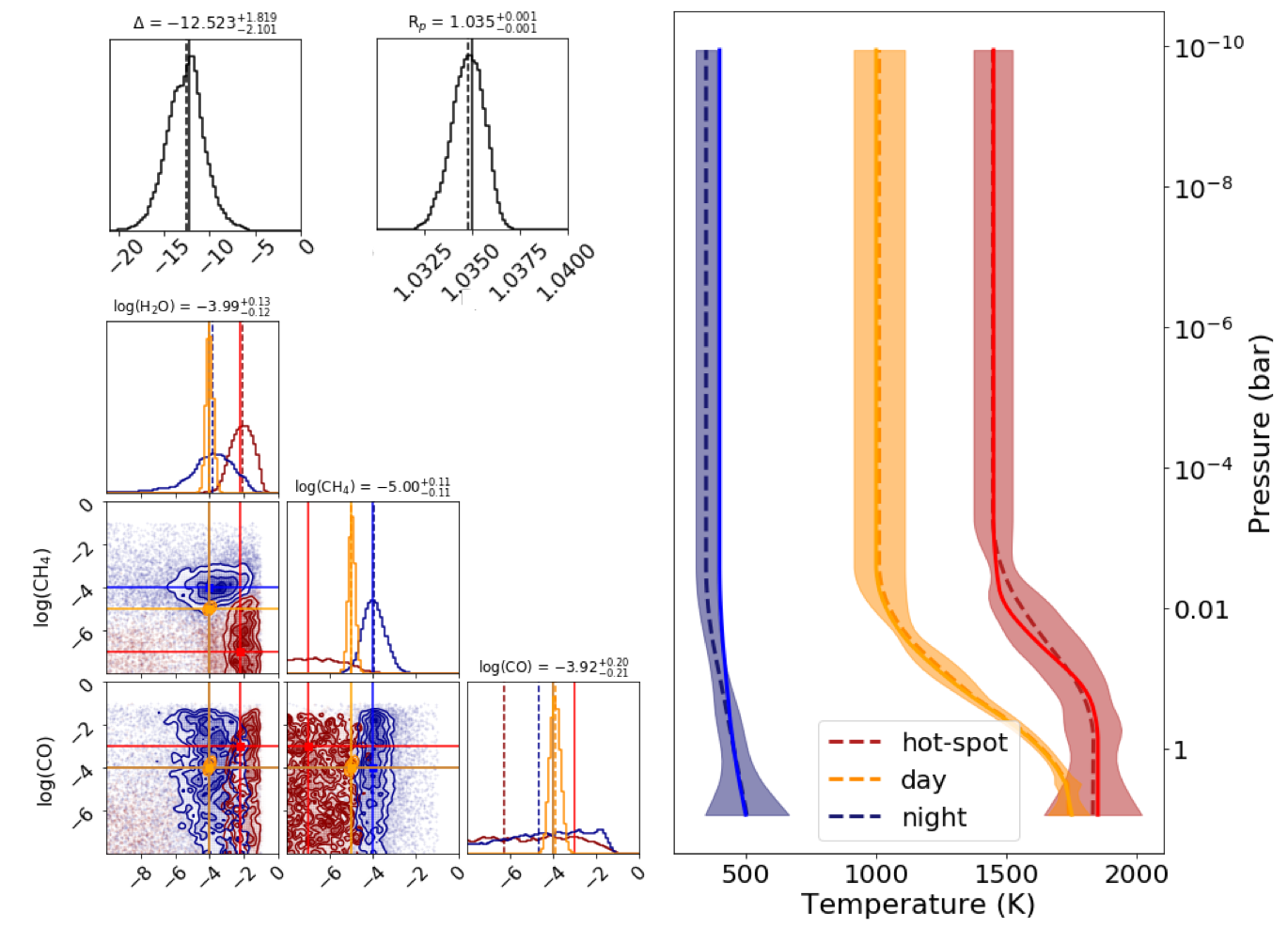}
    \caption{Results of our retrieval on the WASP-43\,b simulated spectra as observed with Ariel. Top: Observed and best fit spectra from phase 0.0625 to 0.5. The observed datapoints are randomly scattered in this figure for visual  were not randomly scattered for the retrieval. Bottom: Posterior distributions (left) and Temperature structure inferred by our retrieval. The solid lines corresponds to the true values. Red: hot-spot; Orange: day-side; Blue: Night-side.  }
    \label{fig:spectra_mock}
\end{figure}
\begin{figure}[H]
\centering
    \includegraphics[width = 0.90\textwidth]{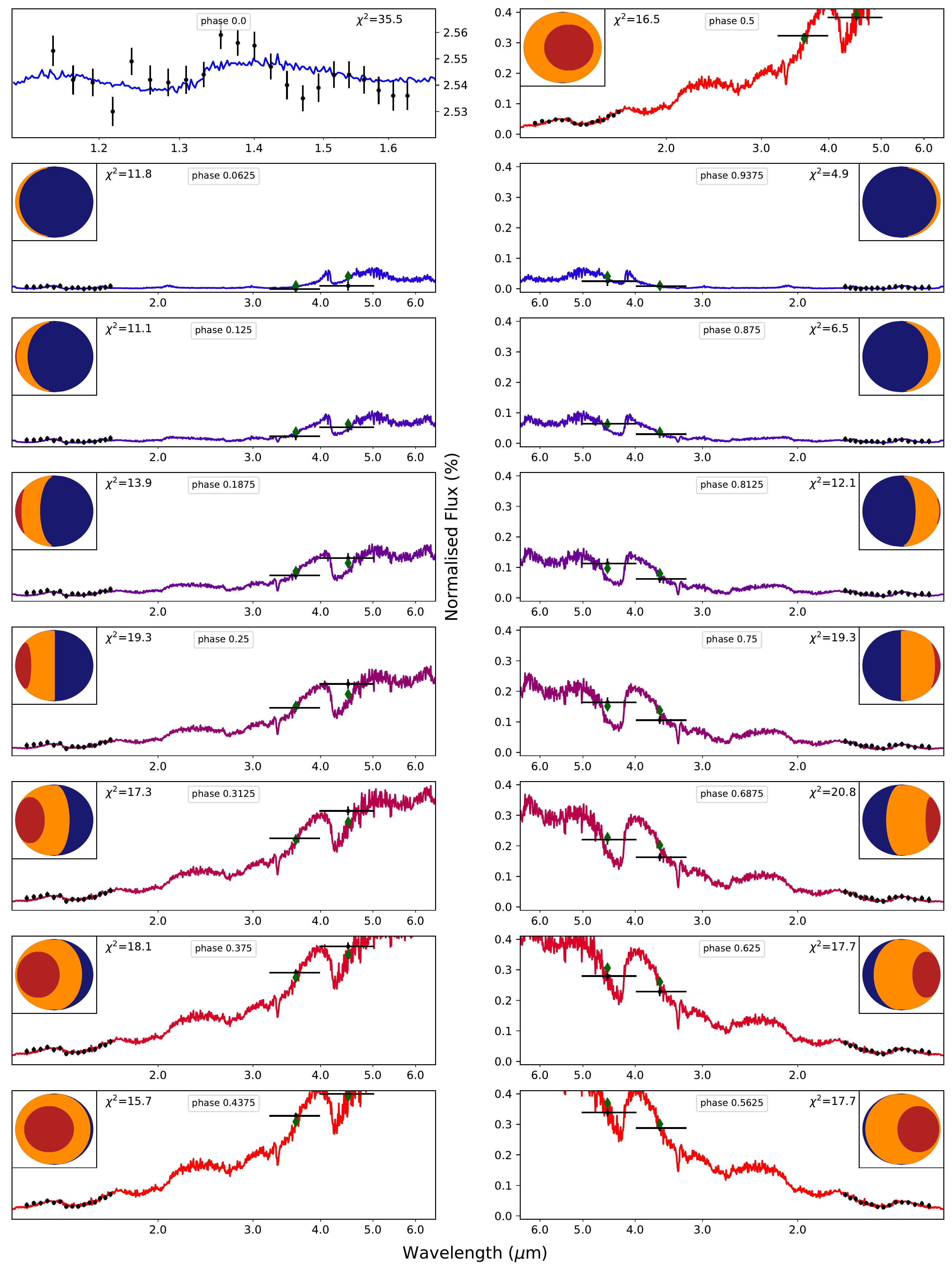}
    \caption{Best fit spectra and geometry of our WASP-43\,b phase-curve retrieval with the Full model. At the top, we show the eclipse (left) and the transit (right). Green diamonds represent the the averaged spitzer bandpasses. The right panels have inverted wavelength axis.}
    \label{fig:spectra}
\end{figure}

\twocolumngrid

\begin{table}\vspace*{1\baselineskip}
\centering
\begin{tabular}{cccc}
\hline\hline
Parameters & Prior bounds & Scale\\
\hline 
metallicity & -1 ; 3 & log  \\
C/O ratio & 0.1 ; 5 & linear  \\
free abundances & -12 ; -1 & log  \\
hot-spot temperature points (K)& 1000 ; 4000 & linear  \\
day temperature points (K)& 700 ; 3000 & linear  \\
night temperature points (K)& 300 ; 1700 & linear  \\
$P_{clouds}$ (Pa)& 7 ; 0 & log  \\
$R_p$ (R$_{jup}$)& 1 ; 1.1 & linear \\ \hline\hline
\end{tabular}
\caption{List of the parameters fitted in the retrieval, their uniform prior bounds and the scaling we used. Note that the parameters metallicity and C/O are activated for the equilibrium runs, while in the free chemistry run we use the free abundances parameters.}\label{tab:priors}
\end{table}

\section{Mock retrieval results}\label{sec:results_mock}

The results of the retrieval on the ad-hoc spectra are presented in Figure \ref{fig:spectra_mock}. The figure only shows the spectra for the phases from 0.0625 to 0.5 but the retrieval also included the phases from 0.5625 to 0.9375. As can be seen, the retrieval manages to produce a good fit of all the phases. The posterior distribution for the chemical species, displayed in the same figure, demonstrate that the retrieval is able to recover the correct abundances for water in all three regions. Methane is recovered on the day-side and the night-side but the molecule is not captured on the hot-spot, which has a lower input abundance of 10$^{-7}$. CO, which is present in high abundance on the hot-spot and the day-side was only retrieved from the day-side. CO possesses a single broadband feature in the Ariel wavelength range \citep{changeat2020_alfnoor} and is therefore more challenging to constrain. This lower signature, associated with the higher input abundance of H$_2$O and the smaller size of the hot-spot as compared to the rest of the day-side, could explain the difficulties of our retrieval to constrain this molecule on the hot-spot. We note that the hot-spot offset is properly recovered with very high accuracy, which indicates the theoretical capabilities of this class of models to directly constrain the geometrical properties of exoplanet atmospheres. For the temperature structure, the retrieval recovers the non-inverted thermal profiles with very good agreement (see Figure \ref{fig:spectra_mock}). We did not observe discrepant behaviour in this retrieval, validating capabilities of our model for retrieval applications. This example also shows the potential of the Ariel Space Telescope for phase curve retrieval studies. Ariel is expected to yield spectra for about 1000 exoplanets in transit and eclipse mainly, however, a significant amount of observing time will be dedicated for phase curves studies as part of the Tier 4 survey \citep{Tinetti_ariel,Edwards_2019}.

\section{Real data results} \label{sec:result}

In this section, we present the results from our unified retrievals on the real phase-curve spectra obtained from \cite{stevenson_w43_1,stevenson_w43_2}. The retrieval setups we use are described in Section \ref{sec:details_retrieval}. For comparison we complement this exploration with a more standard approach, applying our model on the individual spectra in Appendix \ref{apx:append_indiv_spectra}. We also provide some exploratory work on the required complexity when considering free temperature profiles in Appendix \ref{apx:tp_investigations}. The best fit spectra, obtained for each scenario, are compared in Appendix \ref{apx:spectra_compa}.

\subsection{On model comparison} \label{sec:model_compa}

For an easier comparison of our results, we provide the model integrated values in the Spitzer bands and the $\chi^2$ values corresponding to each spectrum. We note that $\chi^2$ values on individual spectra might be difficult to interpret for model comparison in a Bayesian context and should be viewed with caution \citep{edwards1963bayesian}. In particular, the dataset and parameters considered in this work are not expected to be independent, while the tested models are also non-linear \citep{andrae2010dos, gelman2013}. The stated $\chi^2$ value is calculated for the best fit model, which might not be representative of the larger pool of solutions found via our Bayesian inference technique. 
Within a Bayesian framework, the Bayesian evidence, log(E), is used for model comparison and accounts for both goodness of fit and model complexity \citep{rougier2020exact}. When comparing two models, a log difference of 3 indicates a strong preference for the model with highest evidence \citep{Kass1995bayes, jeffreys1998_bayesfactor}. When testing models of increasing complexity, their Bayesian evidence should also increase as long as the additional flexibility is justified. For models of adequate complexity, the Bayesian evidence should plateau around a maximum value. Finally, if the model overfits, the Bayesian evidence should decrease.

\subsection{2-Faces free: Benchmark retrieval} \label{sec:2F_free}

In this run (log(E) = 2238.1), we consider only two regions: day and night-sides. The geometry and the chemistry model (free chemistry) are similar to what has recently been presented in \cite{feng2020_2d}, which allows us for an easier comparison with their findings. A major difference in our model is the parametrisation of the temperature structure. Here, we choose a free temperature profile and we retrieve the temperature structure from the day and night-side independently. Indeed, alternative physics based descriptions, such as the parametrisation from \cite{Guillot_TP_model} assume a 1-dimensional atmosphere experiencing irradiation from 2-sources (upward and downward fluxes). While this well describes a planet day-side with poor atmospheric redistribution, this geometry does not accurately represent the night-side of a tidally locked planet, which are by definition not receiving direct stellar flux. For all models in this paper, except when stated otherwise, we therefore opted for a free parametrisation. The best fit spectra, the geometry and the posteriors for our 2-Faces free chemistry model are described in Appendix \ref{apx:free_scn} (blue runs). These highlights the difficulty of the model to fit the Spitzer points around the secondary eclipse. This is most likely due to the symmetry enforced in this type of geometry. The temperature profile for the day-side mainly decreases with altitude, which is consistent with previous studies \citep{stevenson_w43_1,stevenson_w43_2, Kataria_2015_W43GCM, Irwin_w43b_phase, feng2020_2d}, but we note the preference (large 1$\sigma$ uncertainties) for a thermal inversion above 10$^3$ Pa. We ran complementary retrievals, parametrising the temperature profile with the prescription from \cite{Guillot_TP_model} and found a non-inverted temperature profile  and chemistry (orange runs in Appendix \ref{apx:free_scn}) similar to the findings in \cite{feng2020_2d}. This run, however, led to a particularly low Bayesian evidence of log(E) = 2045.2, potentially highlighting the lack of flexibility in the Guillot profile. We discuss the presence of thermal inversions for this planet later, in the section describing the results of the Full model scenario. On the night-side, the temperature is poorly constrained due to the low flux received at these phases. We note that the model prefers high altitude clouds. In terms of chemical species, this retrieval finds very constrained abundances for water and ammonia (see posteriors). These match the abundances found by the joint-scenario in \cite{feng2020_2d}. As opposed to their results, the free T-p profile run does not recover constraints on CO$_2$, which can be explained by our use of the more flexible temperature structure. Indeed, when the Guillot profile was used, we recovered similar abundance for CO$_2$ (see orange posteriors in Appendix \ref{apx:free_scn}). This retrieval exploration using the 2-Faces free chemistry model confirms the findings presented in \cite{stevenson_w43_2,Irwin_w43b_phase,feng2020_2d} but also highlight different solutions when using more flexible temperature structures (also see Appendix \ref{apx:tp_investigations}). We also ran a complementary retrieval where the chemistry is de-coupled between the day and night-side. This retrieval led to two solutions corresponding to the green (log(E) = 2242.3) and grey (log(E) = 2240.4) runs in Appendix \ref{apx:free_scn}. Since the grey run corresponds to an unphysical high mean molecular weight solution, we focus on the green solution. Due to large day/night  temperature contrast, differences in the chemistry can be expected. This retrieval presents a very different chemistry to the previous models with a much higher water content on the day-side (around 10$^{-2}$) and a much more complex temperature structure. The chemistry recovered here is similar to the ones presented in the next sections using de-coupled equilibrium chemistry models (2-Faces equilibrium and full models). On the night-side, we do not recover any molecule.

\subsection{2-Faces equilibrium: Equilibrium chemistry retrieval}\label{sec:2F_eq}

One major assumption, taken in the previous scenario, is the constant chemistry with altitude and longitude. Previous studies showed that atmospheric chemistry is not constant and that assuming so could lead to observable biases \citep{Agundez_2dchemical_HD209_HD189, venot_chem_HJ,Drummond_2018, Woitke_2018, Stock_2018, Venot_2020, Drummond_2020, changeat2020_alfnoor}. In order to provide a more realistic description of chemical properties, one could either assume more complex parametrisations \citep{ Parmentier_2018,Changeat_2019_2l} or use self consistent chemical models \citep{Woitke_2018, Agundez_2dchemical_HD209_HD189, Stock_2018}. While both approaches are viable with high quality data, the low wavelength coverage of HST favor the more constrained models. We therefore assumed equilibrium chemistry for all three regions using the scheme from \cite{Agundez_2dchemical_HD209_HD189}. Since we assume the atomic elements are evenly spread across the planet atmosphere, this leaves us with only two free parameters: metallicity and C/O ratio. The impacts of these assumptions are explored more in the Discussion section. We present the best fit spectra, chemical profiles, temperature structure and posteriors in Appendix \ref{apx:2f_eq}. As compared with the previous run, the use of equilibrium chemistry allows to better fit the Spitzer photomotric points (log(E) = 2245.8). In this scenario, the retrieved chemistry is well constrained (see posterior distribution). The atmosphere is consistent with a slightly supersolar metallicity (0.76$^{+0.1}_{-0.12}$) and a low C/O ratio (0.25$^{+0.07}_{-0.06}$). In particular, the use of the chemical scheme allows large variations in the abundances of carbon bearing species between the day and night-side. There are linked to the large retrieved temperature differences between the two regions (from 2500K down to 500K) and are matching the findings in our free chemistry run with decoupled chemistry (green run in Appendix \ref{apx:free_scn}). As compared to the 2-Faces free run, an interesting difference is the retrieved abundance for water. Here, we find that the abundance for water remain constant between the two regions (as expected from thermochemical equilibrium) but that it might be 2 orders of magnitude higher than the values found in the free chemistry case (log(H$_2$O) = -4.4). In fact, we find that a very similar solution (with high water abundance) can also be recovered from a free chemistry retrieval when the abundances between the day and night-side regions are de-coupled, clearly indicating the need to account for chemical variations between those two regions. In terms of temperature structure, we find a day-side temperature profile that mainly decreases with altitude and the presence of a thermosphere, similar to the free chemistry run. The night-side is much colder and include high altitude clouds at pressures as high as 10 Pa.

\subsection{Full retrieval}\label{sec:full}

The previous models have difficulties to fit the spectra for the phases near eclipse. This is due to the symmetry imposed when considering only 2-Faces. To increase to complexity, we split the geometry in 3 distinct regions: hot-spot, day-side and night-side. For the run presented here, the model has a fixed hot-spot size of 40 degrees. We also model the planet without including clouds on the hot-spot (the cloud top pressure was fixed at 10$^6$ Pa). This choice is justified by theoretical predictions, which suggests that the hot-day-side of irradiated exoplanets might be cleared due to the strong stellar irradiation \citep{2016_lee, Parmentier_2016,Lines_2018, Helling_2019}. The best fit spectra for this scenario and the corresponding geometries are shown in Figure \ref{fig:spectra}. For this retrieval, we obtain a log evidence of 2277.4, which indicate the relevance of adding this additional region to the model.

We also investigated the same run on the HST spectra only. Both HST+Spitzer and HST only runs are presented in Appendix \ref{apx:complem_full}, which provides a zoomed version of the same run around the HST wavelength range. The HST+Spitzer run is colored while the HST only run is grey. Both fits (with and without the Spitzer photometric points) do not show major differences in the HST wavelengths range and are well fitted to the observed data. Outside the HST wavelength coverage, however, we see that the inclusion of the Spitzer points is leading to large differences in the retrievals. 

\begin{figure*}
\begin{center}
    \includegraphics[width = 0.9\textwidth]{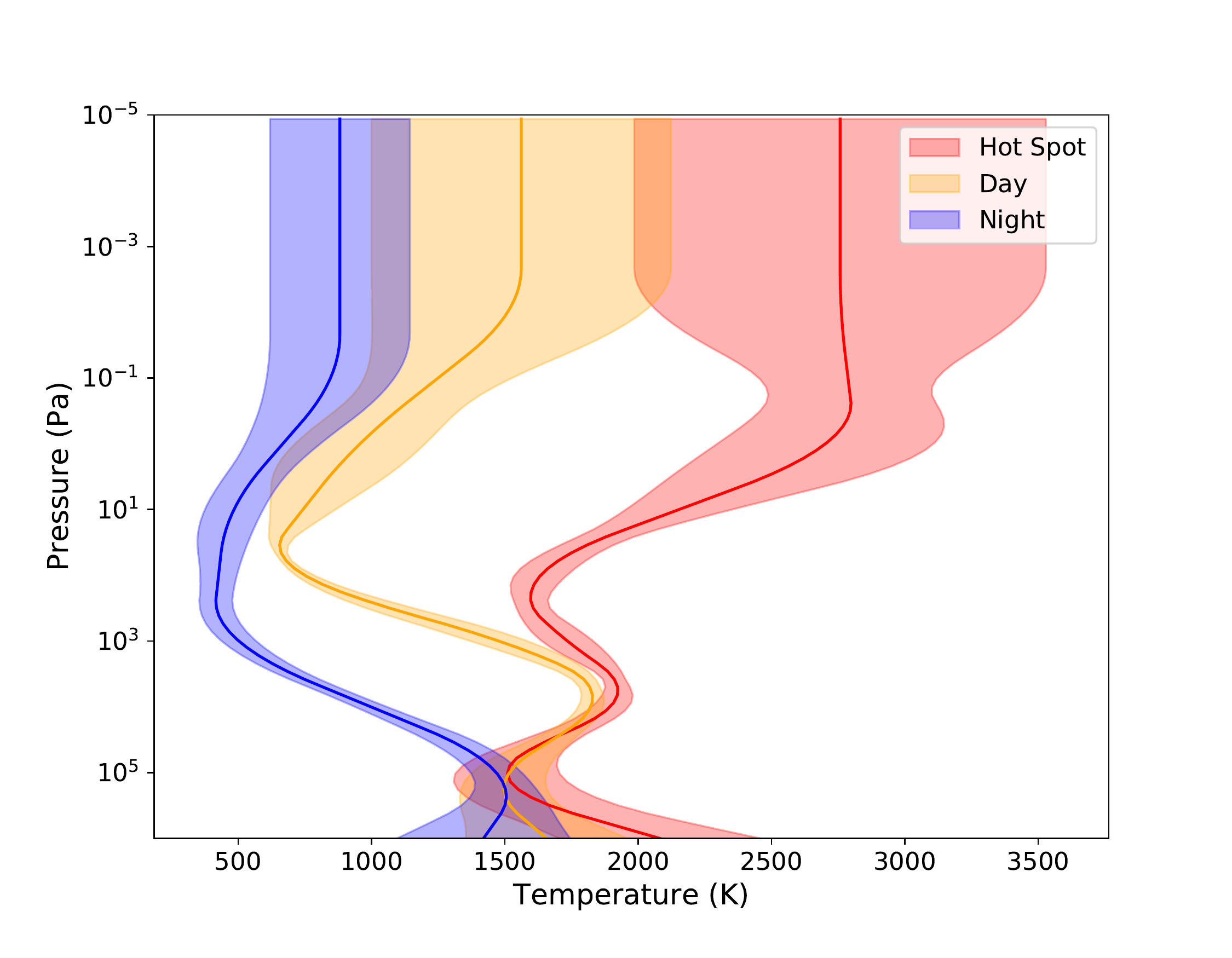}
    \caption{Retrieved mean and 1$\sigma$ temperature structure of WASP-43\,b for the different regions. Hot spot: red; day-side: orange; night-side: blue. The hot-spot presents a significantly higher temperature, especially at the top of the atmosphere.}
    \label{fig:temperature}
\end{center}
\end{figure*}

The temperature structure associated with the HST+Spitzer retrieval is presented in Figure \ref{fig:temperature}, which as expected, clearly displays a hotter day-side and a more strongly irradiated hot-spot.

As in the free temperature 2-Faces runs presented in previous sections, the temperature structure is consistent with the presence of thermal inversions for this planet. This constrasts with previous findings from \cite{Blecic_2014, stevenson_w43_1, Kreidberg_2014}. The retrieved temperature structure indicates the presence of a stratosphere (above 10$^{5}$ Pa) and the presence of a thermosphere (above 10 Pa) on the planet's day-side and hot-spot. We note that the current data does not allow us to probe the highest altitudes (thermosphere) with great accuracy, leading to large 1$\sigma$ uncertainties on the retrieved profile (around $\pm$600K). This is also shown in the contribution functions for each regions, which are plotted in Appendix \ref{apx:contrib_full}. In the full model and for most of the temperature structures explored in Appendix \ref{apx:tp_investigations}, the hot-spot and day-side present a thermal inversion at high altitude, below 1 Pa. This drives the thermal dissociation of all molecules at those altitudes, making the atmosphere transparent. We discuss further the interplay between this high altitude thermal inversion and our chemical model assumptions in the discussion section. For this planet, thermal inversions are predicted \citep{Kataria_2015_W43GCM, Showman_2009} in presence of optical absorbers such as atomic species (K, Na) or metal hydrides and oxides (e.g. AlO, FeH, SiO, TiO, TiH or VO). The presence of aluminium oxide (AlO) in the transmission spectrum of WASP-43\,b has recently been highlighted in \cite{chubb2020aluminium}. This detection, made at the terminator region, is not expected from equilibrium chemistry models at the temperatures and pressures considered \citep{chubb2020aluminium, helling_2020_mineral}. \cite{chubb2020aluminium} highlighted that the presence of this molecule could come from disequilibrium or dynamical processes. The higher temperature on the day-side allows for this molecule and other metal oxides/hydrides in a gaseous form \citep{Woitke_2018}, which could then be transported to the terminator regions \citep{caldas_3deffects,Pluriel_2020}. Earlier ground-based observations of the transit \citep{Chen_2014} were also consistent with optical absorption, which were interpreted as pure Rayleigh scattering or absorption from K/Na or TiO/VO. Their observations of the eclipse suggested poor day-night contrast, as also suggested in our results, and potential high altitude emission on the planet day-side. 

From a theoretical perspective, recent studies have also shown that thermal inversions could occur naturally in the upper atmosphere of hot-Jupiters \citep{Lothringer_2018_invert, lavvas2021impact}. For example, the day-side of irradiated hot-Jupiters might be significantly heated by photochemical hazes and/or  dissociation of the main molecules and the addition of continuum opacity from negative hydrogen (H-). In addition, local thermal inversions are also predicted to occur if sulfur bearing species are present \citep{lavvas2021impact}. While the investigations in \cite{Lothringer_2018_invert} were performed on a F-type star with a larger UV flux than WASP-43 (K-type), their fiducial hot-Jupiter simulations showed that planets at distances from their host star similar to WASP-43\,b (0.1 AU) and with equilibrium temperatures around 1500 K, could display inversions above 100 Pa, regardless of their TiO/VO content. The results presented here might provide a strong observational evidence in favour of these theoretical predictions.

\begin{figure}
\begin{center}
    \includegraphics[width = 0.48\textwidth]{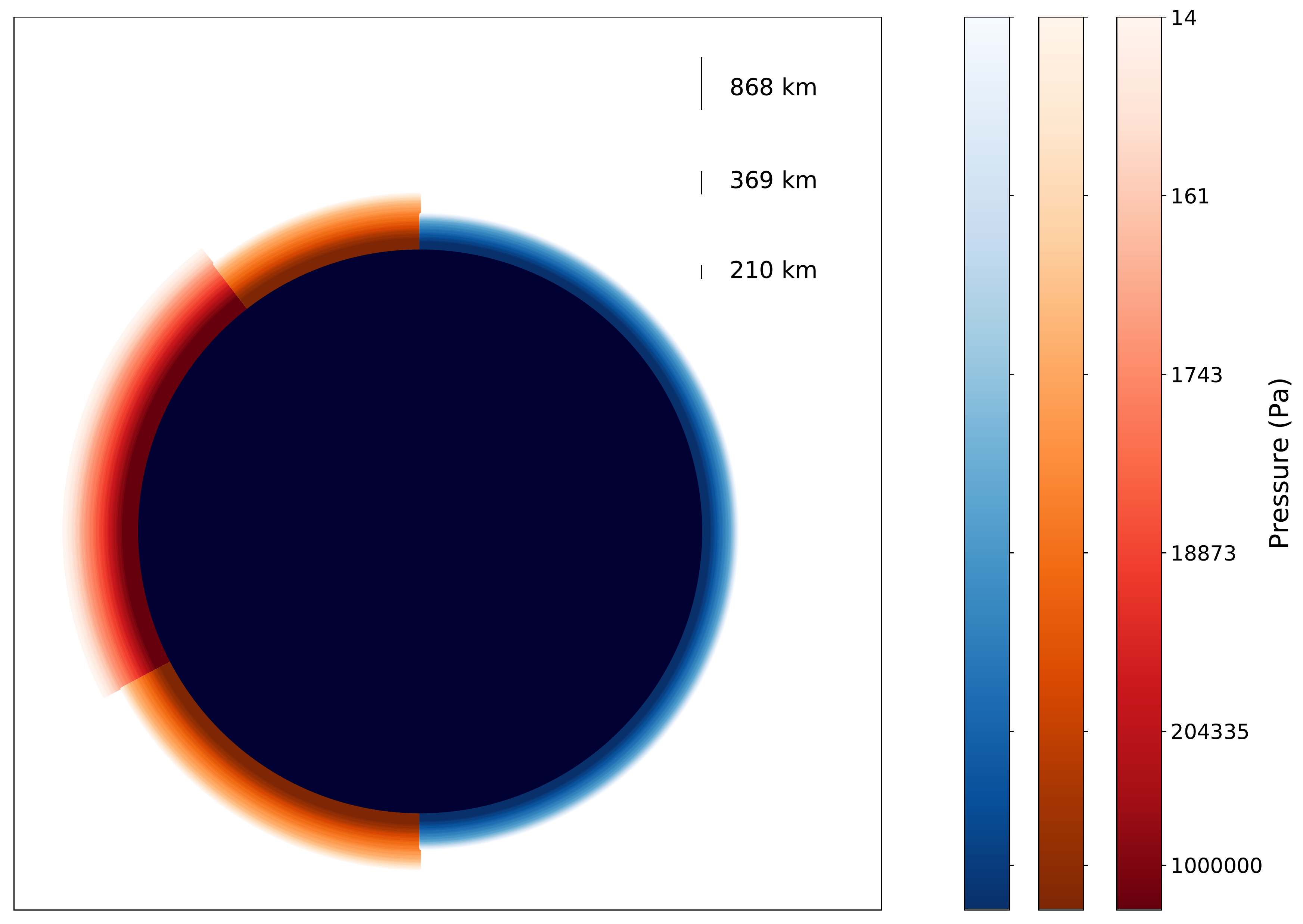}
    \caption{Retrieved atmospheric structure of WASP-43\,b. The red region is the hot-spot (shifted by 12.2 degrees); The orange region is the day-side; The blue region is the night-side. The legend corresponds to the altitude at 5 scale heights. We notice that the hotter temperatures on the day-side lead to a significantly larger atmosphere.}
    \label{fig:structure}
\end{center}
\end{figure}

In Figure \ref{fig:structure}, we also show the atmospheric structure derived from our model. The planet presents an inflated day-side (and hot-spot) with a large scale height difference between the day and the night regions. This conclusion also holds for the models with only 2-Faces, strongly suggesting that using 1-dimensional transmission models to analyse transit spectra might lead to large biases for these types of planets as their current formulations lack the flexibility to properly represent such 3-dimensional effects. A recent study \citep{skaf_2020_ares} already found observational evidences of these effects in Hubble transmission spectra and, as next generation telescopes will be more and more precise, this indicates the importance to consider complementary phase curve studies. For irradiated targets, phase curves will bring additional constraints which will allow us to break these 3-dimensional degeneracies. On the night-side, we retrieve a large temperature decrease with altitude, which can be inferred from the very low emission of the spectra at these phases. Similar results were already highlighted in \cite{stevenson_w43_1, stevenson_w43_2, Irwin_w43b_phase, feng2020_2d}. 

\begin{figure*}
\begin{center}
\begin{minipage}{0.8\textwidth}
    \includegraphics[width = 0.49\textwidth]{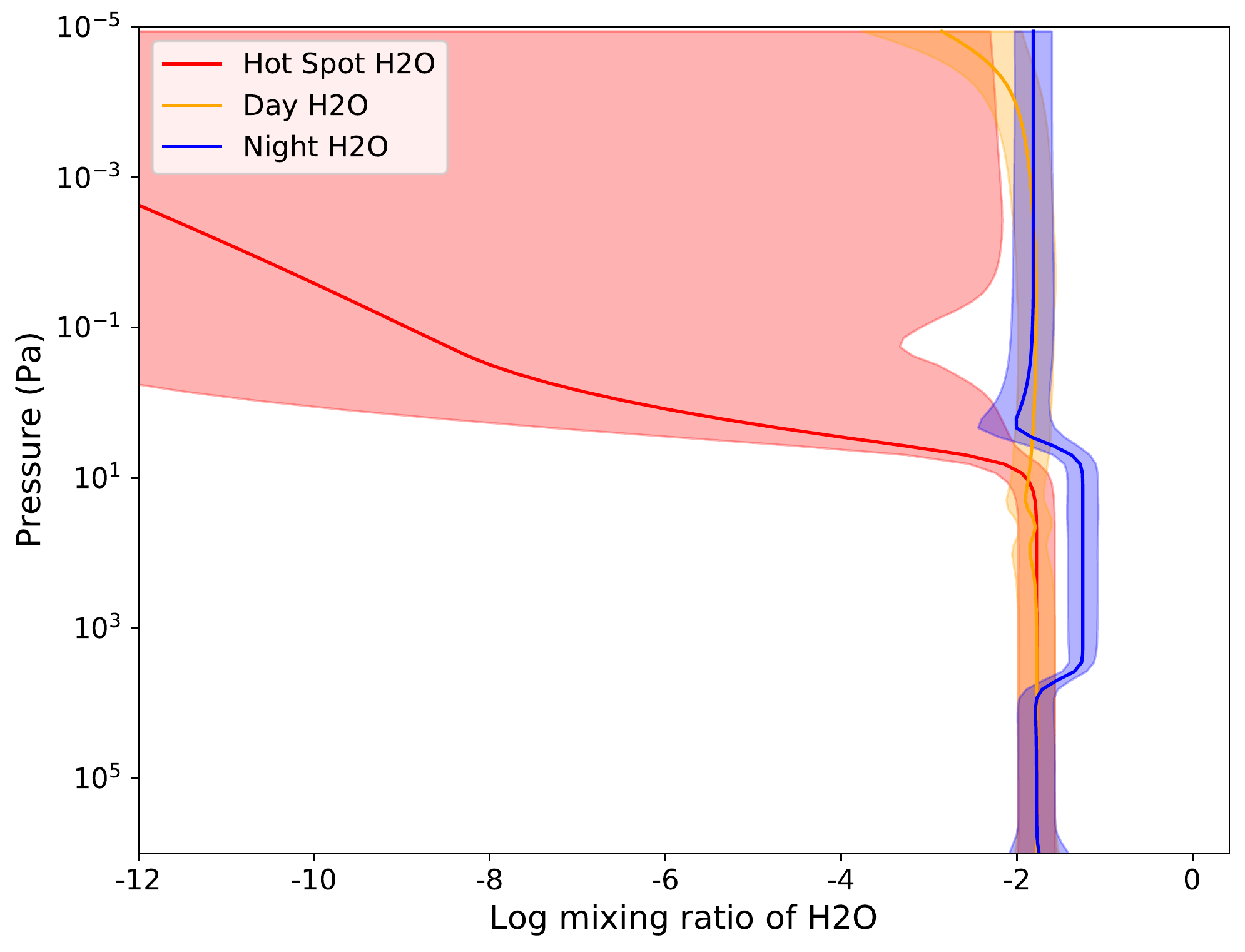}
    \includegraphics[width = 0.49\textwidth]{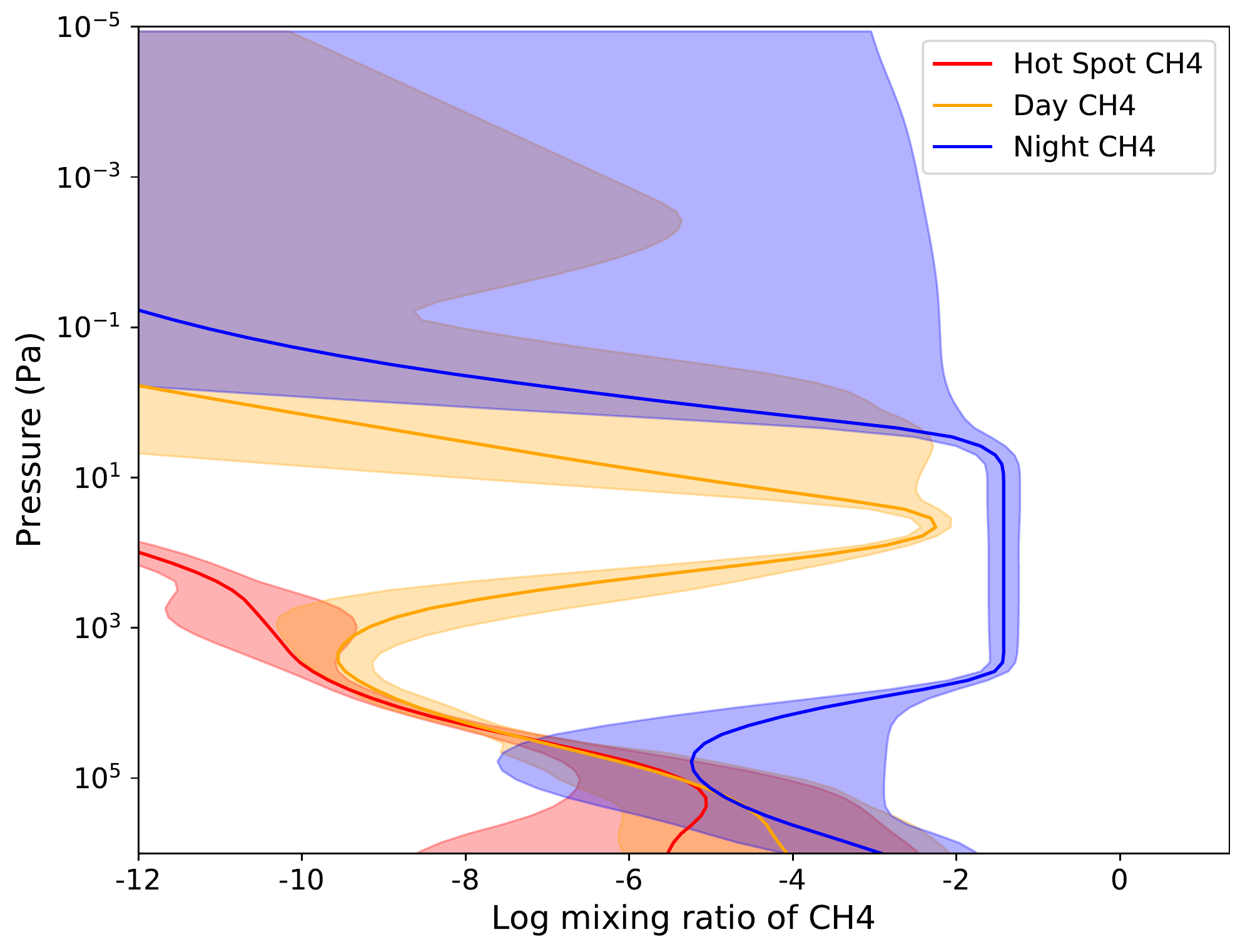} \\
    \includegraphics[width = 0.49\textwidth]{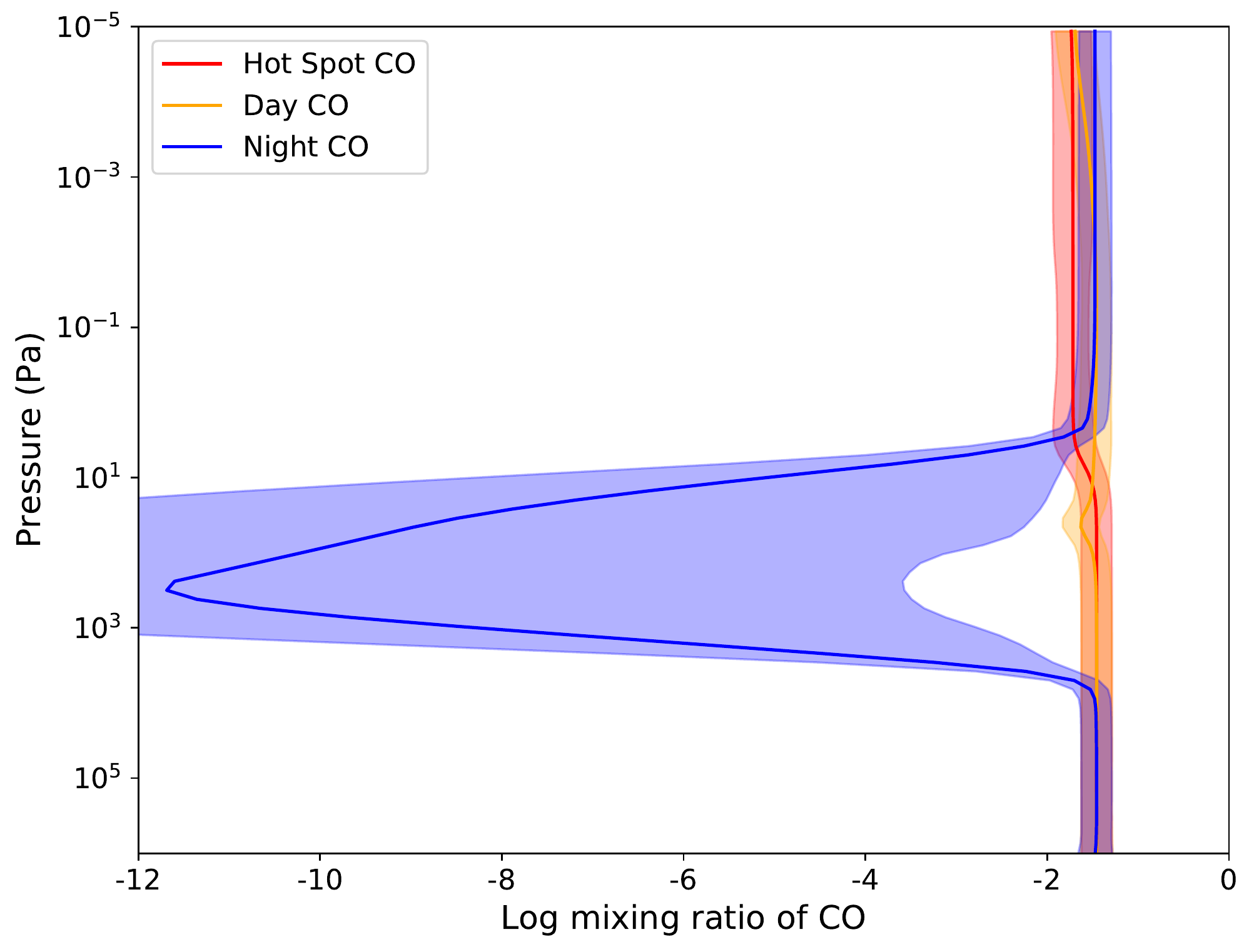}
    \includegraphics[width = 0.49\textwidth]{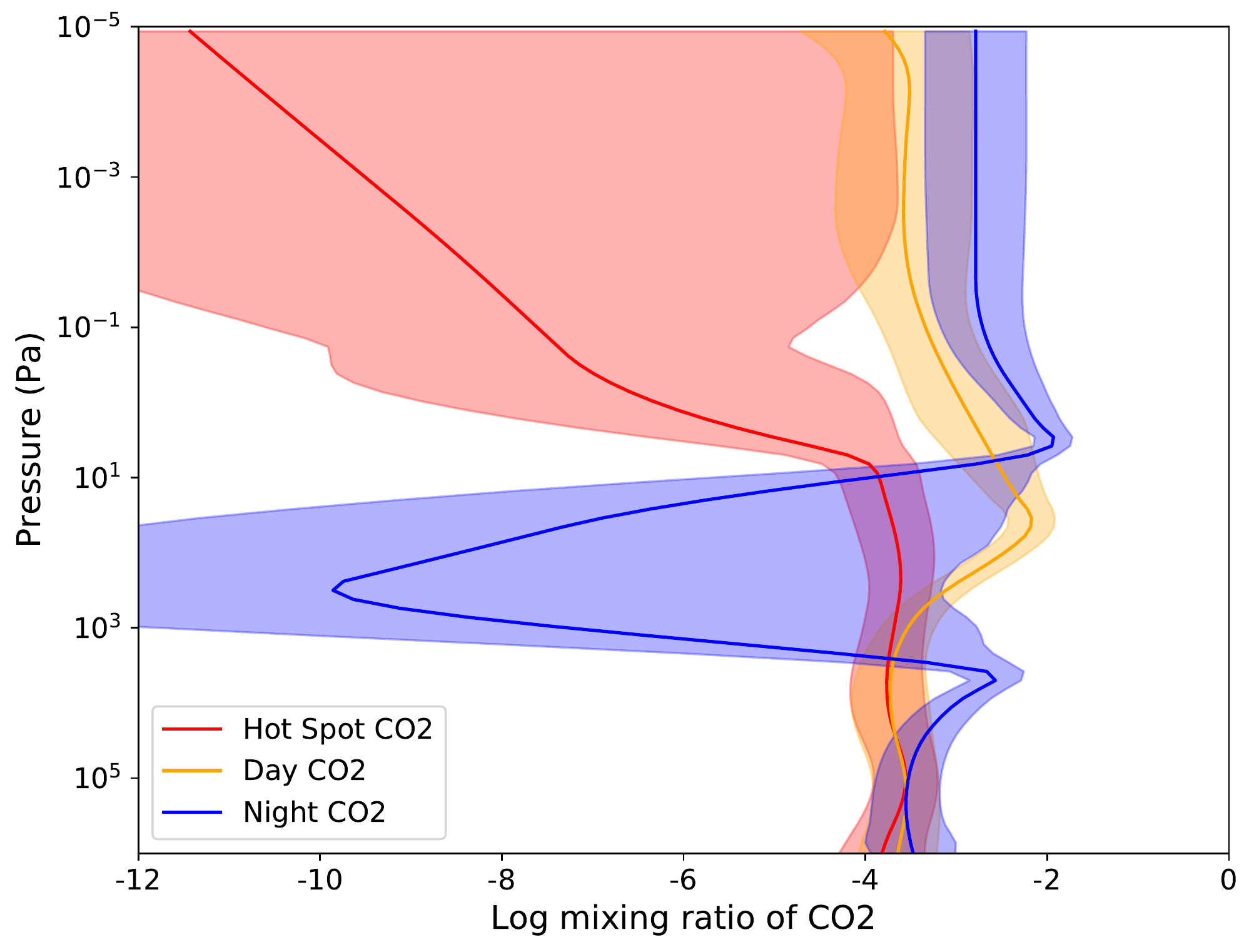} \\
    \includegraphics[width = 0.49\textwidth]{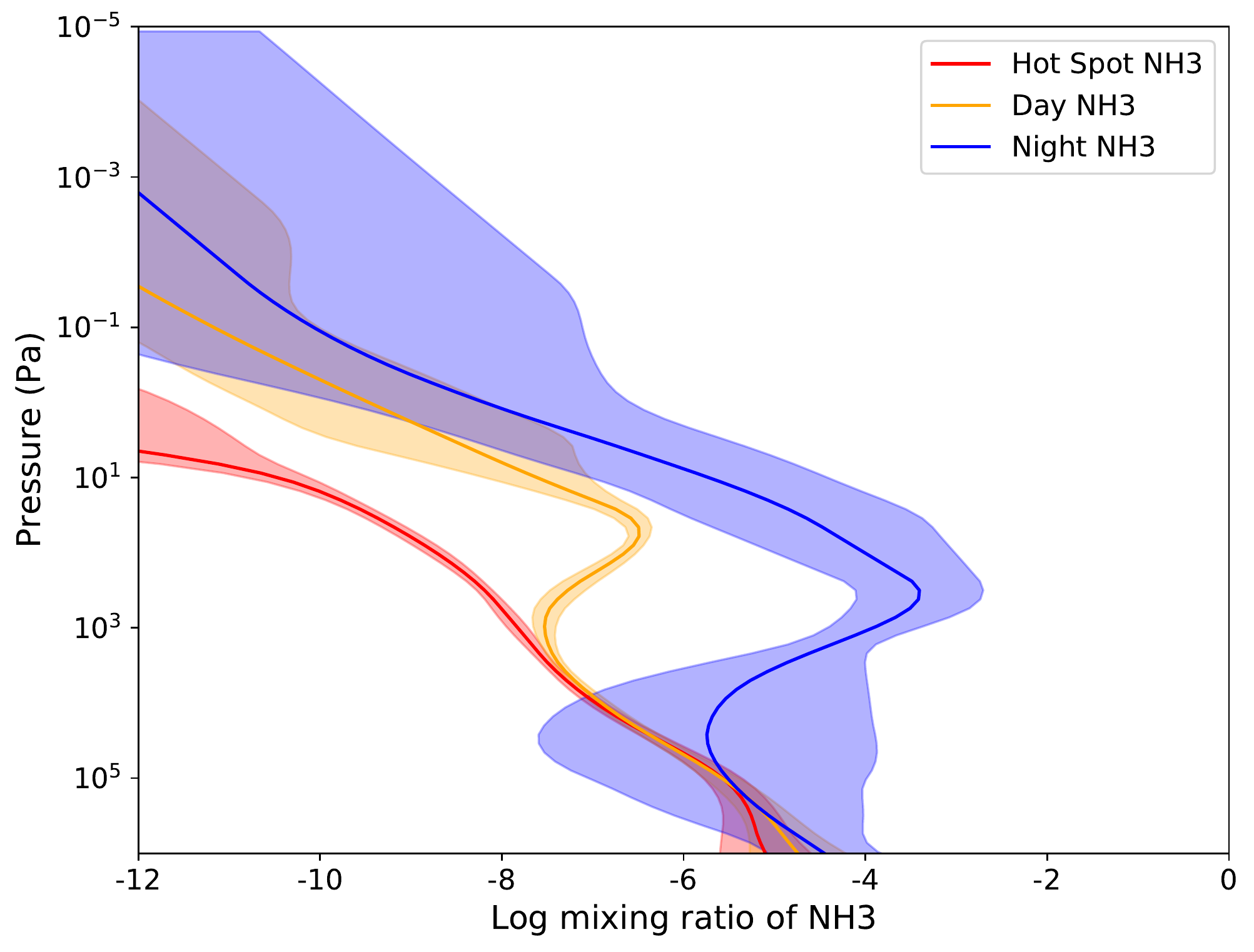}
    \includegraphics[width = 0.49\textwidth]{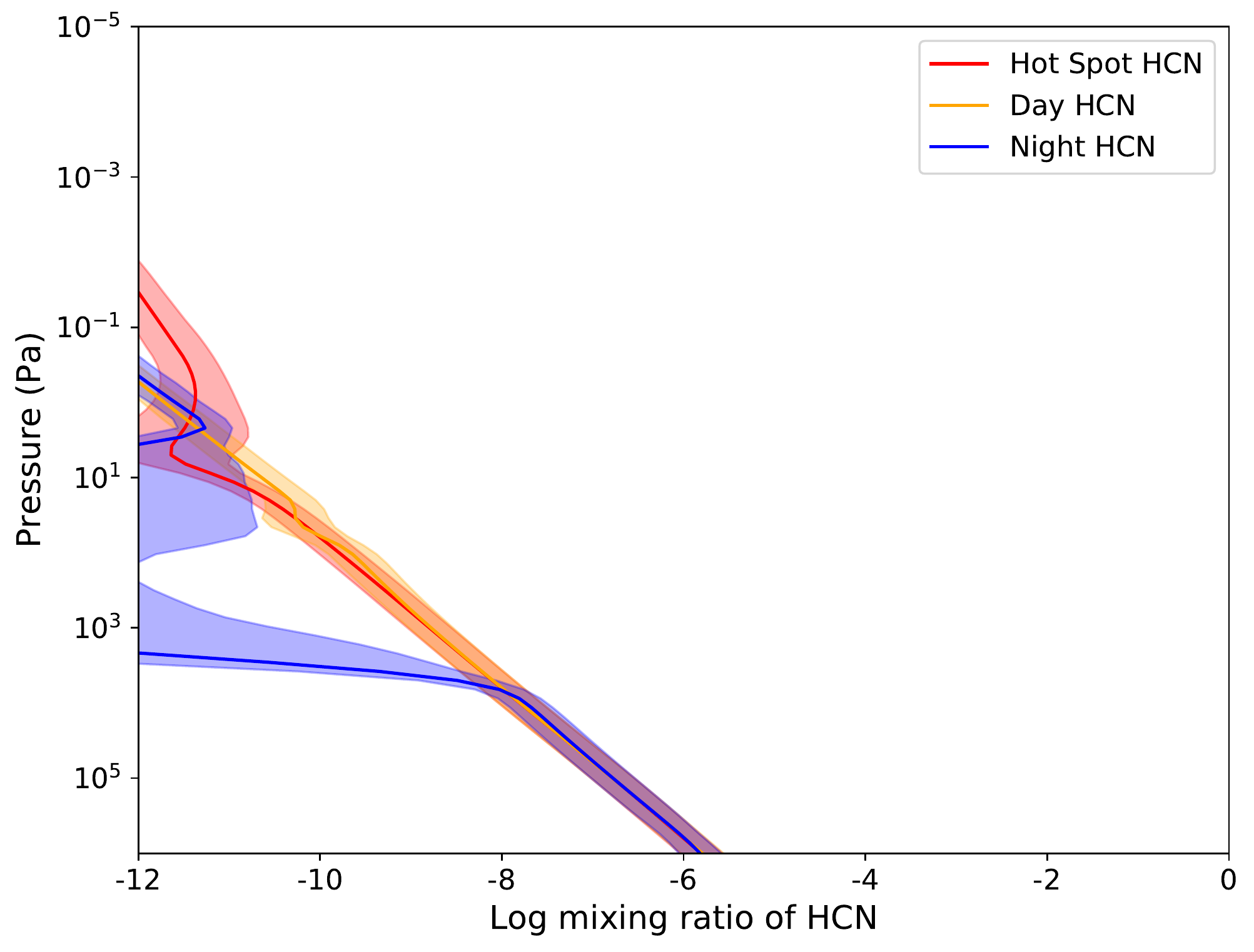}
    \caption{Molecular abundances in the different regions according to the chemical equilibrium scheme. Red: hot-spot; Orange: day-side; Blue: night-side.}
    \label{fig:abundances}
\end{minipage}
\end{center}
\end{figure*}

The abundance profiles predicted by our equilibrium chemistry scheme for this planet are shown in Figure \ref{fig:abundances}. The retrieved metallicity (1.81$^{+0.19}_{-0.17}$) and C/O ratio (0.68$^{+0.11}_{-0.12}$) for this atmosphere, shown in the posterior distribution in Figure \ref{fig:posteriors}, are particularly well constrained. These correspond to a slightly carbon enriched atmosphere with a supersolar metallicity, which contrast with the previous 2-Faces equilibrum retrieval. For WASP-43\,b, Global Climate Models (GCM) have shown that supersolar metallicities provide a better fit to the \cite{stevenson_w43_2} data  \citep{Kataria_2015_W43GCM}. Their work however, explored lower metallicities than considered in our study. They also highlighted that the presence of night-side clouds provides a good explaination of the observed data. Overall, we find that water remains present in all regions of the atmosphere and that its abundance is consistent with our previous 2-Faces equilibrium run.  Carbon converts from mainly CO on the day-side to CH$_4$ on the night-side, with an overall higher abundance of these species as compared with the previous runs. We note that NH$_3$, which will be a very strong absorber in the spectra from the next generation telescopes, such as JWST \citep{Greene_2016}, Twinkle \cite{Edwards_twinkle} and Ariel \citep{Tinetti_ariel}, might reach detectable abundances on the night-side of the planet.

In terms of cloud properties, our full scenario did not recover evidence of opaque absorbing layers on the night-side. Although clouds have been suggested as an explanation for the very low temperature of the night-side, we highlight the fact that all spectra, even at low orbital phases, are showing some prominent water absorption features at 1.4 $\mu$m which explains why clear atmosphere solutions are favoured when enough flexibility is given to the model. While we did not detect direct evidence of night-side clouds, these are not ruled out and since the signal on the night-side phases is lower, the retrieval does not inform us about clouds at pressures higher than $10^4$ Pa. 



Overall, when compared to our previous runs without hot-spot and/or the results from \cite{stevenson_w43_2,Irwin_w43b_phase,feng2020_2d}, we obtain a very different picture. When a hot-spot is added, our retrieval finds that thermal inversions (stratosphere and thermosphere) with a rich chemistry provides the best fit to the combined phase-curve spectra.


\subsection{Section summary}

In the 3 scenarios dealing with read data, we recovered particularly well constrained solutions. By considering all spectra together, the information content contained in our analysis is increased as compared to traditional single spectrum retrievals. This can be seen by comparing the results to our phase curve retrievals on individual spectra (see Appendix \ref{apx:append_indiv_spectra} and Appendix \ref{apx:tp_investigations}) and the results from this section with our unified exploration.
This allows us to extract more complex information as compared to previous studies. In all three cases, we recover a decent fit of the spectra. This is showcased in Appendix \ref{apx:spectra_compa}, where we compare the best fit spectra of our 3 scenarios. We note the increase in complexity between the different scenarios allow to better reflect the observed data, as shown by the increase in log(E), especially in the Spitzer region. For example, the inclusion of the hot-spot with a shift of -12.2 $^{\circ}$ in the Full model improved the capability to fit the slightly higher flux observed from phases 0.0625 to 0.4375 (See also Figure \ref{fig:spectra}). This is expected as the raw phase curve data from WASP-43\,b exhibit day-night contrast and asymmetric emission, features that require a minimum of 3 regions to be properly described. Interesting different interpretations of the data can be made from the solutions recovered by our 3 scenarios. 

More precisely, our 2-Faces free chemistry retrieval indicates similar results to previous studies for this planet \citep{stevenson_w43_2, feng2020_2d} but it also highlights the need for an increased complexity, matching the information content in the spectra. We find a hot day-side with a large temperature decrease and a cool night-side. The retrieved chemistry is also fairly similar, with a relatively low abundance of water. When we introduce variations with altitude and longitude in the chemistry (2-Faces model with equilibrium chemistry), the retrieval prefers models with higher water abundances (2 orders of magnitude larger). This behaviour drives a supersolar metallicity and C/O ratio. In our most complex scenario, when a hot-spot is added, the temperature profile displays a stratosphere with a thermal inversion around 10$^3$ Pa on the day-side and hot-spot. This is associated to a different chemistry with a supersolar metallicity and a slightly higher C/O ratio. The solution also appears stable to changes in the model parameters such as clouds and hot-spot size. In all three scenarios, the best fit T-p profiles on the day-side are consistent with a thermosphere, a high altitude thermal inversion. In essence, these tests demonstrate the potential of phase curve data in describing 3-dimensional effects and the feasibility to extract these information using retrieval frameworks. This also clearly shows the model dependence of data analysis techniques and the need to use models of adapted complexity. 

\begin{table*} 
\centering
\resizebox{0.97\textwidth}{!}{%
\begin{tabular}{|l|l|l|l|l|l|}

\hline
N & Retrievals & Assumptions & Highlighted results & log(E) & \S \\ \hline \hline

\multirow{3}{*}{1} & \multirow{3}{*}{2-Faces free T-p (C)} & C free chemistry & low H$_2$O abundance and NH$_3$ consistent with literature & \multirow{3}{*}{2238.1} & \multirow{3}{*}{\ref{sec:2F_free}} \\
& & DC free T-p & Mostly non-inverted T-p with a potential thermal inversion & & \\
& & DC grey clouds & night-side clouds & & \\
\hline

\multirow{3}{*}{2} & \multirow{3}{*}{2-Faces Guillot T-p (C)} & C free chemistry & low H$_2$O abundance, CO$_2$ and NH$_3$ consistent with \cite{feng2020_2d} & \multirow{3}{*}{2045.2} & \multirow{3}{*}{\ref{sec:2F_free}} \\
& & DC Guillot T-p & non-inverted T-p consistent with \cite{feng2020_2d} & & \\
& & DC grey clouds & No clouds - these were not included in \cite{feng2020_2d} & & \\
\hline

\multirow{3}{*}{3} & \multirow{3}{*}{2-Faces free T-p (DC)} & DC free chemistry & high H$_2$O, CH$_4$ and CO$_2$ abundances (not consistent with 1 and 2) & \multirow{3}{*}{2242.3} & \multirow{3}{*}{\ref{sec:2F_free}} \\
& & DC free T-p & Presence of a stratosphere & & \\
& & DC grey clouds & high-altitude night-side clouds & & \\

\hline

\multirow{3}{*}{4} & \multirow{3}{*}{2-Faces equilibrium} & C eq chemistry & high abundances for H$_2$O, CH$_4$, CO and CO$_2$ (similar to retrieval 3) & \multirow{3}{*}{2245.8} & \multirow{3}{*}{\ref{sec:2F_eq}} \\
& & DC free T-p & mostly decreasing, presence of a thermosphere & & \\
& & DC grey clouds & day and night-side clouds &  &\\

\hline

\multirow{3}{*}{5} & \multirow{3}{*}{Full 40$^{\circ}$} & C eq chemistry & supersolar metallicity & \multirow{3}{*}{2277.4} & \multirow{3}{*}{\ref{sec:full}} \\
& & DC free T-p & presence of stratosphere and thermosphere on day-side & & \\
& & DC but clear HS & no clouds & &\\
\hline

\multirow{3}{*}{6} & \multirow{3}{*}{Full (other HS sizes)} & C eq chemistry & supersolar metallicity (very slight changes from 5) & \multirow{3}{*}{2278.2} & \multirow{3}{*}{\ref{sec:hs_size}} \\
& & DC free T-p & same as 40$^{\circ}$, with a T-p increase for smaller HS & & \\
& & DC but clear HS & no changes in clouds from 5 & &\\

\hline

\multirow{3}{*}{7} & \multirow{3}{*}{Full 40$^{\circ}$ (HS clouds)} & C eq chemistry & supersolar metallicity (same as 5) & \multirow{3}{*}{2277.3} & \multirow{3}{*}{\ref{sec:hs_clouds}} \\
& & DC free T-p & presence of stratosphere and thermosphere on day-side & & \\
& & DC allowed on HS & no clouds day and night-sides but possible on HS & & \\
\hline

\multirow{3}{*}{8} & \multirow{3}{*}{Full 40$^{\circ}$ (HST only)} & C eq chemistry & very high metallicity (nonphysical) & \multirow{3}{*}{2101.6} & \multirow{3}{*}{\ref{sec:full_hst}} \\
& & DC free T-p & presence of thermosphere. Larger T-p uncertainties from 5 & & \\
& & DC but clear HS & no clouds day and night-sides & & \\
\hline

\hline
\hline
\end{tabular}%
}

\caption{Assumptions and main results from the WASP-43\,b retrieval runs presented in this paper  (C: coupled; DC: de-coupled; eq: equilibrium; HS: hot-spot)}
\label{tab:priors}

\label{tab:summary}
\end{table*}

\section{Discussion}

In our retrieval with hot-spot, a large number of assumptions were taken: Hot-spot size, clear hot-spot, addition of the Spitzer data and other model assumptions such as chemistry or temperature parametrisation. As all retrieval analyses are to some degree model dependent, it is always interesting to study the stability of a solution to model assumptions, especially since large differences are observed to the model without hot-spot. In this section, we present complementary runs and illustrate caveats that are relevant to understand the stability of our solution.

\subsection{The impact of the hot-spot size}\label{sec:hs_size}

When performing the retrievals with free hot-spot size and offsets, we encounter solutions which do not match the values recovered in \cite{stevenson_w43_1,stevenson_w43_2}, motivating the need to fix those two values (see Appendix \ref{apx:free_hs_run}). While there might be observational reasons to fix the hot-spot offset (we use the peak emission in the phase from \cite{stevenson_w43_1}), the hot-spot size cannot be constrained easily from prior analyses. For our baseline run it was set to 40$^{\circ}$ following suggestions from 3-dimensional studies \citep{Kataria_2015_W43GCM,Irwin_w43b_phase, helling_2020_mineral} but other values might, in fact, also explain the observed data. The impact of the hot-spot size, can be investigated by running complementary retrievals with various hot-spot sizes. We considered the cases with 30$^{\circ}$ and 50$^{\circ}$ fixed hot-spot sizes. For the chemistry and the clouds (see Appendix \ref{apx:post_full_alpha}) comparisons of the 3 simulations indicate very similar results. The temperature structure also remains similar in most of the atmosphere, but there are a few differences in the top part of the hot-spot and the day-side where the retrieved temperature increases with smaller hot-spot. As only little differences appear, it explains the difficulties to recover reliable information on the geometry of this planet with current data. Using similar techniques to the one presented in this paper, we however believe that data from the next generation telescopes, with a larger wavelength coverage and a higher signal-to-noise ratio, might be sensitive enough to infer the hot-spot geometry directly. Section \ref{sec:results_mock} provides an example where the hot-spot shift is directly captured in the retrieval of Ariel data.

\subsection{Model with hot-spot clouds}\label{sec:hs_clouds}

In our baseline model, we represented the planet without including clouds on the hot-spot (the cloud top pressure was fixed to 10$^6$ Pa). This choice is justified by theoretical predictions from \cite{2016_lee,Parmentier_2016,Lines_2018, Helling_2019}, which suggests that the hot day-side of irradiated exoplanets might be cleared up due to strong stellar irradiation from the host stars. 

When clouds are included on the hot-spot, we recovered a clear hot-spot solution, but the posterior solution also includes a cloudy hot-spot scenario with clouds up to 10$^3$ Pa (see Appendix \ref{apx:full_cloudy}). As a result, the corresponding temperature profile in the cloudy hot-spot solution presents larger uncertainties for higher pressures, which are in this case enabling a wider range of solutions without necessarily stratospheric thermal inversions. While opaque clouds are less likely present on the day-side of WASP-43\,b, recent theoretical models of this planet from \cite{helling_2020_mineral} suggest that this region could host some mineral clouds with large particle size as well as photo-chemically driven hydrocarbon hazes. With a much stronger day-side emission, the information contained in the phase curve data is greater for this regions, thus potentially allowing us to detect the first evidences confirming these predictions.   


\subsection{Importance of the Spitzer data for this analysis}\label{sec:full_hst}

It is known that combining data from multiple instrument can lead to strong degeneracies \citep{Yip_lightcurve, Irwin_w43b_phase,pluriel2020_ares,feng2020_2d, Yip_W96, Changeat_k11} in retrieval results. While we believe the use of HST along with Spitzer provides a large wavelength coverage, necessary to simultaneously infer atmospheric properties in emission spectroscopy (temperature/clouds/chemistry), the stability of the solution is investigated by reproducing our retrieval on the HST data only. Appendix \ref{apx:complem_full} provides the spectra centered around the HST wavelength range for our HST+Spitzer retrieval (coloured spectra) and our HST only case (grey). The runs are also extended to the Spitzer region in Figure \ref{fig:spectra_HST_Spz} of Appendix \ref{apx:complem_full}. Both fits (with and without the Spitzer photometric points) do not show major differences in the HST wavelengths range and are well fitted to the observed data. Outside the HST wavelength coverage, however, we see that the inclusion of the Spitzer points is leading to large differences in the retrievals predictions. 

For the thermal structure, we found similar results (see Appendix \ref{apx:complem_full}), with larger error on the retrieved profiles for the HST only run. This behaviour is expected from the decrease in information content since HST only covers a short wavelength range. The addition of the Spitzer data leads to observational constraints on the carbon bearing species: CH$_4$ at 3.6$\mu$m; CO and CO$_2$ at 4.5$\mu$m. The chemistry, in the HST only case, pushed towards very high metallicities (log(m) $>$ 2.0 for the posterior distribution in Appendix \ref{apx:complem_full}). We believe the addition of the Spitzer points is required to simultaneously retrieve the temperature and chemistry profiles in emission. The addition of the 3.6 $\mu$m and 4.5 $\mu$m Spitzer points provide valuable information to constrain the carbon-bearing species, thus impacting the retrieved chemistry greatly. We highlight, however, that the Spitzer datapoints are sensitive to the reduction employed. For the particular case of WASP-43\,b, there are now five independent studies \citep{stevenson_w43_2,Mendon_a_2018,morello_w43_phase,May_2020,bell2020comprehensive} that obtained very different reductions, potentially introducing biases to our results. In addition, most models use simple sine functions to reduce phase-curve data, and alternative physically motivated approaches might provide different results \citep{Louden_2018_spider}.

\subsection{Other model choices}

In this work, we assumed a particular geometry when computing the phase curve emission. The solutions found with this new model contrast with previous analyses \citep{stevenson_w43_2,Irwin_w43b_phase,feng2020_2d} and highlight new possibilities for the atmosphere of WASP-43\,b. The choices made here aimed to achieve a balance between our current understanding of the physical properties of irradiated planets and the complexity of the analysed data.  However, as of today, there is no obvious way to assess the amount of information that can be extracted from an exoplanet spectrum or a set of spectra. For example, we demonstrated that the analysis of phase curve data with unified phase curve models allows for the extraction of a more complete and complex picture than standard techniques. However, there is no guarantee that our model is complex enough or best represents the planet geometry. In other terms, when fitting exoplanet data, one always recovers a model dependent solution and in our case, other 3-dimensional geometries might be more accurate or relevant than what we presented. In the case of phase curves, we believe complementary work (testing more complex/simple geometries, comparing with simulated data) must be performed to completely understand the model dependent errors introduced by our choices. We highlight however that, the technique presented in this work is very general and can be adapted to other 3-dimensional geometries. We intend to evolve this model to reflect the advancements in our understanding of 3-dimensional effects.

Similarly, this comment also applies for physical and chemical assumptions, which might lead to biases too. In our most complex run we assumed the planet chemistry is in thermo-chemical equilibrium and that the chemistry is coupled between the different regions. While suggested as a good approximation for hot-Jupiter planets in the temperature ranges of WASP-43\,b, this may not be the case and previous studies already demonstrated the dangers of assuming equilibrium chemistry for exoplanets undergoing disequilibrium processes \citep{Line_2013,Rocchetto_biais_JWST,Blumenthal_2018, mendonca_2018_diseq, Steinrueck_2019, Changeat_2019_2l,Venot_2020,changeat2020_alfnoor,Drummond_2020,al-refaie_taurex3.1}. In particular, the thermal inversion observed in the full model for the pressures lower than 1 Pa, is driven by the need to increase the temperature and thermally dissociate all the molecules, which makes the atmosphere transparent there. In fact, if other mechanisms are able to remove the molecules at those altitudes, the retrieval would most likely not require such strong thermal inversions, which we believe is linked to the equilibrium chemistry assumption. Such processes could for example come from photo-dissociation. In addition to this, the current equilibrium scheme we use does not include exotic molecules such as TiO, VO, AlO or H-. This might also lead to biases and have an important impact on the predicted molecular abundances, especially as the results from this analysis suggests the presence of thermal inversions on the day-side of this planet. The definitive detection of such optical absorbers would provide further evidence to verify the thermal inversion hypothesis. Similarly, clouds were assumed as fully opaque grey opacities, which is a simplification that will have to be challenged when considering higher quality spectra from next generation telescopes \citep{bohren_2008_absorption,Lee_haze_model,Heng_2013,Powell_2018_clouds,Powell_2019_clouds,barstow2020_unveiling, helling_2020_mineral}. For WASP-43\,b, reflected light, which is not included here, is shown to have an important impact on the energy budget of this planet \citep{Keating_2017_reflected}. Finally, our exploration of the parameter space was performed using the MultiNest routine \citep{Feroz_multinest}, which is well established in the exoplanet community. However, the recovered solutions might also depend on the sampling algorithms used, for example alternative nested sampling algorithms are described in \cite{Handley_2015, Speagle_2020, buchner2021ultranest}, the chosen settings and the convergence criteria. We tested our full scenario with an increased number of live points (2500) to verify the stability of our solution to this parameter. The posterior distribution, shown in Appendix \ref{apx:full_2500}, does not indicate different results as compared to the 500 live points run for this example, but we note that retrieval abilities to separate multi-modal solutions might greatly depend on this parameter.

\section{Conclusion} \label{sec:conc}

Phase curve data associated with a unified analysis such as the one presented in this paper, undeniably provide a more complete picture of exoplanet atmospheres than conventional methods. By using a new type of simplified representation for the exoplanet WASP-43\,b, we demonstrated the feasibility of building dedicated techniques for the study of these complex datasets. This technique provides access to an unprecedented level of detail which, along with the upcoming increase in data quality by next generation space based instruments, has the potential to revolutionise our understanding of exoplanet atmospheric physics. However, by testing different assumptions for the geometry, the chemistry and the thermal structure of WASP-43\,b in our model, we demonstrated the impact of model assumptions to the result recovered (see summary in Table \ref{tab:summary}). With the upcoming JWST and Ariel next generation space telescopes, this highlights the importance of studying these model dependent behaviour to ensure an optimal extraction of the higher quality spectral data.

\section*{Acknowledgements}

We thank the referee for providing thoughtful suggestions that greatly improved our manuscript.

This project has received funding from the European Research Council (ERC) under the European Union's Horizon 2020 research and innovation programme (grant agreement No 758892, ExoAI) and under the European Union's Seventh Framework Programme (FP7/2007-2013)/ ERC grant agreement numbers 617119 (ExoLights). Furthermore, we acknowledge funding by the Science and Technology Funding Council (STFC) grants: ST/K502406/1, ST/P000282/1, ST/P002153/1, ST/S002634/1 and ST/T001836/1.

This work utilised the OzSTAR national facility at Swinburne University of Technology. The OzSTAR program receives funding in part from the Astronomy National Collaborative Research Infrastructure Strategy (NCRIS) allocation provided by the Australian Government. This work utilised the Cambridge Service for Data Driven Discovery (CSD3), part of which is operated by the University of Cambridge Research Computing on behalf of the STFC DiRAC HPC Facility (www.dirac.ac.uk). The DiRAC component of CSD3 was funded by BEIS capital funding via STFC capital grants ST/P002307/1 and ST/R002452/1 and STFC operations grant ST/R00689X/1. DiRAC is part of the National e-Infrastructure.

\section*{Data Availability}

The data analysed in this work are available through the NASA MAST HST archive (\url{https://archive.stsci.edu/}) programmes 13665 and 14682. The molecular line lists used are available from the ExoMol website (\url{www.exomol.com}).

\clearpage

\bibliographystyle{aasjournal}
\bibliography{main}

\begin{thebibliography}{}
\expandafter\ifx\csname natexlab\endcsname\relax\def\natexlab#1{#1}\fi
\providecommand{\url}[1]{\href{#1}{#1}}

\bibitem[{Abel {et~al.}(2011)Abel, Frommhold, Li, \& Hunt}]{abel_h2-h2}
Abel, M., Frommhold, L., Li, X., \& Hunt, K.~L. 2011, The Journal of Physical
  Chemistry A, 115, 6805

\bibitem[{Abel {et~al.}(2012)Abel, Frommhold, Li, \& Hunt}]{abel_h2-he}
---. 2012, The Journal of chemical physics, 136, 044319

\bibitem[{{Ag{\'u}ndez} {et~al.}(2014){Ag{\'u}ndez}, {Parmentier}, {Venot},
  {Hersant}, \& {Selsis}}]{Agundez_2dchemical_HD209_HD189}
{Ag{\'u}ndez}, M., {Parmentier}, V., {Venot}, O., {Hersant}, F., \& {Selsis},
  F. 2014, \aap, 564, A73

\bibitem[{{Al-Refaie} {et~al.}(2020){Al-Refaie}, {Changeat}, {Venot},
  {Waldmann}, \& {Tinetti}}]{al-refaie_taurex3.1}
{Al-Refaie}, A.~F., {Changeat}, Q., {Venot}, O., {Waldmann}, I.~P., \&
  {Tinetti}, G. 2020, in prep

\bibitem[{Al-Refaie {et~al.}(2019)Al-Refaie, Changeat, Waldmann, \&
  Tinetti}]{al-refaie_taurex3}
Al-Refaie, A.~F., Changeat, Q., Waldmann, I.~P., \& Tinetti, G. 2019, TauREx
  III: A fast, dynamic and extendable framework for retrievals, , ,
  arXiv:1912.07759

\bibitem[{Andrae {et~al.}(2010)Andrae, Schulze-Hartung, \&
  Melchior}]{andrae2010dos}
Andrae, R., Schulze-Hartung, T., \& Melchior, P. 2010, Dos and don'ts of
  reduced chi-squared, , , arXiv:1012.3754

\bibitem[{Arcangeli {et~al.}(2019)Arcangeli, Désert, Parmentier, Stevenson,
  Bean, Line, Kreidberg, Fortney, \& Showman}]{Arcangeli_2019_W18}
Arcangeli, J., Désert, J.-M., Parmentier, V., {et~al.} 2019, Astronomy \&
  Astrophysics, 625, A136.
\newblock \url{http://dx.doi.org/10.1051/0004-6361/201834891}

\bibitem[{Barber {et~al.}(2013)Barber, Strange, Hill, Polyansky, Mellau,
  Yurchenko, \& Tennyson}]{Barber_2013_HCN}
Barber, R.~J., Strange, J.~K., Hill, C., {et~al.} 2013, Monthly Notices of the
  Royal Astronomical Society, 437, 1828–1835.
\newblock \url{http://dx.doi.org/10.1093/mnras/stt2011}

\bibitem[{Barstow(2020)}]{barstow2020_unveiling}
Barstow, J.~K. 2020, Unveiling cloudy exoplanets: the influence of cloud model
  choices on retrieval solutions, , , arXiv:2002.02945

\bibitem[{Barstow {et~al.}(2020)Barstow, Changeat, Garland, Line, Rocchetto, \&
  Waldmann}]{Barstow_2020_compar}
Barstow, J.~K., Changeat, Q., Garland, R., {et~al.} 2020, Monthly Notices of
  the Royal Astronomical Society, 493, 4884–4909.
\newblock \url{http://dx.doi.org/10.1093/mnras/staa548}

\bibitem[{Barstow \& Heng(2020)}]{barstow_2020outstanding}
Barstow, J.~K., \& Heng, K. 2020, Outstanding Challenges of Exoplanet
  Atmospheric Retrievals, , , arXiv:2003.14311

\bibitem[{Barton {et~al.}(2017)Barton, Hill, Yurchenko, Tennyson, Dudaryonok,
  \& Lavrentieva}]{barton_h2o}
Barton, E.~J., Hill, C., Yurchenko, S.~N., {et~al.} 2017, Journal of
  Quantitative Spectroscopy and Radiative Transfer, 187, 453

\bibitem[{Beatty {et~al.}(2019)Beatty, Marley, Gaudi, Colón, Fortney, \&
  Showman}]{Beatty_2019}
Beatty, T.~G., Marley, M.~S., Gaudi, B.~S., {et~al.} 2019, The Astronomical
  Journal, 158, 166.
\newblock \url{http://dx.doi.org/10.3847/1538-3881/ab33fc}

\bibitem[{Bell {et~al.}(2019)Bell, Zhang, Cubillos, Dang, Fossati, Todorov,
  Cowan, Deming, Zellem, Stevenson, \& et~al.}]{Bell_2019}
Bell, T.~J., Zhang, M., Cubillos, P.~E., {et~al.} 2019, Monthly Notices of the
  Royal Astronomical Society, 489, 1995–2013.
\newblock \url{http://dx.doi.org/10.1093/mnras/stz2018}

\bibitem[{Bell {et~al.}(2020)Bell, Dang, Cowan, Bean, Désert, Fortney,
  Keating, Kempton, Kreidberg, Line, Mansfield, Parmentier, Stevenson, Swain,
  \& Zellem}]{bell2020comprehensive}
Bell, T.~J., Dang, L., Cowan, N.~B., {et~al.} 2020, A Comprehensive Reanalysis
  of Spitzer's 4.5 $\mu$m Phase Curves, and the Phase Variations of the
  Ultra-hot Jupiters MASCARA-1b and KELT-16b, , , arXiv:2010.00687

\bibitem[{Blecic {et~al.}(2014)Blecic, Harrington, Madhusudhan, Stevenson,
  Hardy, Cubillos, Hardin, Bowman, Nymeyer, Anderson, \& et~al.}]{Blecic_2014}
Blecic, J., Harrington, J., Madhusudhan, N., {et~al.} 2014, The Astrophysical
  Journal, 781, 116.
\newblock \url{http://dx.doi.org/10.1088/0004-637X/781/2/116}

\bibitem[{Blumenthal {et~al.}(2018)Blumenthal, Mandell, Hébrard, Batalha,
  Cubillos, Rugheimer, \& Wakeford}]{Blumenthal_2018}
Blumenthal, S.~D., Mandell, A.~M., Hébrard, E., {et~al.} 2018, The
  Astrophysical Journal, 853, 138.
\newblock \url{http://dx.doi.org/10.3847/1538-4357/aa9e51}

\bibitem[{Bohren \& Huffman(2008)}]{bohren_2008_absorption}
Bohren, C.~F., \& Huffman, D.~R. 2008, Absorption and scattering of light by
  small particles (John Wiley \& Sons)

\bibitem[{{Bonomo} {et~al.}(2017){Bonomo}, {Desidera}, {Benatti}, {Borsa},
  {Crespi}, {Damasso}, {Lanza}, {Sozzetti}, {Lodato}, {Marzari}, {Boccato},
  {Claudi}, {Cosentino}, {Covino}, {Gratton}, {Maggio}, {Micela}, {Molinari},
  {Pagano}, {Piotto}, {Poretti}, {Smareglia}, {Affer}, {Biazzo}, {Bignamini},
  {Esposito}, {Giacobbe}, {H{\'e}brard}, {Malavolta}, {Maldonado}, {Mancini},
  {Martinez Fiorenzano}, {Masiero}, {Nascimbeni}, {Pedani}, {Rainer}, \& {Scand
  ariato}}]{bonomo_2017}
{Bonomo}, A.~S., {Desidera}, S., {Benatti}, S., {et~al.} 2017, \aap, 602, A107

\bibitem[{Buchner(2021)}]{buchner2021ultranest}
Buchner, J. 2021, UltraNest -- a robust, general purpose Bayesian inference
  engine, , , arXiv:2101.09604

\bibitem[{{Buchner} {et~al.}(2014){Buchner}, {Georgakakis}, {Nandra}, {Hsu},
  {Rangel}, {Brightman}, {Merloni}, {Salvato}, {Donley}, \&
  {Kocevski}}]{buchner_pymultinest}
{Buchner}, J., {Georgakakis}, A., {Nandra}, K., {et~al.} 2014, \aap, 564, A125

\bibitem[{{Caldas} {et~al.}(2019){Caldas}, {Leconte}, {Selsis}, {Waldmann},
  {Bord{\'e}}, {Rocchetto}, \& {Charnay}}]{caldas_3deffects}
{Caldas}, A., {Leconte}, J., {Selsis}, F., {et~al.} 2019, \aap, 623, A161

\bibitem[{Changeat \& Al-Refaie(2020)}]{changeat_2020_phase1}
Changeat, Q., \& Al-Refaie, A. 2020, The Astrophysical Journal, 898, 155.
\newblock \url{http://dx.doi.org/10.3847/1538-4357/ab9b82}

\bibitem[{Changeat {et~al.}(2020)Changeat, Al-Refaie, Mugnai, Edwards,
  Waldmann, Pascale, \& Tinetti}]{changeat2020_alfnoor}
Changeat, Q., Al-Refaie, A., Mugnai, L.~V., {et~al.} 2020, The Astronomical
  Journal, 160, 80.
\newblock \url{http://dx.doi.org/10.3847/1538-3881/ab9a53}

\bibitem[{Changeat \& Edwards(2021)}]{Changeat_2021_K9}
Changeat, Q., \& Edwards, B. 2021, The Astrophysical Journal Letters, 907, L22.
\newblock \url{http://dx.doi.org/10.3847/2041-8213/abd84f}

\bibitem[{{Changeat} {et~al.}(2020){Changeat}, {Edwards}, {Al-Refaie},
  {Morvan}, {Tsiaras}, {Waldmann}, \& {Tinetti}}]{Changeat_k11}
{Changeat}, Q., {Edwards}, B., {Al-Refaie}, A., {et~al.} 2020, AJ

\bibitem[{Changeat {et~al.}(2019)Changeat, Edwards, Waldmann, \&
  Tinetti}]{Changeat_2019_2l}
Changeat, Q., Edwards, B., Waldmann, I.~P., \& Tinetti, G. 2019, The
  Astrophysical Journal, 886, 39.
\newblock \url{http://dx.doi.org/10.3847/1538-4357/ab4a14}

\bibitem[{Changeat {et~al.}(2020)Changeat, Keyte, Waldmann, \&
  Tinetti}]{changeat2019_impact}
Changeat, Q., Keyte, L., Waldmann, I.~P., \& Tinetti, G. 2020, The
  Astrophysical Journal, 896, 107.
\newblock \url{http://dx.doi.org/10.3847/1538-4357/ab8f8b}

\bibitem[{Chen {et~al.}(2014)Chen, van Boekel, Wang, Nikolov, Fortney, Seemann,
  Wang, Mancini, \& Henning}]{Chen_2014}
Chen, G., van Boekel, R., Wang, H., {et~al.} 2014, Astronomy \& Astrophysics,
  563, A40.
\newblock \url{http://dx.doi.org/10.1051/0004-6361/201322740}

\bibitem[{Chubb {et~al.}(2020)Chubb, Min, Kawashima, Helling, \&
  Waldmann}]{chubb2020aluminium}
Chubb, K.~L., Min, M., Kawashima, Y., Helling, C., \& Waldmann, I. 2020,
  Astronomy \& Astrophysics, 639, A3.
\newblock \url{http://dx.doi.org/10.1051/0004-6361/201937267}

\bibitem[{Chubb {et~al.}(2021)Chubb, Rocchetto, Yurchenko, Min, Waldmann,
  Barstow, Mollière, Al-Refaie, Phillips, \& Tennyson}]{Chubb_2021_exomol}
Chubb, K.~L., Rocchetto, M., Yurchenko, S.~N., {et~al.} 2021, Astronomy \&
  Astrophysics, 646, A21.
\newblock \url{http://dx.doi.org/10.1051/0004-6361/202038350}

\bibitem[{Cox(2015)}]{cox_allen_rayleigh}
Cox, A.~N. 2015, Allen’s astrophysical quantities (Springer)

\bibitem[{Deming \& Knutson(2020)}]{deming_2020highlights}
Deming, D., \& Knutson, H. 2020, Highlights of Exoplanetary Science from
  Spitzer, , , arXiv:2005.11331

\bibitem[{Demory {et~al.}(2016)Demory, Gillon, de~Wit, Madhusudhan, Bolmont,
  Heng, Kataria, Lewis, Hu, Krick, \& et~al.}]{Demory_2016}
Demory, B.-O., Gillon, M., de~Wit, J., {et~al.} 2016, Nature, 532, 207–209.
\newblock \url{http://dx.doi.org/10.1038/nature17169}

\bibitem[{Drummond {et~al.}(2018)Drummond, Mayne, Baraffe, Tremblin, Manners,
  Amundsen, Goyal, \& Acreman}]{Drummond_2018}
Drummond, B., Mayne, N.~J., Baraffe, I., {et~al.} 2018, Astronomy \&
  Astrophysics, 612, A105.
\newblock \url{http://dx.doi.org/10.1051/0004-6361/201732010}

\bibitem[{Drummond {et~al.}(2020)Drummond, Hébrard, Mayne, Venot, Ridgway,
  Changeat, Tsai, Manners, Tremblin, Abraham, \& et~al.}]{Drummond_2020}
Drummond, B., Hébrard, E., Mayne, N.~J., {et~al.} 2020, Astronomy \&
  Astrophysics, 636, A68.
\newblock \url{http://dx.doi.org/10.1051/0004-6361/201937153}

\bibitem[{Edwards {et~al.}(2019)Edwards, Mugnai, Tinetti, Pascale, \&
  Sarkar}]{Edwards_2019}
Edwards, B., Mugnai, L., Tinetti, G., Pascale, E., \& Sarkar, S. 2019, The
  Astronomical Journal, 157, 242.
\newblock \url{http://dx.doi.org/10.3847/1538-3881/ab1cb9}

\bibitem[{Edwards {et~al.}(2018)Edwards, Rice, Zingales, Tessenyi, Waldmann,
  Tinetti, Pascale, Savini, \& Sarkar}]{Edwards_twinkle}
Edwards, B., Rice, M., Zingales, T., {et~al.} 2018, Experimental Astronomy,
  doi:10.1007/s10686-018-9611-4

\bibitem[{Edwards {et~al.}(2020)Edwards, Changeat, Baeyens, Tsiaras, Al-Refaie,
  Taylor, Yip, Bieger, Blain, Gressier, \& et~al.}]{Edwards_2020}
Edwards, B., Changeat, Q., Baeyens, R., {et~al.} 2020, The Astronomical
  Journal, 160, 8.
\newblock \url{http://dx.doi.org/10.3847/1538-3881/ab9225}

\bibitem[{Edwards {et~al.}(1963)Edwards, Lindman, \&
  Savage}]{edwards1963bayesian}
Edwards, W., Lindman, H., \& Savage, L.~J. 1963, Psychological review, 70, 193

\bibitem[{Esteves {et~al.}(2013)Esteves, De~Mooij, \&
  Jayawardhana}]{Esteves_2013}
Esteves, L.~J., De~Mooij, E. J.~W., \& Jayawardhana, R. 2013, The Astrophysical
  Journal, 772, 51.
\newblock \url{http://dx.doi.org/10.1088/0004-637X/772/1/51}

\bibitem[{Feng {et~al.}(2020)Feng, Line, \& Fortney}]{feng2020_2d}
Feng, Y.~K., Line, M.~R., \& Fortney, J.~J. 2020, 2D Retrieval Frameworks for
  Hot Jupiter Phase Curves, , , arXiv:2006.11442

\bibitem[{Feng {et~al.}(2016)Feng, Line, Fortney, Stevenson, Bean, Kreidberg,
  \& Parmentier}]{feng_2016}
Feng, Y.~K., Line, M.~R., Fortney, J.~J., {et~al.} 2016, The Astrophysical
  Journal, 829, 52.
\newblock \url{http://dx.doi.org/10.3847/0004-637X/829/1/52}

\bibitem[{{Feroz} {et~al.}(2009){Feroz}, {Hobson}, \&
  {Bridges}}]{Feroz_multinest}
{Feroz}, F., {Hobson}, M.~P., \& {Bridges}, M. 2009, \mnras, 398, 1601

\bibitem[{Fletcher {et~al.}(2018)Fletcher, Gustafsson, \&
  Orton}]{fletcher_h2-h2}
Fletcher, L.~N., Gustafsson, M., \& Orton, G.~S. 2018, The Astrophysical
  Journal Supplement Series, 235, 24

\bibitem[{{Gandhi} \& {Madhusudhan}(2018)}]{Gandhi_retrieval}
{Gandhi}, S., \& {Madhusudhan}, N. 2018, \mnras, 474, 271

\bibitem[{Gelman(2013)}]{gelman2013}
Gelman, A. 2013, Electron. J. Statist., 7, 2595.
\newblock \url{https://doi.org/10.1214/13-EJS854}

\bibitem[{{Gordon} {et~al.}(2016){Gordon}, {Rothman}, {Wilzewski}, {Kochanov},
  {Hill}, {Tan}, \& {Wcislo}}]{gordon}
{Gordon}, I., {Rothman}, L.~S., {Wilzewski}, J.~S., {et~al.} 2016, in
  AAS/Division for Planetary Sciences Meeting Abstracts, Vol.~48, AAS/Division
  for Planetary Sciences Meeting Abstracts \#48, 421.13

\bibitem[{{Greene} {et~al.}(2016){Greene}, {Line}, {Montero}, {Fortney},
  {Lustig-Yaeger}, \& {Luther}}]{Greene_2016}
{Greene}, T.~P., {Line}, M.~R., {Montero}, C., {et~al.} 2016, \apj, 817, 17

\bibitem[{{Guillot}(2010)}]{Guillot_TP_model}
{Guillot}, T. 2010, \aap, 520, A27

\bibitem[{Handley {et~al.}(2015)Handley, Hobson, \& Lasenby}]{Handley_2015}
Handley, W.~J., Hobson, M.~P., \& Lasenby, A.~N. 2015, Monthly Notices of the
  Royal Astronomical Society, 453, 4385–4399.
\newblock \url{http://dx.doi.org/10.1093/mnras/stv1911}

\bibitem[{Harris {et~al.}(2006)Harris, Tennyson, Kaminsky, Pavlenko, \&
  Jones}]{Harris_2006_HCN}
Harris, G.~J., Tennyson, J., Kaminsky, B.~M., Pavlenko, Y.~V., \& Jones, H.
  R.~A. 2006, Monthly Notices of the Royal Astronomical Society, 367,
  400–406.
\newblock \url{http://dx.doi.org/10.1111/j.1365-2966.2005.09960.x}

\bibitem[{Hellier {et~al.}(2011)Hellier, Anderson, Collier~Cameron, Gillon,
  Jehin, Lendl, Maxted, Pepe, Pollacco, Queloz, \& et~al.}]{Hellier_2011}
Hellier, C., Anderson, D.~R., Collier~Cameron, A., {et~al.} 2011, Astronomy \&
  Astrophysics, 535, L7.
\newblock \url{http://dx.doi.org/10.1051/0004-6361/201117081}

\bibitem[{Helling(2019)}]{Helling_2019}
Helling, C. 2019, Annual Review of Earth and Planetary Sciences, 47, 583–606.
\newblock \url{http://dx.doi.org/10.1146/annurev-earth-053018-060401}

\bibitem[{Helling {et~al.}(2020)Helling, Kawashima, Graham, Samra, Chubb, Min,
  Waters, \& Parmentier}]{helling_2020_mineral}
Helling, C., Kawashima, Y., Graham, V., {et~al.} 2020, Mineral cloud and
  hydrocarbon haze particles in the atmosphere of the hot Jupiter JWST target
  WASP-43b, , , arXiv:2005.14595

\bibitem[{Heng \& Demory(2013)}]{Heng_2013}
Heng, K., \& Demory, B.-O. 2013, The Astrophysical Journal, 777, 100.
\newblock \url{http://dx.doi.org/10.1088/0004-637X/777/2/100}

\bibitem[{Hill {et~al.}(2013)Hill, Yurchenko, \& Tennyson}]{hill_xsec}
Hill, C., Yurchenko, S.~N., \& Tennyson, J. 2013, Icarus, 226, 1673

\bibitem[{{Hou Yip} {et~al.}(2020){Hou Yip}, {Changeat}, {Edwards}, {Morvan},
  {Chubb}, {Tsiaras}, {Waldmann}, \& {Tinetti}}]{Yip_W96}
{Hou Yip}, K., {Changeat}, Q., {Edwards}, B., {et~al.} 2020, ApJ

\bibitem[{{Hou Yip} {et~al.}(2018){Hou Yip}, {Waldmann}, {Tsiaras}, \&
  {Tinetti}}]{Yip_lightcurve}
{Hou Yip}, K., {Waldmann}, I.~P., {Tsiaras}, A., \& {Tinetti}, G. 2018,
  submitted, arXiv:1811.04686

\bibitem[{{Irwin} {et~al.}(2019){Irwin}, {Parmentier}, {Taylor}, {Barstow},
  {Aigrain}, {Lee}, \& {Garland }}]{Irwin_w43b_phase}
{Irwin}, P. G.~J., {Parmentier}, V., {Taylor}, J., {et~al.} 2019, arXiv
  e-prints, arXiv:1909.03233

\bibitem[{{Irwin} {et~al.}(2008){Irwin}, {Teanby}, {de Kok}, {Fletcher},
  {Howett}, {Tsang}, {Wilson}, {Calcutt}, {Nixon}, \& {Parrish}}]{irwin2008}
{Irwin}, P.~G.~J., {Teanby}, N.~A., {de Kok}, R., {et~al.} 2008, \jqsrt, 109,
  1136

\bibitem[{Jeffreys(1998)}]{jeffreys1998_bayesfactor}
Jeffreys, H. 1998, The theory of probability (OUP Oxford)

\bibitem[{Kass \& Raftery(1995)}]{Kass1995bayes}
Kass, R.~E., \& Raftery, A.~E. 1995, Journal of the american statistical
  association, 90, 773

\bibitem[{Kataria {et~al.}(2015)Kataria, Showman, Fortney, Stevenson, Line,
  Kreidberg, Bean, \& Désert}]{Kataria_2015_W43GCM}
Kataria, T., Showman, A.~P., Fortney, J.~J., {et~al.} 2015, The Astrophysical
  Journal, 801, 86.
\newblock \url{http://dx.doi.org/10.1088/0004-637X/801/2/86}

\bibitem[{{Keating} \& {Cowan}(2017)}]{Keating_2017_reflected}
{Keating}, D., \& {Cowan}, N.~B. 2017, \apjl, 849, L5

\bibitem[{Knutson {et~al.}(2012)Knutson, Lewis, Fortney, Burrows, Showman,
  Cowan, Agol, Aigrain, Charbonneau, Deming, \& et~al.}]{Knutson_2012}
Knutson, H.~A., Lewis, N., Fortney, J.~J., {et~al.} 2012, The Astrophysical
  Journal, 754, 22.
\newblock \url{http://dx.doi.org/10.1088/0004-637X/754/1/22}

\bibitem[{Kreidberg {et~al.}(2014)Kreidberg, Bean, Désert, Line, Fortney,
  Madhusudhan, Stevenson, Showman, Charbonneau, McCullough, \&
  et~al.}]{Kreidberg_2014}
Kreidberg, L., Bean, J.~L., Désert, J.-M., {et~al.} 2014, The Astrophysical
  Journal, 793, L27.
\newblock \url{http://dx.doi.org/10.1088/2041-8205/793/2/L27}

\bibitem[{Kreidberg {et~al.}(2018)Kreidberg, Line, Parmentier, Stevenson,
  Louden, Bonnefoy, Faherty, Henry, Williamson, Stassun, \&
  et~al.}]{Kreidberg_2018}
Kreidberg, L., Line, M.~R., Parmentier, V., {et~al.} 2018, The Astronomical
  Journal, 156, 17.
\newblock \url{http://dx.doi.org/10.3847/1538-3881/aac3df}

\bibitem[{Lavvas \& Arfaux(2021)}]{lavvas2021impact}
Lavvas, P., \& Arfaux, A. 2021, Impact of photochemical hazes and gases on
  exoplanet atmospheric thermal structure, , , arXiv:2102.05763

\bibitem[{{Lee} {et~al.}(2016){Lee}, {Dobbs-Dixon}, {Helling}, {Bognar}, \&
  {Woitke}}]{2016_lee}
{Lee}, G., {Dobbs-Dixon}, I., {Helling}, C., {Bognar}, K., \& {Woitke}, P.
  2016, \aap, 594, A48

\bibitem[{Lee {et~al.}(2013)Lee, Heng, \& Irwin}]{Lee_haze_model}
Lee, J.-M., Heng, K., \& Irwin, P. G.~J. 2013, The Astrophysical Journal, 778,
  97.
\newblock \url{https://doi.org/10.1088%2F0004-637x%2F778%2F2%2F97}

\bibitem[{Li {et~al.}(2015)Li, Gordon, Rothman, Tan, Hu, Kassi, Campargue, \&
  Medvedev}]{li_co_2015}
Li, G., Gordon, I.~E., Rothman, L.~S., {et~al.} 2015, The Astrophysical Journal
  Supplement Series, 216, 15

\bibitem[{Line \& Parmentier(2016)}]{Line_2016}
Line, M.~R., \& Parmentier, V. 2016, The Astrophysical Journal, 820, 78.
\newblock \url{http://dx.doi.org/10.3847/0004-637X/820/1/78}

\bibitem[{Line \& Yung(2013)}]{Line_2013}
Line, M.~R., \& Yung, Y.~L. 2013, The Astrophysical Journal, 779, 3.
\newblock \url{http://dx.doi.org/10.1088/0004-637X/779/1/3}

\bibitem[{{Line} {et~al.}(2013){Line}, {Wolf}, {Zhang}, {Knutson}, {Kammer},
  {Ellison}, {Deroo}, {Crisp}, \& {Yung}}]{chimera}
{Line}, M.~R., {Wolf}, A.~S., {Zhang}, X., {et~al.} 2013, \apj, 775, 137

\bibitem[{Lines {et~al.}(2018)Lines, Mayne, Boutle, Manners, Lee, Helling,
  Drummond, Amundsen, Goyal, Acreman, \& et~al.}]{Lines_2018}
Lines, S., Mayne, N.~J., Boutle, I.~A., {et~al.} 2018, Astronomy \&
  Astrophysics, 615, A97.
\newblock \url{http://dx.doi.org/10.1051/0004-6361/201732278}

\bibitem[{Lothringer {et~al.}(2018)Lothringer, Barman, \&
  Koskinen}]{Lothringer_2018_invert}
Lothringer, J.~D., Barman, T., \& Koskinen, T. 2018, The Astrophysical Journal,
  866, 27.
\newblock \url{http://dx.doi.org/10.3847/1538-4357/aadd9e}

\bibitem[{{Louden} \& {Kreidberg}(2018)}]{Louden_2018_spider}
{Louden}, T., \& {Kreidberg}, L. 2018, \mnras, 477, 2613

\bibitem[{MacDonald {et~al.}(2020)MacDonald, Goyal, \&
  Lewis}]{MacDonald_2020_cold}
MacDonald, R.~J., Goyal, J.~M., \& Lewis, N.~K. 2020, The Astrophysical
  Journal, 893, L43.
\newblock \url{http://dx.doi.org/10.3847/2041-8213/ab8238}

\bibitem[{Madhusudhan(2019)}]{Madhusudhan_2019_review}
Madhusudhan, N. 2019, Annual Review of Astronomy and Astrophysics, 57,
  617–663.
\newblock \url{http://dx.doi.org/10.1146/annurev-astro-081817-051846}

\bibitem[{{Madhusudhan} \& {Seager}(2009)}]{Madhu_retrieval_method}
{Madhusudhan}, N., \& {Seager}, S. 2009, \apj, 707, 24

\bibitem[{{Mai} \& {Line}(2019)}]{Mai_2019}
{Mai}, C., \& {Line}, M.~R. 2019, \apj, 883, 144

\bibitem[{Mant {et~al.}(2018)Mant, Yachmenev, Tennyson, \&
  Yurchenko}]{2018_Mant_C2H4}
Mant, B.~P., Yachmenev, A., Tennyson, J., \& Yurchenko, S.~N. 2018, Monthly
  Notices of the Royal Astronomical Society, 478, 3220–3232.
\newblock \url{http://dx.doi.org/10.1093/mnras/sty1239}

\bibitem[{May \& Stevenson(2020)}]{May_2020}
May, E.~M., \& Stevenson, K.~B. 2020, The Astronomical Journal, 160, 140.
\newblock \url{http://dx.doi.org/10.3847/1538-3881/aba833}

\bibitem[{{Mendon{\c{c}}a} {et~al.}(2018){Mendon{\c{c}}a}, {Tsai}, {Malik},
  {Grimm}, \& {Heng}}]{mendonca_2018_diseq}
{Mendon{\c{c}}a}, J.~M., {Tsai}, S.-m., {Malik}, M., {Grimm}, S.~L., \& {Heng},
  K. 2018, \apj, 869, 107

\bibitem[{Mendonça {et~al.}(2018)Mendonça, Malik, Demory, \&
  Heng}]{Mendon_a_2018}
Mendonça, J.~M., Malik, M., Demory, B.-O., \& Heng, K. 2018, The Astronomical
  Journal, 155, 150.
\newblock \url{http://dx.doi.org/10.3847/1538-3881/aaaebc}

\bibitem[{Min {et~al.}(2020)Min, Ormel, Chubb, Helling, \&
  Kawashima}]{min2020arcis}
Min, M., Ormel, C.~W., Chubb, K., Helling, C., \& Kawashima, Y. 2020, The ARCiS
  framework for Exoplanet Atmospheres: Modelling Philosophy and Retrieval, , ,
  arXiv:2006.12821

\bibitem[{Mollière {et~al.}(2019)Mollière, Wardenier, van Boekel, Henning,
  Molaverdikhani, \& Snellen}]{Mollire_petitrad}
Mollière, P., Wardenier, J.~P., van Boekel, R., {et~al.} 2019, Astronomy \&
  Astrophysics, 627, A67.
\newblock \url{http://dx.doi.org/10.1051/0004-6361/201935470}

\bibitem[{{Morello} {et~al.}(2019){Morello}, {Danielski}, {Dickens},
  {Tremblin}, \& {Lagage}}]{morello_w43_phase}
{Morello}, G., {Danielski}, C., {Dickens}, D., {Tremblin}, P., \& {Lagage},
  P.~O. 2019, \aj, 157, 205

\bibitem[{{Mugnai} {et~al.}(2019){Mugnai}, {Pascale}, \&
  {Edwards}}]{mugnai_Arielrad}
{Mugnai}, L., {Pascale}, E., \& {Edwards}, B. 2019, in prep

\bibitem[{Parmentier \& Crossfield(2018)}]{Parmentier_2018}
Parmentier, V., \& Crossfield, I. J.~M. 2018, Handbook of Exoplanets,
  1419–1440.
\newblock \url{http://dx.doi.org/10.1007/978-3-319-55333-7_116}

\bibitem[{Parmentier {et~al.}(2016)Parmentier, Fortney, Showman, Morley, \&
  Marley}]{Parmentier_2016}
Parmentier, V., Fortney, J.~J., Showman, A.~P., Morley, C., \& Marley, M.~S.
  2016, The Astrophysical Journal, 828, 22.
\newblock \url{http://dx.doi.org/10.3847/0004-637X/828/1/22}

\bibitem[{Placek {et~al.}(2017)Placek, Angerhausen, \& Knuth}]{Placek_2017}
Placek, B., Angerhausen, D., \& Knuth, K.~H. 2017, The Astronomical Journal,
  154, 154.
\newblock \url{http://dx.doi.org/10.3847/1538-3881/aa880d}

\bibitem[{Pluriel {et~al.}(2020{\natexlab{a}})Pluriel, Zingales, Leconte, \&
  Parmentier}]{Pluriel_2020}
Pluriel, W., Zingales, T., Leconte, J., \& Parmentier, V. 2020{\natexlab{a}},
  Astronomy \& Astrophysics, 636, A66.
\newblock \url{http://dx.doi.org/10.1051/0004-6361/202037678}

\bibitem[{Pluriel {et~al.}(2020{\natexlab{b}})Pluriel, Whiteford, Edwards,
  Changeat, Yip, Baeyens, Al-Refaie, Bieger, Blain, Gressier, Guilluy, Jaziri,
  Kiefer, Modirrousta-Galian, Morvan, Mugnai, Poveda, Skaf, Zingales, Wright,
  Charnay, Drossart, Leconte, Tsiaras, Venot, Waldmann, \&
  Beaulieu}]{pluriel2020_ares}
Pluriel, W., Whiteford, N., Edwards, B., {et~al.} 2020{\natexlab{b}}, ARES III:
  Unveiling the Two Faces of KELT-7 b with HST WFC3, , , arXiv:2006.14199

\bibitem[{Polyansky {et~al.}(2018)Polyansky, Kyuberis, Zobov, Tennyson,
  Yurchenko, \& Lodi}]{polyansky_h2o}
Polyansky, O.~L., Kyuberis, A.~A., Zobov, N.~F., {et~al.} 2018, Monthly Notices
  of the Royal Astronomical Society, 480, 2597

\bibitem[{Powell {et~al.}(2019)Powell, Louden, Kreidberg, Zhang, Gao, \&
  Parmentier}]{Powell_2019_clouds}
Powell, D., Louden, T., Kreidberg, L., {et~al.} 2019, The Astrophysical
  Journal, 887, 170.
\newblock \url{http://dx.doi.org/10.3847/1538-4357/ab55d9}

\bibitem[{Powell {et~al.}(2018)Powell, Zhang, Gao, \&
  Parmentier}]{Powell_2018_clouds}
Powell, D., Zhang, X., Gao, P., \& Parmentier, V. 2018, The Astrophysical
  Journal, 860, 18.
\newblock \url{http://dx.doi.org/10.3847/1538-4357/aac215}

\bibitem[{{Rocchetto} {et~al.}(2016){Rocchetto}, {Waldmann}, {Venot}, {Lagage},
  \& {Tinetti}}]{Rocchetto_biais_JWST}
{Rocchetto}, M., {Waldmann}, I.~P., {Venot}, O., {Lagage}, P.~O., \& {Tinetti},
  G. 2016, \apj, 833, 120

\bibitem[{{Rodgers}(2000)}]{rodger_retrievals}
{Rodgers}, C.~D. 2000, {Inverse Methods for Atmospheric Sounding - Theory and
  Practice}, Inverse Methods for Atmospheric Sounding - Theory and Practice.
  Series: Series on Atmospheric Oceanic and Planetary Physics, , ,
  doi:10.1142/9789812813718

\bibitem[{Rothman {et~al.}(2010)Rothman, Gordon, Barber, Dothe, Gamache,
  Goldman, Perevalov, Tashkun, \& Tennyson}]{rothman_hitremp_2010}
Rothman, L., Gordon, I., Barber, R., {et~al.} 2010, Journal of Quantitative
  Spectroscopy and Radiative Transfer, 111, 2139

\bibitem[{{Rothman} \& {Gordon}(2014)}]{rothman}
{Rothman}, L.~S., \& {Gordon}, I.~E. 2014, in 13th International HITRAN
  Conference, June 2014, Cambridge, Massachusetts, USA

\bibitem[{Rougier \& Priebe(2020)}]{rougier2020exact}
Rougier, J., \& Priebe, C. 2020, The exact form of the 'Ockham factor' in model
  selection, , , arXiv:1906.11592

\bibitem[{Seager \& Deming(2010)}]{Seager_2010}
Seager, S., \& Deming, D. 2010, Annual Review of Astronomy and Astrophysics,
  48, 631–672.
\newblock \url{http://dx.doi.org/10.1146/annurev-astro-081309-130837}

\bibitem[{Showman {et~al.}(2009)Showman, Fortney, Lian, Marley, Freedman,
  Knutson, \& Charbonneau}]{Showman_2009}
Showman, A.~P., Fortney, J.~J., Lian, Y., {et~al.} 2009, The Astrophysical
  Journal, 699, 564–584.
\newblock \url{http://dx.doi.org/10.1088/0004-637X/699/1/564}

\bibitem[{Sing(2018)}]{sing_2018observational}
Sing, D.~K. 2018, Observational Techniques With Transiting Exoplanetary
  Atmospheres, , , arXiv:1804.07357

\bibitem[{Skaf {et~al.}(2020)Skaf, Bieger, Edwards, Changeat, Morvan, Kiefer,
  Blain, Zingales, Poveda, Al-Refaie, Baeyens, Gressier, Guilluy, Jaziri,
  Modirrousta-Galian, Mugnai, Pluriel, Whiteford, Wright, Yip, Charnay,
  Leconte, Drossart, Tsiaras, Venot, Waldmann, \& Beaulieu}]{skaf_2020_ares}
Skaf, N., Bieger, M.~F., Edwards, B., {et~al.} 2020, ARES II: Characterising
  the Hot Jupiters WASP-127 b, WASP-79 b and WASP-62 b with HST, , ,
  arXiv:2005.09615

\bibitem[{Speagle(2020)}]{Speagle_2020}
Speagle, J.~S. 2020, Monthly Notices of the Royal Astronomical Society, 493,
  3132–3158.
\newblock \url{http://dx.doi.org/10.1093/mnras/staa278}

\bibitem[{Steinrueck {et~al.}(2019)Steinrueck, Parmentier, Showman, Lothringer,
  \& Lupu}]{Steinrueck_2019}
Steinrueck, M.~E., Parmentier, V., Showman, A.~P., Lothringer, J.~D., \& Lupu,
  R.~E. 2019, The Astrophysical Journal, 880, 14.
\newblock \url{http://dx.doi.org/10.3847/1538-4357/ab2598}

\bibitem[{{Stevenson} {et~al.}(2014){Stevenson}, {D{\'e}sert}, {Line}, {Bean},
  {Fortney}, {Showman}, {Kataria}, {Kreidberg}, {McCullough}, {Henry},
  {Charbonneau}, {Burrows}, {Seager}, {Madhusudhan}, {Williamson}, \&
  {Homeier}}]{stevenson_w43_1}
{Stevenson}, K.~B., {D{\'e}sert}, J.-M., {Line}, M.~R., {et~al.} 2014, Science,
  346, 838

\bibitem[{{Stevenson} {et~al.}(2017){Stevenson}, {Line}, {Bean}, {D{\'e}sert},
  {Fortney}, {Showman}, {Kataria}, {Kreidberg}, \& {Feng}}]{stevenson_w43_2}
{Stevenson}, K.~B., {Line}, M.~R., {Bean}, J.~L., {et~al.} 2017, \aj, 153, 68

\bibitem[{Stock {et~al.}(2018)Stock, Kitzmann, Patzer, \&
  Sedlmayr}]{Stock_2018}
Stock, J.~W., Kitzmann, D., Patzer, A. B.~C., \& Sedlmayr, E. 2018, Monthly
  Notices of the Royal Astronomical Society, doi:10.1093/mnras/sty1531.
\newblock \url{http://dx.doi.org/10.1093/mnras/sty1531}

\bibitem[{Taylor {et~al.}(2020)Taylor, Parmentier, Irwin, Aigrain, Lee, \&
  Krissansen-Totton}]{taylor_2020}
Taylor, J., Parmentier, V., Irwin, P. G.~J., {et~al.} 2020, Monthly Notices of
  the Royal Astronomical Society, 493, 4342–4354.
\newblock \url{http://dx.doi.org/10.1093/mnras/staa552}

\bibitem[{Tennyson {et~al.}(2016)Tennyson, Yurchenko, Al-Refaie, Barton, Chubb,
  Coles, Diamantopoulou, Gorman, Hill, Lam, Lodi, McKemmish, Na, Owens,
  Polyansky, Rivlin, Sousa-Silva, Underwood, Yachmenev, \&
  Zak}]{Tennyson_exomol}
Tennyson, J., Yurchenko, S.~N., Al-Refaie, A.~F., {et~al.} 2016, Journal of
  Molecular Spectroscopy, 327, 73 , new Visions of Spectroscopic Databases,
  Volume II.
\newblock
  \url{http://www.sciencedirect.com/science/article/pii/S0022285216300807}

\bibitem[{Tennyson {et~al.}(2020)Tennyson, Yurchenko, Al-Refaie, Clark, Chubb,
  Conway, Dewan, Gorman, Hill, Lynas-Gray, \& et~al.}]{Tennyson_2020_exomol}
---. 2020, Journal of Quantitative Spectroscopy and Radiative Transfer, 255,
  107228.
\newblock \url{http://dx.doi.org/10.1016/j.jqsrt.2020.107228}

\bibitem[{{Tinetti} {et~al.}(2013){Tinetti}, {Encrenaz}, \&
  {Coustenis}}]{Tinetti_2013}
{Tinetti}, G., {Encrenaz}, T., \& {Coustenis}, A. 2013, \aapr, 21, 63

\bibitem[{{Tinetti} {et~al.}(2018){Tinetti}, {Drossart}, {Eccleston},
  {Hartogh}, {Heske}, {Leconte}, {Micela}, {Ollivier}, {Pilbratt}, {Puig}, \&
  et~al.}]{Tinetti_ariel}
{Tinetti}, G., {Drossart}, P., {Eccleston}, P., {et~al.} 2018, Experimental
  Astronomy, doi:10.1007/s10686-018-9598-x

\bibitem[{{Venot} {et~al.}(2012){Venot}, {H{\'e}brard}, {Ag{\'u}ndez},
  {Dobrijevic}, {Selsis}, {Hersant}, {Iro}, \& {Bounaceur}}]{venot_chem_HJ}
{Venot}, O., {H{\'e}brard}, E., {Ag{\'u}ndez}, M., {et~al.} 2012, \aap, 546,
  A43

\bibitem[{Venot {et~al.}(2020)Venot, Parmentier, Blecic, Cubillos, Waldmann,
  Changeat, Moses, Tremblin, Crouzet, Gao, \& et~al.}]{Venot_2020}
Venot, O., Parmentier, V., Blecic, J., {et~al.} 2020, The Astrophysical
  Journal, 890, 176.
\newblock \url{http://dx.doi.org/10.3847/1538-4357/ab6a94}

\bibitem[{{Waldmann} {et~al.}(2015{\natexlab{a}}){Waldmann}, {Rocchetto},
  {Tinetti}, {Barton}, {Yurchenko}, \& {Tennyson}}]{Waldmann_taurex2}
{Waldmann}, I.~P., {Rocchetto}, M., {Tinetti}, G., {et~al.} 2015{\natexlab{a}},
  \apj, 813, 13

\bibitem[{{Waldmann} {et~al.}(2015{\natexlab{b}}){Waldmann}, {Tinetti},
  {Rocchetto}, {Barton}, {Yurchenko}, \& {Tennyson}}]{Waldmann_taurex1}
{Waldmann}, I.~P., {Tinetti}, G., {Rocchetto}, M., {et~al.} 2015{\natexlab{b}},
  \apj, 802, 107

\bibitem[{Wilzewski {et~al.}(2016)Wilzewski, Gordon, Kochanov, Hill, \&
  Rothman}]{2016_WILZEWSKI_C2H2}
Wilzewski, J.~S., Gordon, I.~E., Kochanov, R.~V., Hill, C., \& Rothman, L.~S.
  2016, Journal of Quantitative Spectroscopy and Radiative Transfer, 168, 193 .
\newblock
  \url{http://www.sciencedirect.com/science/article/pii/S0022407315002988}

\bibitem[{Woitke {et~al.}(2018)Woitke, Helling, Hunter, Millard, Turner,
  Worters, Blecic, \& Stock}]{Woitke_2018}
Woitke, P., Helling, C., Hunter, G.~H., {et~al.} 2018, Astronomy \&
  Astrophysics, 614, A1.
\newblock \url{http://dx.doi.org/10.1051/0004-6361/201732193}

\bibitem[{Yurchenko {et~al.}(2011)Yurchenko, Barber, \&
  Tennyson}]{Yurchenko_2011_NH3}
Yurchenko, S.~N., Barber, R.~J., \& Tennyson, J. 2011, Monthly Notices of the
  Royal Astronomical Society, 413, 1828–1834.
\newblock \url{http://dx.doi.org/10.1111/j.1365-2966.2011.18261.x}

\bibitem[{{Yurchenko} \& {Tennyson}(2014)}]{exomol_ch4}
{Yurchenko}, S.~N., \& {Tennyson}, J. 2014, Monthly Notices of the Royal
  Astronomical Society, 440, 1649

\bibitem[{Zellem {et~al.}(2014)Zellem, Lewis, Knutson, Griffith, Showman,
  Fortney, Cowan, Agol, Burrows, Charbonneau, \& et~al.}]{Zellem_hd209_phase}
Zellem, R.~T., Lewis, N.~K., Knutson, H.~A., {et~al.} 2014, The Astrophysical
  Journal, 790, 53.
\newblock \url{http://dx.doi.org/10.1088/0004-637X/790/1/53}

\bibitem[{{Zhang} {et~al.}(2019){Zhang}, {Chachan}, {Kempton}, \&
  {Knutson}}]{Zhang_platon}
{Zhang}, M., {Chachan}, Y., {Kempton}, E. M.~R., \& {Knutson}, H.~A. 2019,
  \pasp, 131, 034501

\bibitem[{Zhang {et~al.}(2018)Zhang, Knutson, Kataria, Schwartz, Cowan,
  Showman, Burrows, Fortney, Todorov, Desert, \& et~al.}]{Zhang_2018}
Zhang, M., Knutson, H.~A., Kataria, T., {et~al.} 2018, The Astronomical
  Journal, 155, 83.
\newblock \url{http://dx.doi.org/10.3847/1538-3881/aaa458}

\end{thebibliography}


\renewcommand{\floatpagefraction}{.9}%

\appendix 
\onecolumngrid
\section{The phase dependent emission model.}\label{apx:phase_model}

\begin{figure}[H]
\centering
    \includegraphics[width = 0.7\textwidth]{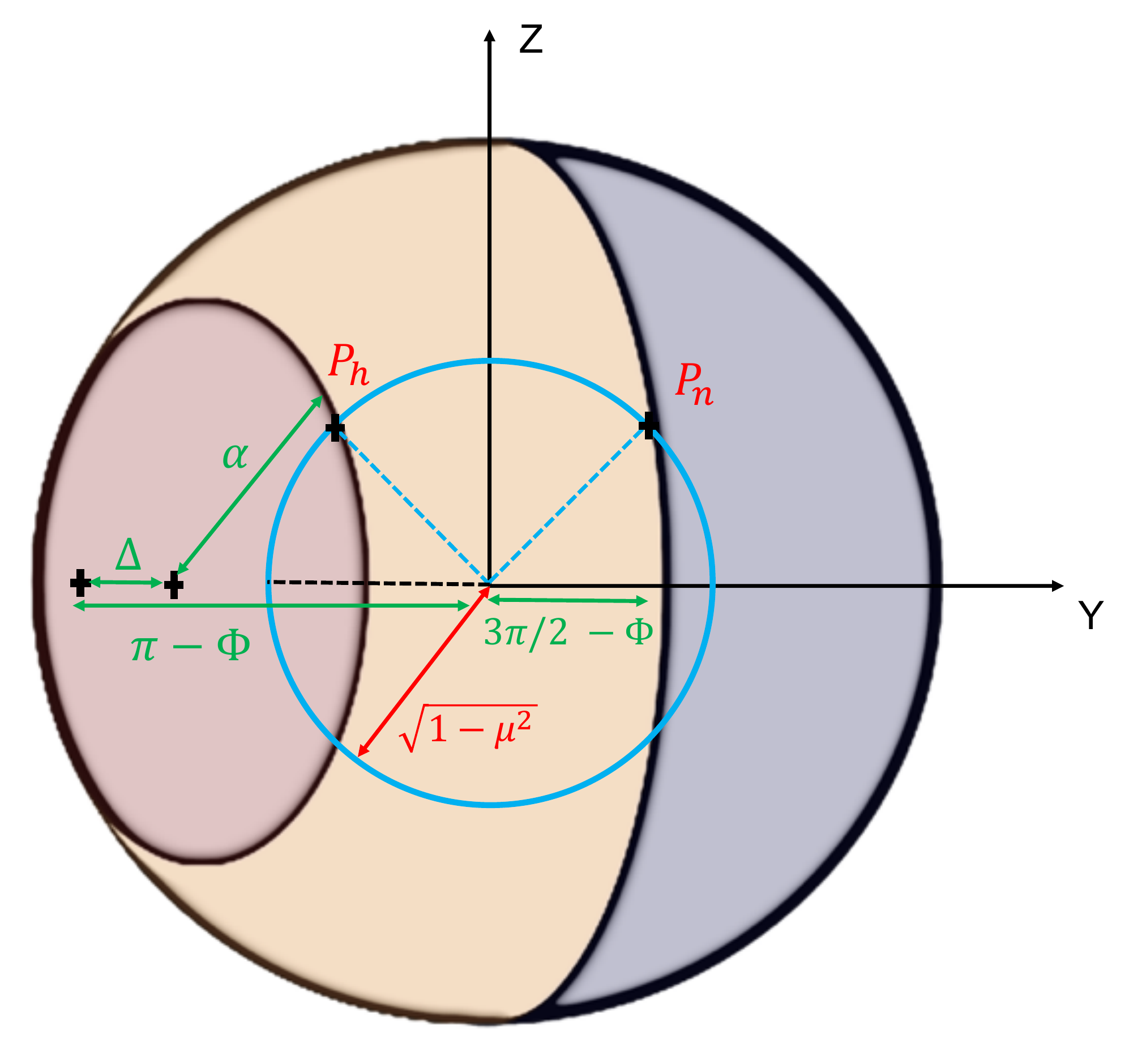}
    \caption{Diagram of the phase curve geometry used in this paper. We show the 3 regions in shaded colors; Red: hot-spot; Orange: day-side; Blue: night-side. The labels in green corresponds to 3-dimensional angles emerging from the centre of the planet sphere. The red labels are 2-dimensional projections in the (y,z) plane, which is perpendicular to the line of sight (tangent to the celestial sphere). The blue circle corresponds to the 2 dimensional Gaussian quadrature integration circles, also projected onto the celestial sphere plane. $\Phi$ is the phase angle considered, $\Delta$ parameterise the hot-spot offset, $\alpha$ defines the size of the hot-spot region. The integration circle is defined by its viewing angle $\mu$=cos($\theta$). The coordinates of P$_h$ and P$_n$, the points of intersection with the integration circle, are the quantities to determine.}
    \label{fig:diagram_geometry}
\end{figure}

The phase dependent emission model uses the principles developed in \cite{changeat_2020_phase1}. The new geometry accounts for high day-night temperature contrasts and presence of a hot-spot by simulating the planet using 3 separated homogeneous regions: hot-spot, day-side and night-side (see Figure \ref{fig:diagram_geometry}).

For each region, the angle dependent specific intensity at the top of the atmosphere is given by:
\begin{equation}\label{eq:em_taurex_top}
	I_{\lambda}(0, \mu) = B_{\lambda}(T_{surf}) e^{-\frac{\tau_{surf}}{\mu}} + \frac{1}{\mu} \int_{\tau_0}^{\tau_{surf}} B_{\lambda}(T_{\tau}) e^{-\frac{\tau}{\mu}} d\tau,
\end{equation}

where $\mu$ is the viewing angle, T$_{surf}$ is the surface temperature, $\tau$ is the optical depth (labelled with $surf$ for the surface and 0 for the top of the atmosphere) and B is the Plank function \citep{Waldmann_taurex2}. The integral over $\tau$ is performed over $N_{layers}$ layers equally spaced in logarithmic pressure.

To obtain the total specific intensity, Equation \ref{eq:em_taurex_top} must be integrated over $\mu$. This step is numerically done using the Gaussian quadrature integration technique, which consist in integrating the planet emission over $N_{G}$ integration circles at discretised viewing angles $\mu_i$. In the standard emission case, the total specific intensity at the top of the atmosphere is given by:

\begin{equation}\label{eq:emission_tot}
	I_{\lambda} =2\pi \sum_i^{N_{G}} \omega_i \mu_i I_{\lambda}(0, \mu_i),
\end{equation}
where $\omega_i$ are the quadrature weights.

In the case of a planet with regions and a phase dependent emission additional weight coefficients can be introduced to account for the contribution of each region to each Gaussian quadrature term in the final emission. Thus, Equation \ref{eq:emission_tot} becomes:
\begin{equation}\label{eq:final_I}
\begin{split}
    I_{\lambda}(\Phi) &= 2 \pi \sum_i^{N_G} \left( I_{\lambda, i}^h C^h_i(\Phi) + I_{\lambda, i}^d C^d_i(\Phi) + I_{\lambda, i}^n C^n_i(\Phi) \right) \omega_i \mu_i,
\end{split}
\end{equation}
where $\Phi$ is the phase angle considered. $I_{\lambda, i}^h$, $I_{\lambda, i}^d$ and $I_{\lambda, i}^n$ are the day, terminator and night intensities at the top of the atmosphere for the Gaussian point $\mu_i$. For each phase considered and for a given geometry, the C coefficients (hot-spot: C$^h$; day-side: C$^d$; night-side C$^n$) allow to maps the weight of each region to each Gaussian quadrature integration circle. The total number of Gaussian quadrature points is labelled $N_G$.

Since the coefficients C are phase dependent, Equation \ref{eq:final_I} gives the specific intensity for a given phase angle $\Phi$. When another phase angle is evaluated, $I_{\lambda, i}^h$, $I_{\lambda, i}^d$ and $I_{\lambda, i}^n$ don't need to be re-computed and only the values of the C coefficients, which have been pre-computed, are updated.

These coefficients corresponds to the intersections between the Gaussian quadrature integration circles and the different regions boundaries (we note those P$_h$ and P$_n$ in Figure \ref{fig:diagram_geometry}). The specific intensity of each region is calculated using the base emission model from TauREx\,3 \citep{al-refaie_taurex3}. The coefficients are projected onto the planet plane (orthogonal to the line of sight). In 3D, the hot-spot and terminator boundaries are considered as circles on the surface of the planet sphere. These are therefore described by intersections between planes (defined by the chosen hot-spot and terminator positions) and the sphere (for the planet). In addition to this, the Gaussian integration circles are equivalent in 3D to cylinder of circular bases and directions parallel to the line of sight. In Cartesian coordinates ($x,y,z$) with $x$ pointing towards the observer, the problem is therefore equivalent to solving the following set of equations : \\ 
- Region Plane : $Ax + By + Cz + D = 0$ \\
- Planet Sphere: $x^2 + y^2 + z^2 -1 = 0$ \\
- Gaussian integration Cylinder: $y^2 + z^2 - 1 + \mu^2 = 0$ 

Solving these equations for ($x, y, z$) leads to 4 possible solutions for the point of intersection $P_{int}$ that can be formulated using Expression \ref{eq:phasecurve_core}.

\noindent\makebox[\linewidth]{\rule{0.82\paperwidth}{0.4pt}}
\begin{equation}\label{eq:phasecurve_core}
\begin{matrix}
    \begin{pmatrix}
           x \\ \\
           y \\ \\
           z
    \end{pmatrix} =

    \begin{pmatrix}
    \mp \mu \\ \\
    \frac{\pm(AB\mu - BD \pm' C \sqrt{-A^2 \mu^2 \pm 2 A \mu D - B^2 \mu^2 + B^2 - C^2 \mu^2 + C^2 - D^2})}{B^2 + C^2} \\ \\
    \frac{\pm(AC\mu +CD \pm' B \sqrt{-A^2 \mu^2 \pm 2 A \mu D - B^2 \mu^2 + B^2 - C^2 \mu^2 + C^2 - D^2})}{B^2 + C^2}
\end{pmatrix}
\end{matrix}
\end{equation}
\noindent\makebox[\linewidth]{\rule{0.82\paperwidth}{0.4pt}}

The coefficients A, B, C, D define the plane equation, hence the position of the hot-spot and the terminator boundaries. These are defined as function of $\theta$ the inclination angle (assumed equal to 0 in this study), $\Phi$ the phase angle, $\Delta$ the hot-spot shift and $\alpha$ the hot-spot size angle and are unique to each phase angle. Their expressions are summarised in Table \ref{tab:coeffs}. \\

\begin{table}
\centering
\begin{tabular}{cccc}
\hline\hline
Coefficient & Hot-spot boundary & Terminator\\
\hline 
A & cos($\theta$) cos($\Phi - \Delta$) & cos($-\theta$) cos($\Phi - \pi$)   \\
B & cos($\theta$) sin($\Phi - \Delta$) & cos($-\theta$) sin($\Phi - \pi$)   \\
C & sin($\theta$) & sin($-\theta$)   \\
D & cos($\alpha$) & cos($\pi /2$)   \\
 \hline\hline
\end{tabular}
\caption{List of the A, B, C and D coefficients depending on the boundary considered (hot-spot and terminator) for the Equation \ref{eq:phasecurve_core} .}\label{tab:coeffs}
\end{table}
Finally, the coefficients $C^h$ and $C^n$ can be calculated from Equation \ref{eq:phasecurve_core} as the angles from the $y$ axis to the intersection points $P_{h}$ and $P_{n}$. For a point of intersection $P_{int}$ of coordinates ($y_{int},z_{int}$), the angle is given by:
\begin{equation}
    C^{int} = \mathrm{arctan}\left( \frac{z_{int}}{y_{int}}\right) / \pi.
\end{equation}
The last coefficient is calculated using the remaining difference:
\begin{equation}
    C^d_i = 1 - C^h_i - C^n_i.
\end{equation}
These equations can be used immediately in most cases. However, some particular situations, corresponding to the cases of ill definitions of Equation \ref{eq:phasecurve_core}, need to be accounted individually: \\
- When the integration circle is fully inside a region R, the intersection coefficients C$^h_i$ and C$^n_i$ don't exist. Then the corresponding coefficient C$^R_i$ = 1 and the two other ones are 0. \\
- If only one coefficient C$^R_i$ exist, then the integration circle is shared with only 2 regions and the other coefficient is equal to 1 - C$^R_i$. 

Finally, the observe planet specific intensity can be computed  for each phase using Equation \ref{eq:final_I}. The observed flux ratio $\Delta_{\Phi}$ at a phase $\Phi$ is then computed using:
\begin{equation}
    \Delta_{\Phi} = \frac{F_p}{F_s} = \frac{I \times R_p^2}{I_s \times R_s^2},
\end{equation}
where $R_p$ is the planetary radius, $I_s$ the stellar intensity and $R_s$ is the radius of the star.


\clearpage

\section{Transmission model}\label{apx:trans_model}

The transmission model considers that the stellar light is filtered through both the day-side and night-side. This consideration accounts for the 3D effects explored in \cite{caldas_3deffects,Pluriel_2020}, where large day-night temperature contrasts on tidally locked irradiated exoplanets create sudden geometrical changes at the planet limb. In their work, they highlighted that the more traditional assumption of 1D atmosphere leads to strong biases in the retrieved transmission spectra. In order to account for these effects, we assumed that the atmosphere can be described by two contributing regions (the day and night-sides in our model). For this study the light rays seen in transit are therefore filtered through the day-side region for half of their path and through the night-side region for the other half. Retrievals from \cite{Irwin_w43b_phase,stevenson_w43_1,stevenson_w43_2} and 3D GCM from \cite{Kataria_2015_W43GCM} have shown that the planet WASP-43\,b presents a large day-night contrast, thus confirming the relevance of this treatment. In practice, we compute both day and night properties and combine them using:
\begin{equation}
   A(\lambda) = 2 \int_z \left( (R_p + z)(1 -e^{-\tau_1(\lambda,z)} \times e^{-\tau_2(\lambda,z)}) \right)dz,
\end{equation}
where A is the atmosphere contribution, $\tau_1$ is the optical depth along the line of sight of the day-side and $\tau_2$ is the one from the night-side. 

The final transit depth $\Delta_{|}$ is calculated by:
\begin{equation}
    \Delta_{|} = \frac{R_p^2 + A(\lambda)}{R_s^2}.
\end{equation}
This description allows us to account for day and night-side properties seen in transmission. The transmission spectrum, therefore provides some constraints linking both hemispheres. For planets where only the transmission spectrum is observed, {it might be difficult to break the} degeneracies arising from 3-dimensional effects at the limb. In a full phase curve analysis however, the day-side is informed from the various emission models and properties, meaning that the degeneracy is diminished and that night-side and terminator properties can be inferred from the residual information in the transmission spectrum.

\clearpage

\section{Retrievals on individual phases}\label{apx:append_indiv_spectra}

\begin{figure}[H]
\centering
    \includegraphics[width = 0.99\textwidth]{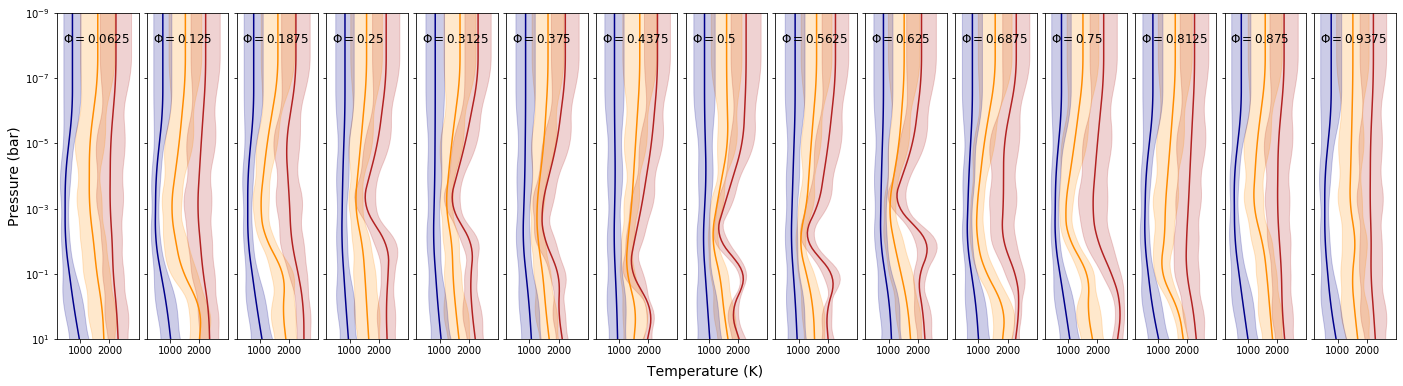}
    \includegraphics[width = 0.99\textwidth]{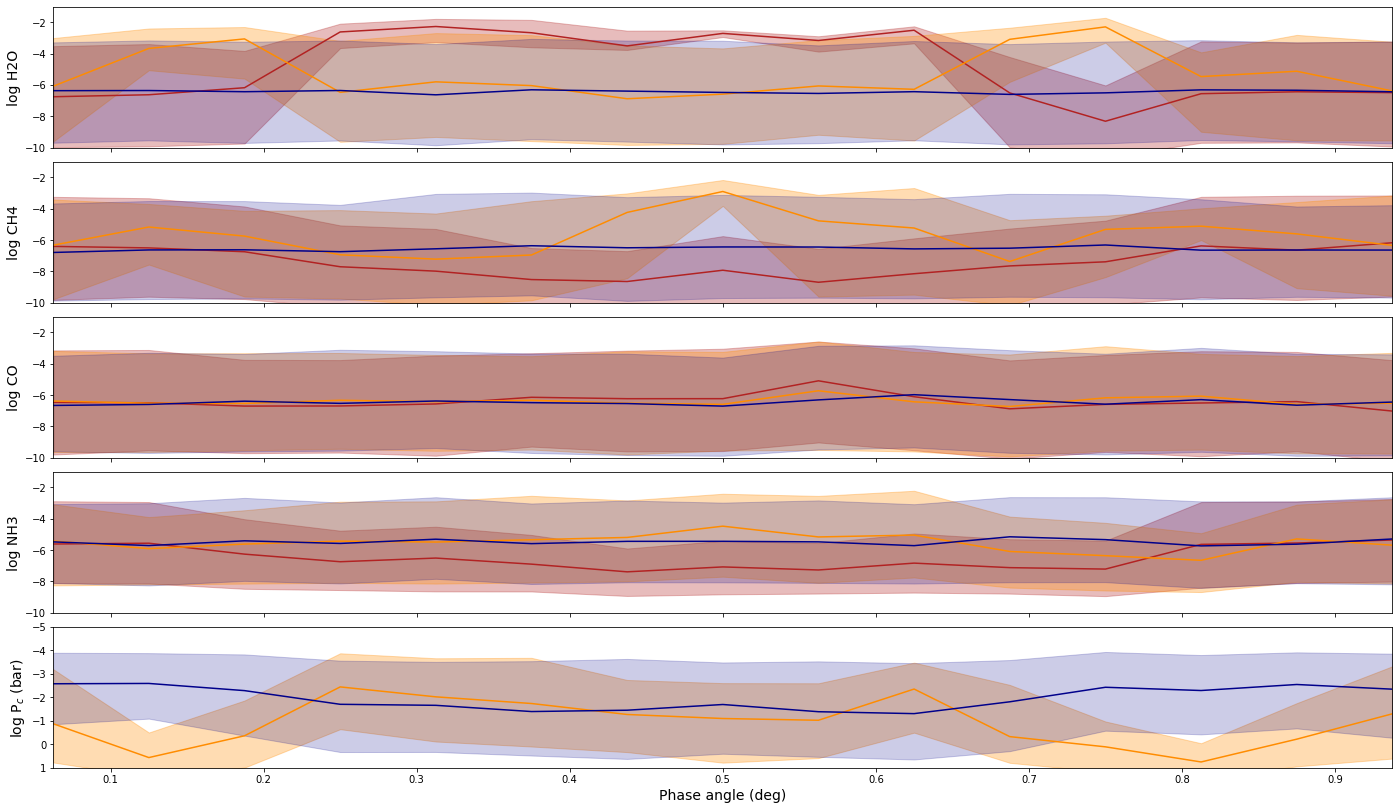}
    \caption{Temperature profiles (top) and retrieved chemical/clouds values (bottom) as function of phase for the individual spectra retrievals. The temperature profiles and retrieved chemistry are aligned for phase angles between the top and bottom panels.}
    \label{fig:ret_individual}
\end{figure}

We explore the information content of each phase by running our phase curve retrieval model separately for each spectrum. This way, we assess whether our combined phase curve retrieval strategy allows us to extract more information from this dataset. To avoid biases, we use free parametric models for both chemistry (constant with altitude) and temperature descriptions (NPoint profile described in the Method section). The molecules considered are H$_2$O, CH$_4$, CO and NH$_3$ and clouds are added to the day and night-sides. The retrieved thermal profiles, chemical abundances and cloud top pressures are presented in Figure \ref{fig:ret_individual}.

For the temperature profile, the expected behaviour is observed. We see that the observations close to the transit are not constraining the day and hot-spot profiles but tighten the constraints on the night-side profile, while the observations closer to eclipse give more information regarding the day and hot-spot temperature structure. For most of the profiles, the temperatures are decreasing with altitude, but we note that some phases around 0.5 show a weak thermal inversion at the hot-spot. The retrieved chemistry indicates the presence of water vapour in both the hot spot (dominant for phases between 0.25 and 0.625) and the day-side (dominant outside). Interestingly, methane is only detected at the day-side in one spectrum (phase 0.5), which is surprising and might indicate remaining systematic errors in this spectrum. Additional constraints on other molecules and the night-side region remain poor due to the low contribution of this region. For the clouds, these are not detected in either the day-side and the night-side, but stronger constraints can be obtained for the day-side at intermediate phases due to the greater flux. Overall, our phase curve retrievals on the individual spectra provide indications that 2-dimensional effects can already be captured in emission spectra from HST (eclipses or phase curves) since the model was successfully able to separate the chemical and temperature structures for all three considered region. This results was already discussed in previous work from \cite{taylor_2020, feng2020_2d}, where the possible biases of using simpler 1-dimensional retrieval models are explored. The extraction of a consistent picture between these different retrievals, which are not considering the spectra together, provides us indications that the whole dataset might be consistent and adapted for our unified retrieval exploration.
\clearpage

\section{Retrieval exploration of various free temperature-pressure models.}\label{apx:tp_investigations}

Here, we explore the required complexity when using free thermal profiles to extract information from HST and Spitzer spectra. The results of this Appendix can serve as rough guideline but are only relevant for the considered WASP-43\,b dataset as other planet spectra might have significantly different information content. In the first place, we evaluate the required complexity for a single phase, using 1D retrievals of the day-side spectrum. This exercise is done with and without the Spitzer data to show the benefits of adding a larger wavelength coverage. Then we perform a unified phase-curve exploration. For each case, we evaluate 4 different models. The first 3 models have 3, 4 and 5 temperature-pressure nodes and for each point, we attempt to retrieve both temperature and pressure. Because including more fully free points dramatically increases the required number of samples (above 10 millions) to reach an accurate sampling of the parameter space, the last model includes 7 temperature points at fixed pressures (10$^6$, 10$^5$,10$^4$, 10$^3$, 10$^2$, 1 and 0.01 Pa). For a low number of nodes ($<5$), retrieving both pressure and temperature is required as fixing a particular node pressure would be arbitrary. When a larger number of nodes ($>5$) is used, fixing the pressure has less consequences as all regions of the atmosphere are well sampled. We note that the number of free parameters of the different models is respectively 5, 7, 9 and 7, meaning that the 4-point profile and the 7-point with fixed pressure nodes have the same number of parameters. As those profiles present different characteristics, it is might be difficult to define the absolute concept of `complexity'. For this exploration, we assume chemical equilibrium in all models and only retrieve the metallicity and C/O ratios. \\

{\bf $\bullet$ 1D retrievals on the day-side HST spectrum} \\

\begin{figure}[H]
\centering
    \includegraphics[width = 0.86\textwidth]{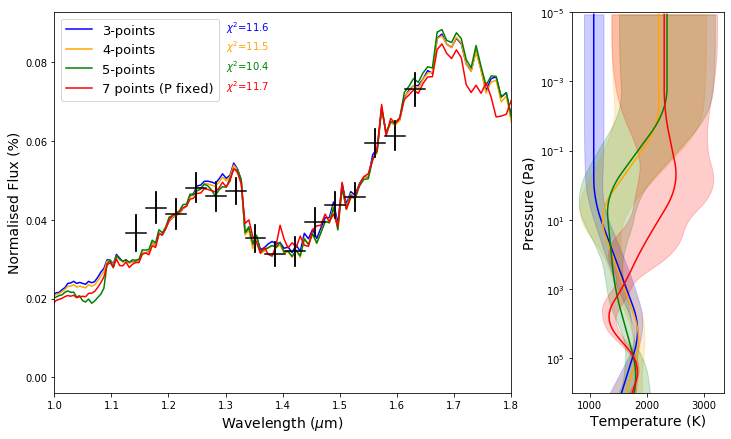}
    \caption{Retrieved spectra (left) and temperature profiles (right) of different 1D retrievals when assuming different free temperature-pressure profiles. The datapoints are for the phase 0.5 of the HST data.}
    \label{fig:spectra_HST1D_tp_explor}
\end{figure}
\begin{figure}
\centering
    \includegraphics[width = 0.45\textwidth]{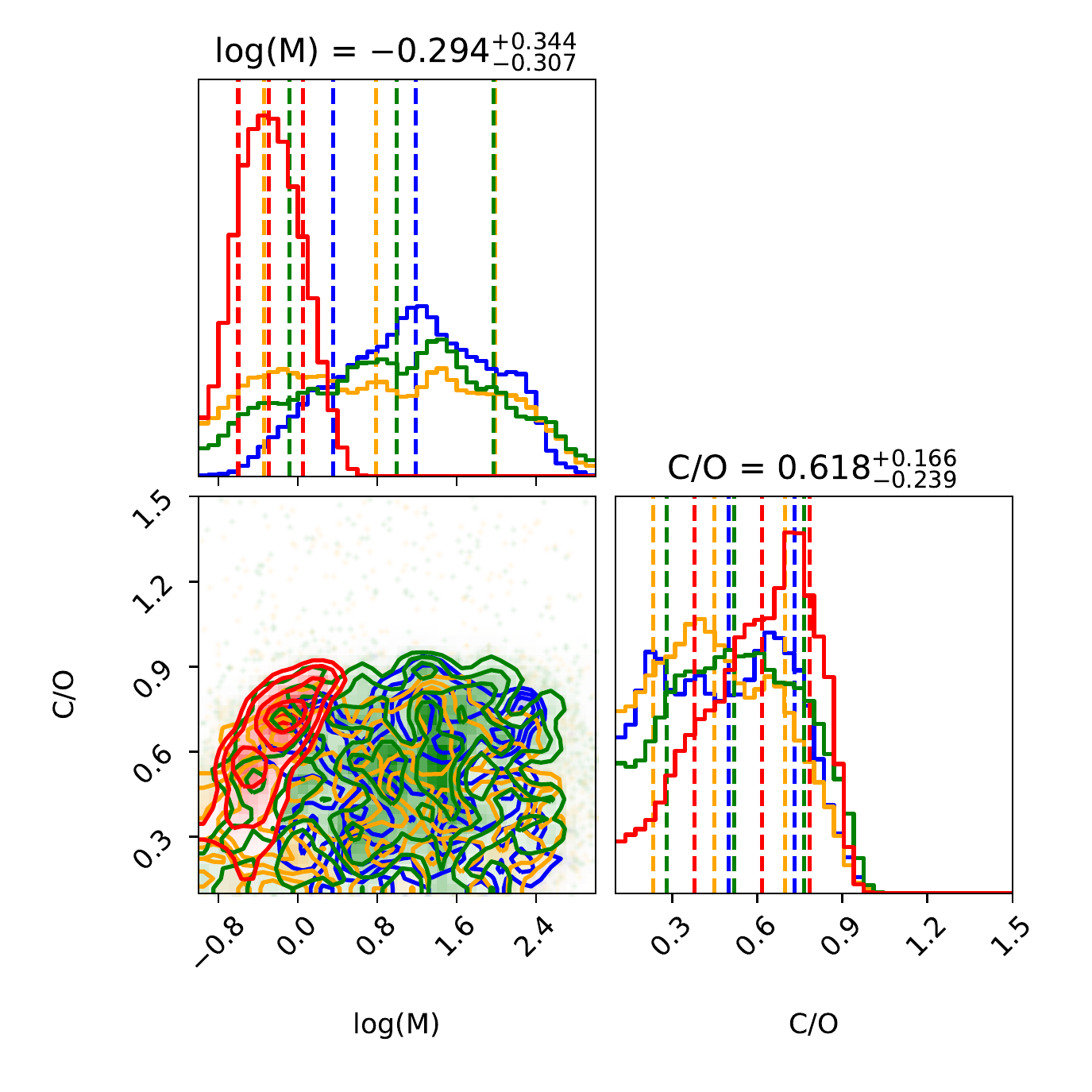}
    \includegraphics[width = 0.45\textwidth]{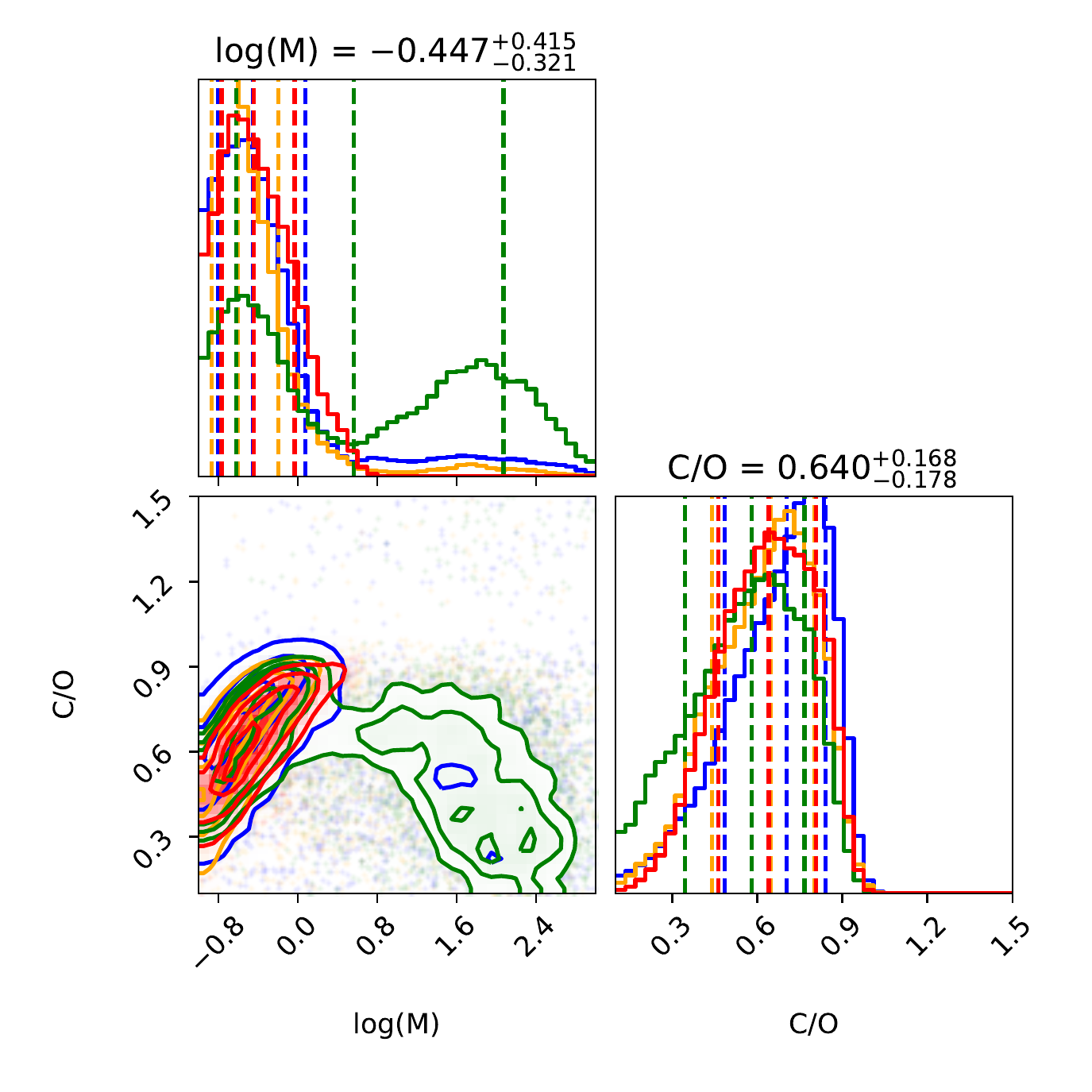}
    \caption{Posterior distributions of the 1D retrievals on the phase 0.5 when assuming different free temperature-pressure profiles. The color scheme is the same as Figure \ref{fig:spectra_HST1D_tp_explor}. Left: HST only; Right: HST+Spitzer.}
    \label{fig:post_1D_tp_explor}
\end{figure}

The retrievals assuming different temperature-pressure strucdtures on the HST only dataset are presented in Figures \ref{fig:spectra_HST1D_tp_explor} and \ref{fig:post_1D_tp_explor}. The different scenarios lead to very similar log(E): 121.1, 121.2, 122.2 and 120.7 for respectively the 3-point, the 4-point, the 5-point and the 7-point with fixed pressure for the nodes. These values are all within a small range and since the Bayesian evidence naturally penalises for model complexity, it indicates that all the tested models are good candidates to explain the data. Taking a conservative approach, we however favour the simplest model that best explains the data (Occam’s razor principle), which is the 3-point temperature profile. We note, however, that all the models of higher complexity present some hints of thermal inversions in the upper atmosphere with large uncertainties. \\

{\bf $\bullet$ 1D retrievals on the day-side HST+Spitzer spectrum} \\

\begin{figure}
\centering
    \includegraphics[width = 0.86\textwidth]{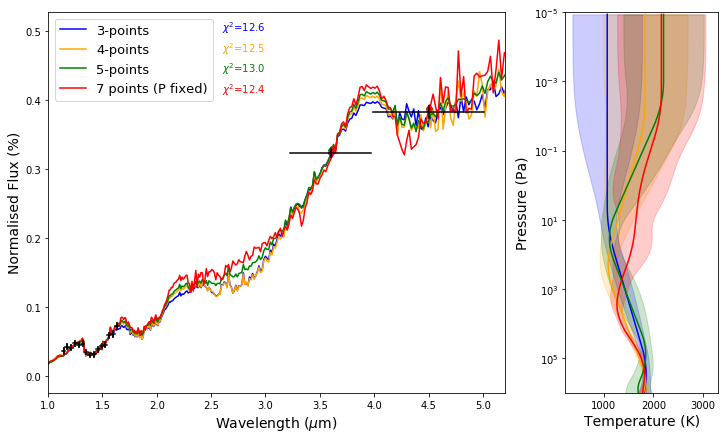}
    \caption{Retrieved spectra (left) and temperature profiles (right) of different 1D retrievals when assuming different free temperature-pressure profiles. The datapoints are for the phase 0.5 of the HST+Spitzer data.}
    \label{fig:spectra_Spz1D_tp_explor}
\end{figure}

When considering the HST+Spitzer data, we find a similar conclusion (see Figures \ref{fig:post_1D_tp_explor} and \ref{fig:spectra_Spz1D_tp_explor}). All the models are within the same range of log(E) with values of respectively 134.0, 133.9, 132.1 and 133.6 for the tested models. Again, this does not allow to conclude on a particular model, but a conservative approach would be to select the `simplest model': 3-point model.    \\

{\bf $\bullet$ Unified phase curve retrievals on HST only} \\

\begin{figure}
\centering
    \includegraphics[width = 0.86\textwidth]{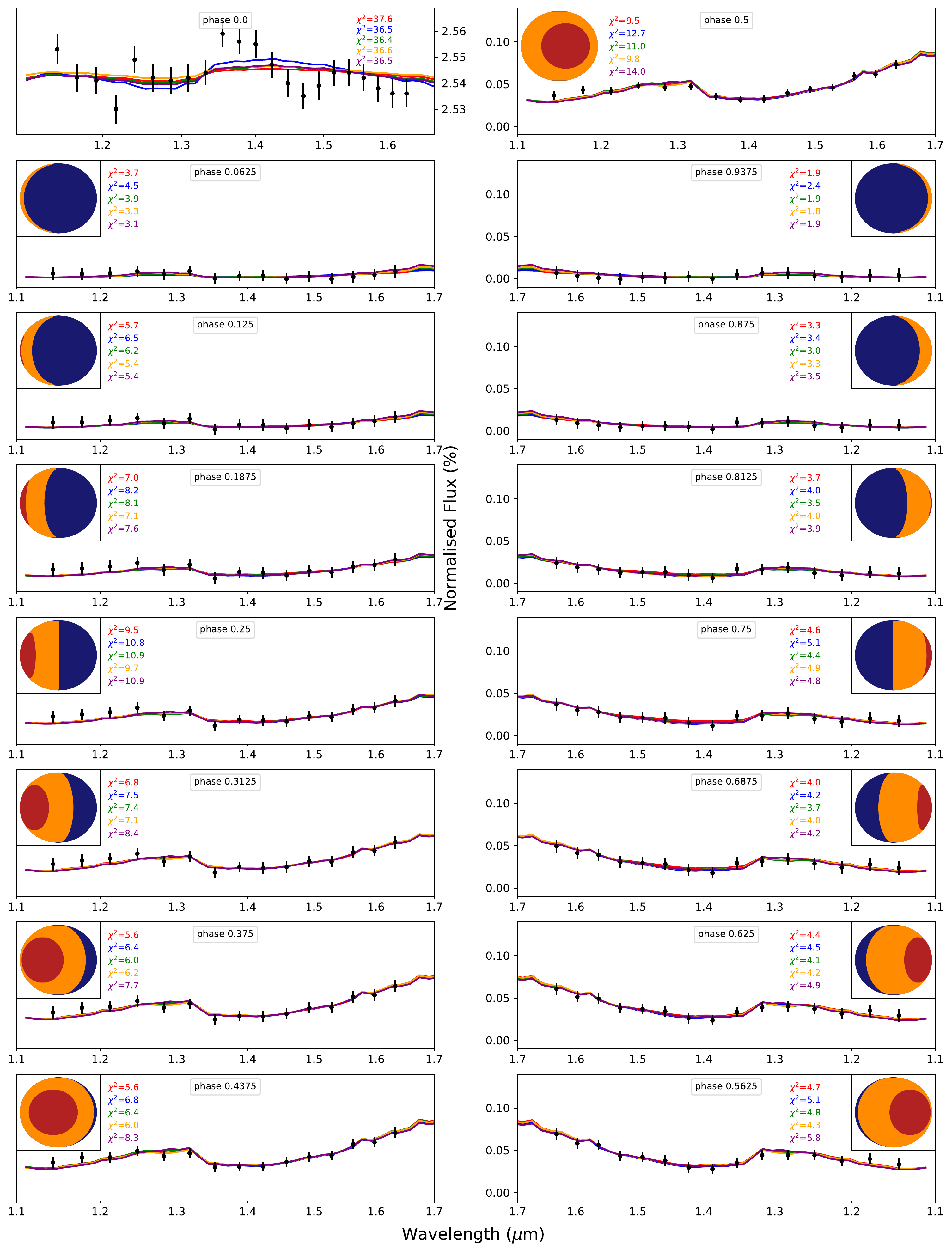}
    \caption{Retrieved spectra from unified retrievals on the HST data when assuming different free temperature-pressure profiles. Blue: 3-point; Green: 4-point; Orange: 5-point (informed); Purple: 5-point (no inversion); Red: 7-point (fixed P)}
    \label{fig:spectra_HSTUnified_tp_explor}
\end{figure}

\begin{figure}
\centering
    \includegraphics[width = 0.86\textwidth]{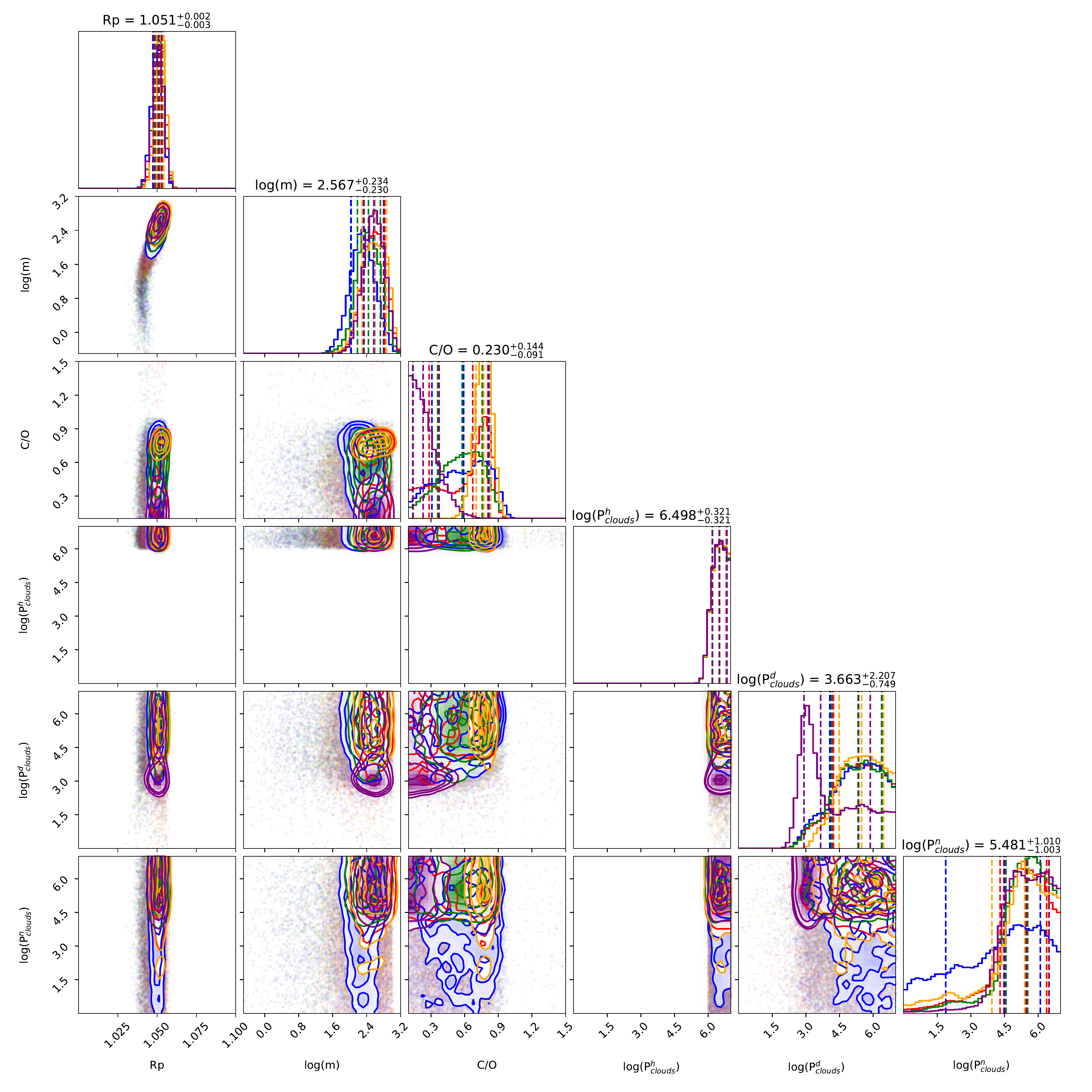}
    \caption{Posteriors from unified retrievals on the HST data when assuming different free temperature-pressure profiles. Blue: 3-point; Green: 4-point; Orange: 5-point (informed); Purple: 5-point (no inversion); Red: 7-point (fixed P)}
    \label{fig:post_HSTUnified_tp_explor}
\end{figure}

\begin{figure}
\centering
    \includegraphics[width = 0.4\textwidth]{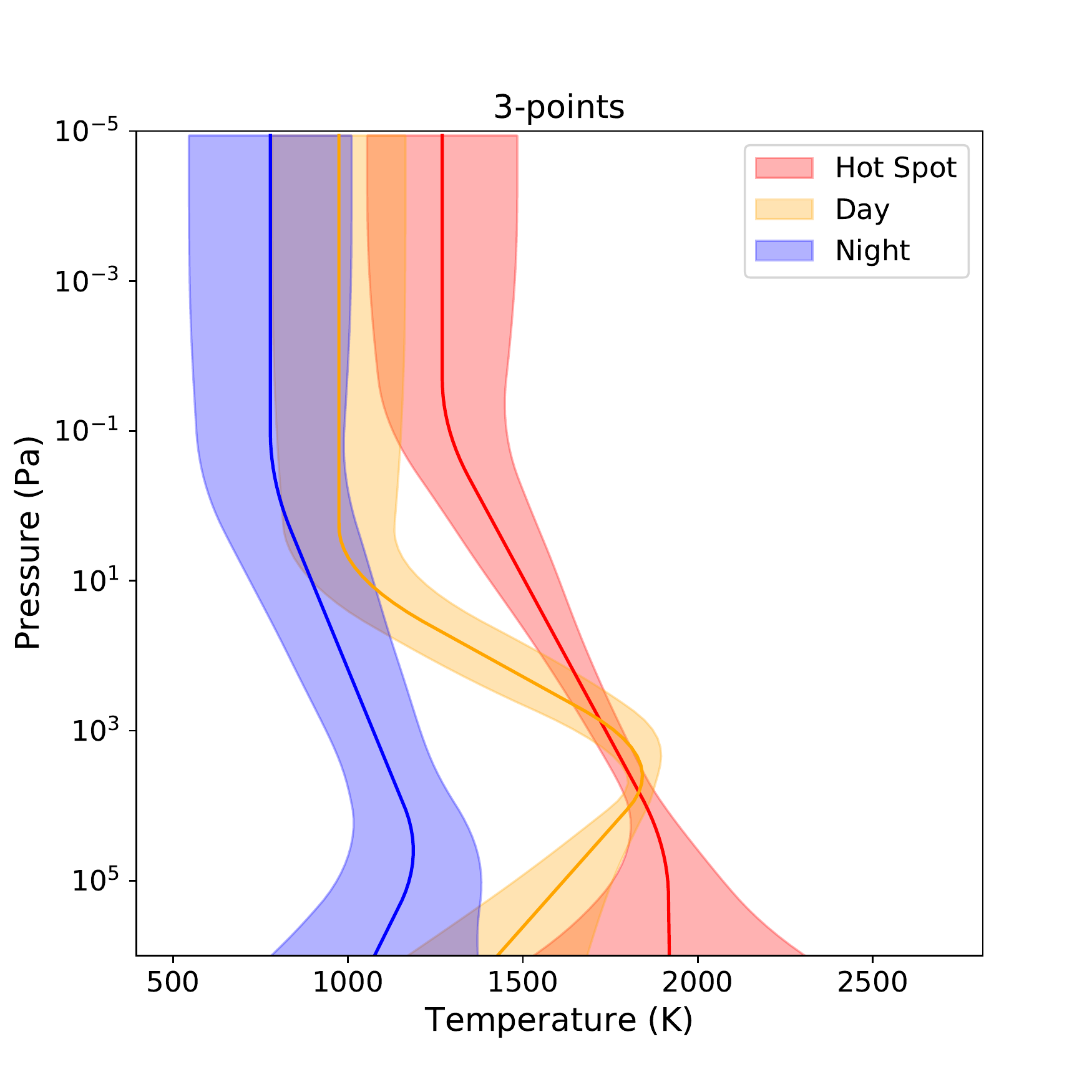}
    \includegraphics[width = 0.4\textwidth]{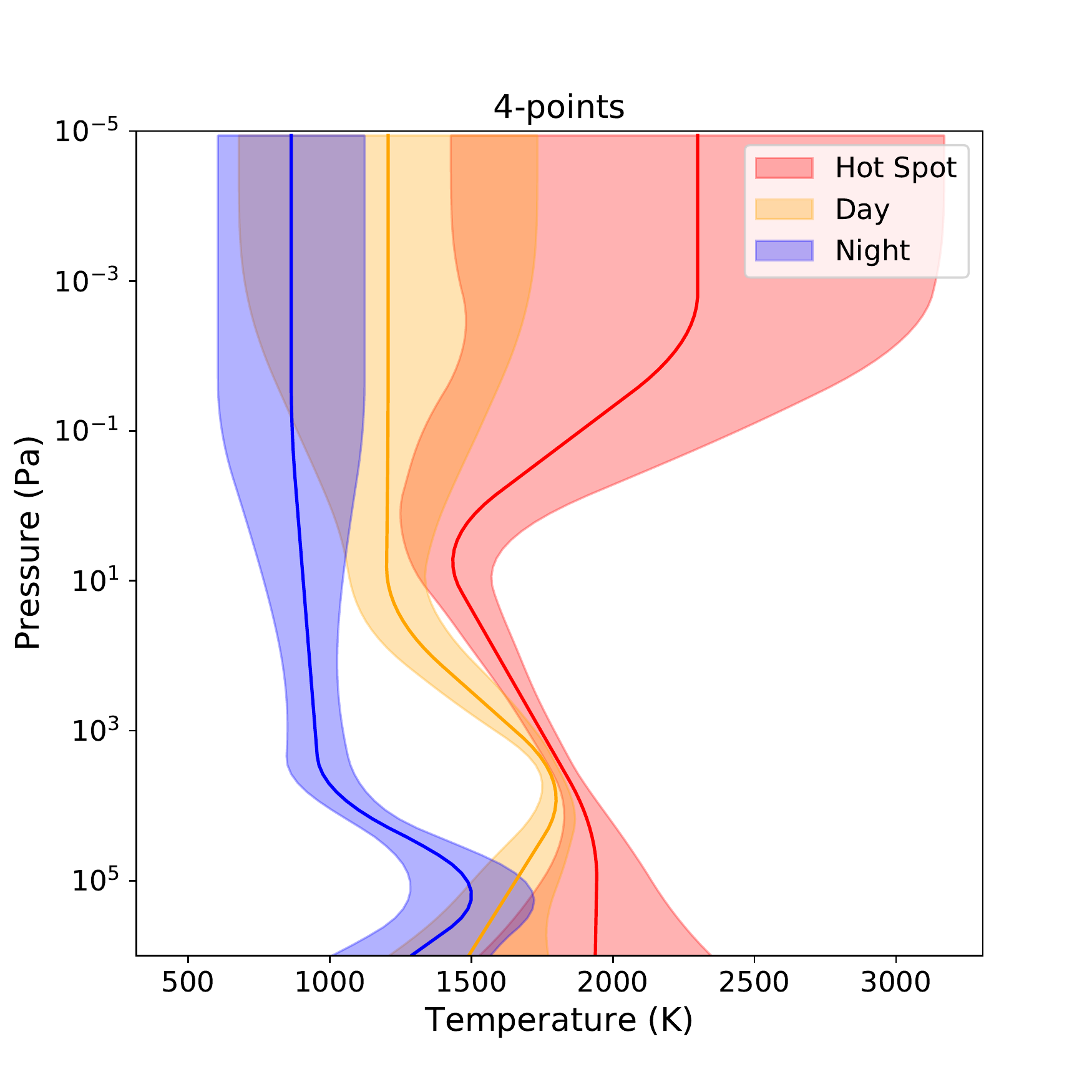}
    \includegraphics[width = 0.4\textwidth]{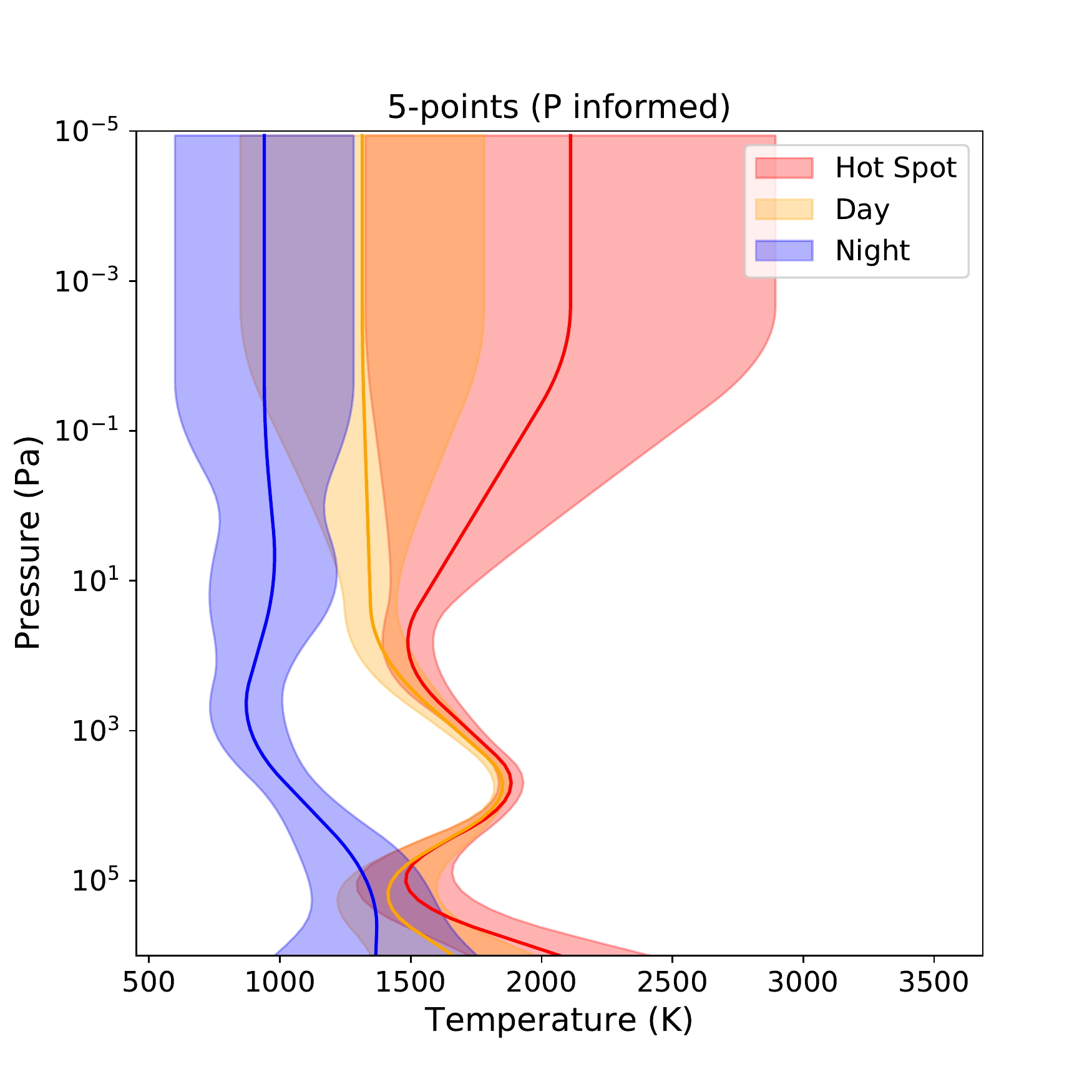}
    \includegraphics[width = 0.4\textwidth]{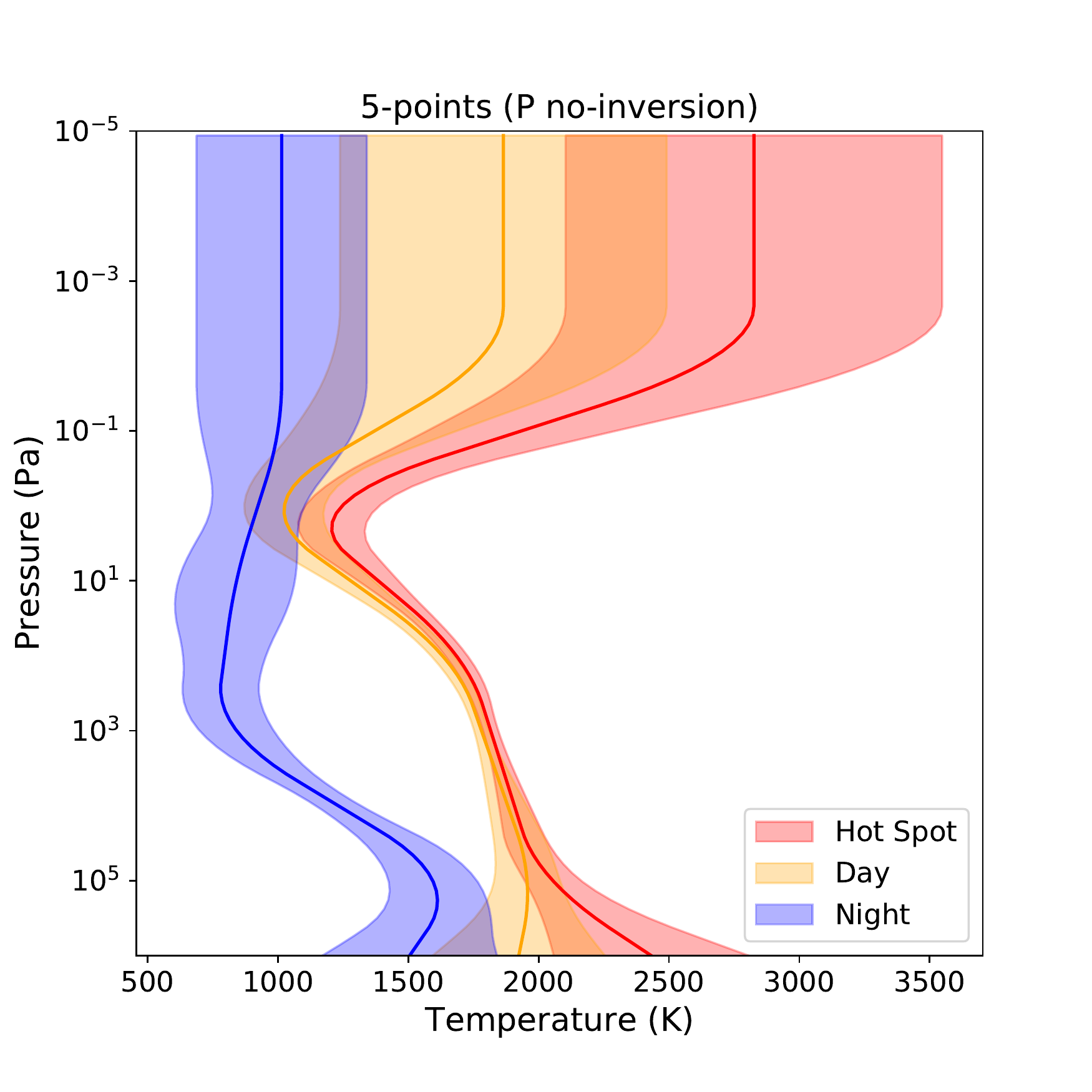}
    \includegraphics[width = 0.4\textwidth]{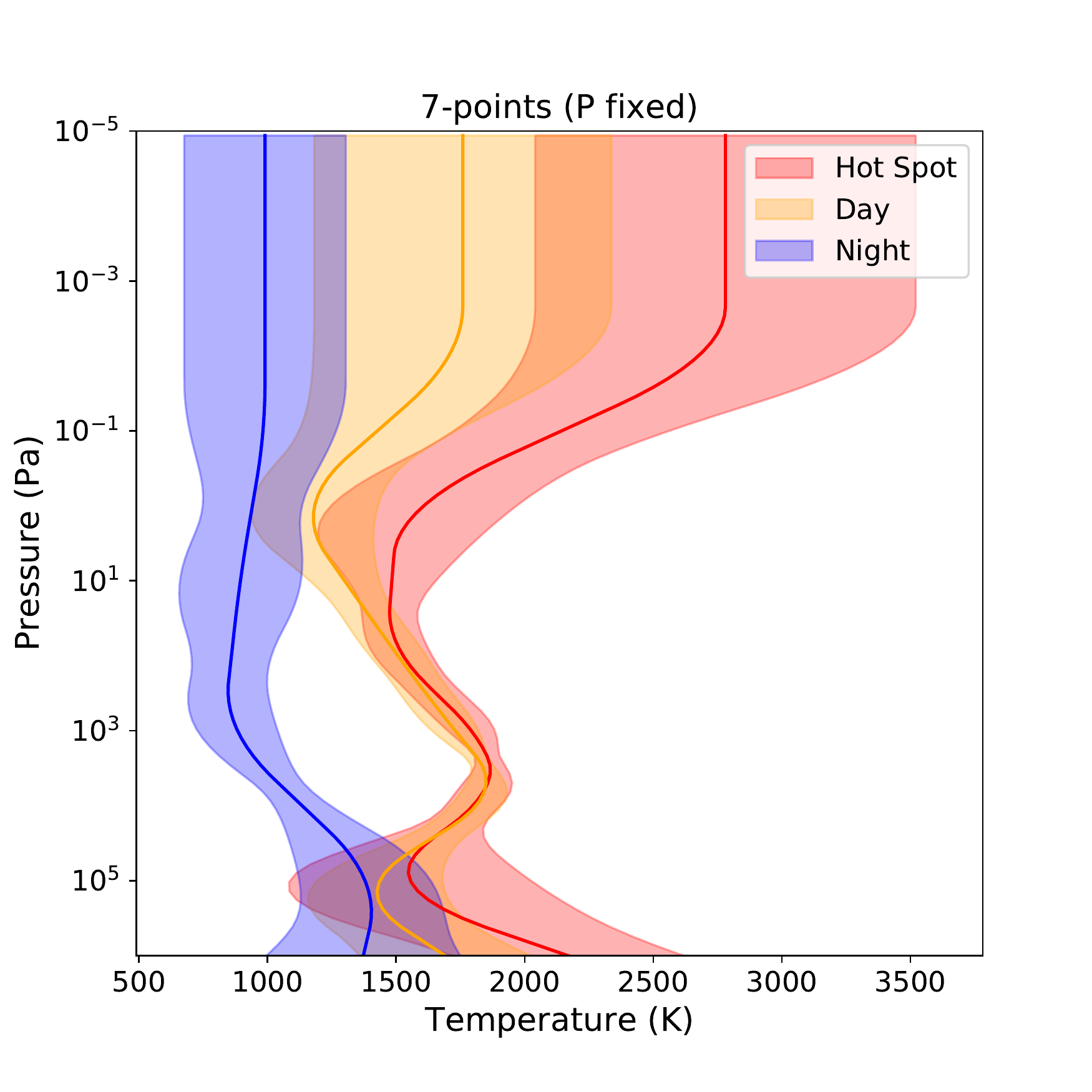}
    \caption{Retrieved temperature profiles from unified retrievals on the HST data when assuming different free temperature-pressure profiles. Note that the pressure of the temperature nodes is fixed for the 5-point and 7-point scenarios.}
    \label{fig:tp_HSTUnified_tp_explor}
\end{figure}

The same exercise was also repeated with the phase curve model to assess the required complexity for the thermal profiles. Since the emission of each region is capture by multiple spectra, one can expect the information content of each region to increase as compared to an individual spectrum retrieval. When performing the unified retrievals on HST data, the 5-point temperature profiles required many model evaluations ($>$ 15 millions) to reach the imposed convergence criteria (evidence tolerance of 0.5). We therefore choose to fix the pressure points for those profiles as is done for the 7-point profile. We however explored two choices for the fixed pressure points of the 5-point model that were informed by the results of the 7-point model. In the 5-point (informed) scenario, we fixed the pressure points to 10$^6$, 10$^5$, 100, 1 and 0.01 Pa. This setup allows to follow the solution found when using 7 points. In the 5-point (no inversion), we attempt to remove the inversion seen around 10$^4$Pa by setting the 5 pressure points to 10$^6$, 10$^5$, 10$^4$, 100 and 0.01 Pa. The results of those retrievals are shown in Figures \ref{fig:spectra_HSTUnified_tp_explor}, \ref{fig:post_HSTUnified_tp_explor} and \ref{fig:tp_HSTUnified_tp_explor}. The  3-point profile obtained log(E) = 2100.2, the 4-point profile obtained log(E) = 2102.4, the 5-point (informed) obtained log(E) = 2103.9, the 5-point (no inversion) obtained log(E) = 2100.0 and the 7-point profile with the fixed pressure node obtained log(E) = 2101.6. While a slight preference can be seen for the 5-point (informed) model, all the tested models obtained evidences that are roughly in the same range, making the choice of a particular model difficult. We note that all temperature profiles, except the simplest, exhibit a thermal inversion at high altitude. When testing the different 5-point models, we obtained a noticeable difference in the log evidence ($\Delta$E = 3.7 for the informed model). This difference highlights that fixing the pressure points of a free thermal profile can lead to user dependant behaviours when a small number of nodes is used. In all models, we find a super-solar metallicity and depending on the model, the C/O ratio can be sub-solar to solar.  \\ 

{\bf $\bullet$ Unified phase curve retrievals on HST+Spitzer} \\

\begin{figure}
\centering
    \includegraphics[width = 0.86\textwidth]{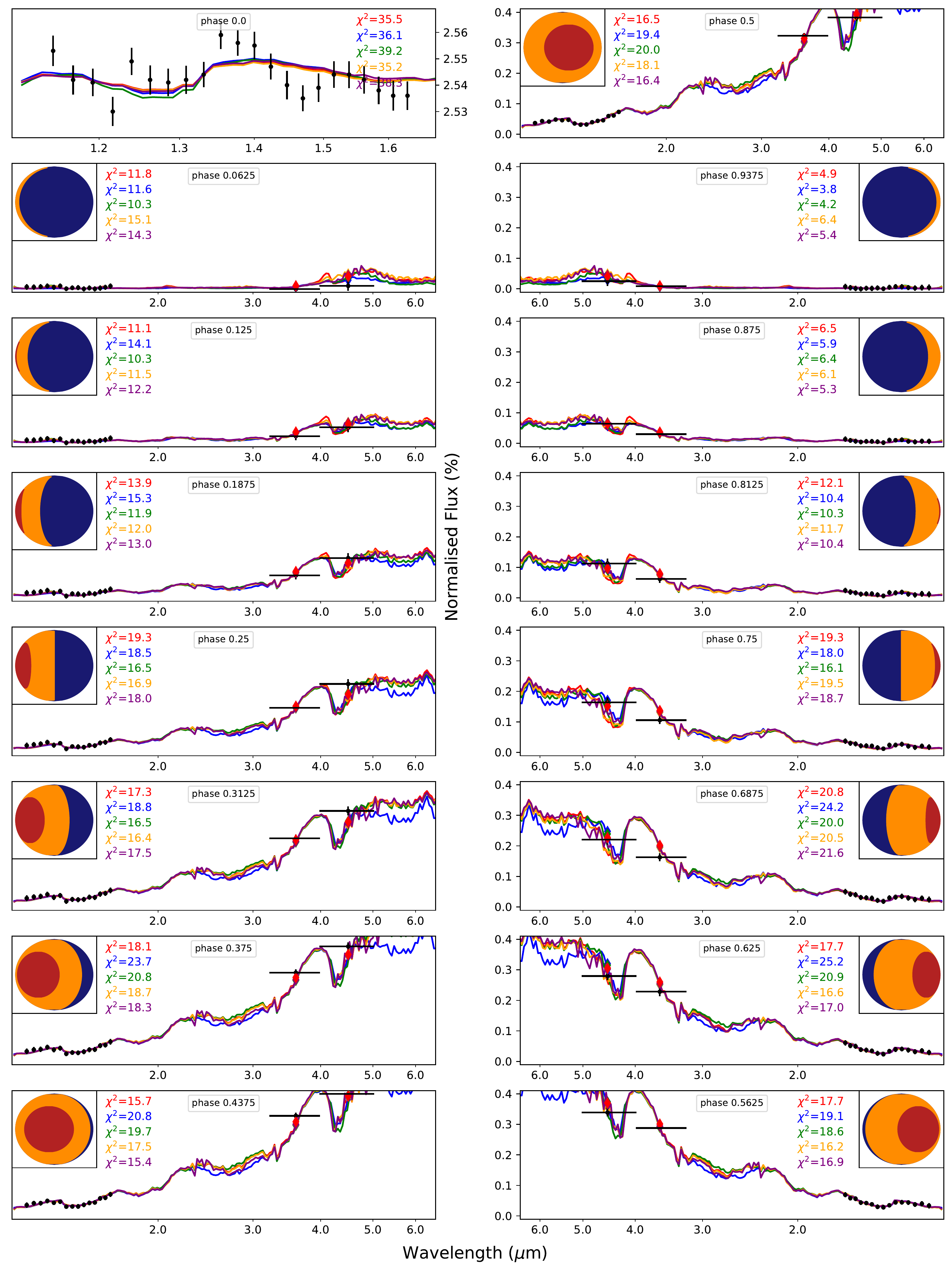}
    \caption{Retrieved spectra from unified retrievals on the HST+Spitzer data when assuming different free temperature-pressure profiles. Blue: 3-point; Green: 4-point; Orange: 5-point (informed); Purple: 5-point (no inversion); Red: 7-point (fixed P)}
    \label{fig:spectra_SpzUnified_tp_explor}
\end{figure}

\begin{figure}
\centering
    \includegraphics[width = 0.86\textwidth]{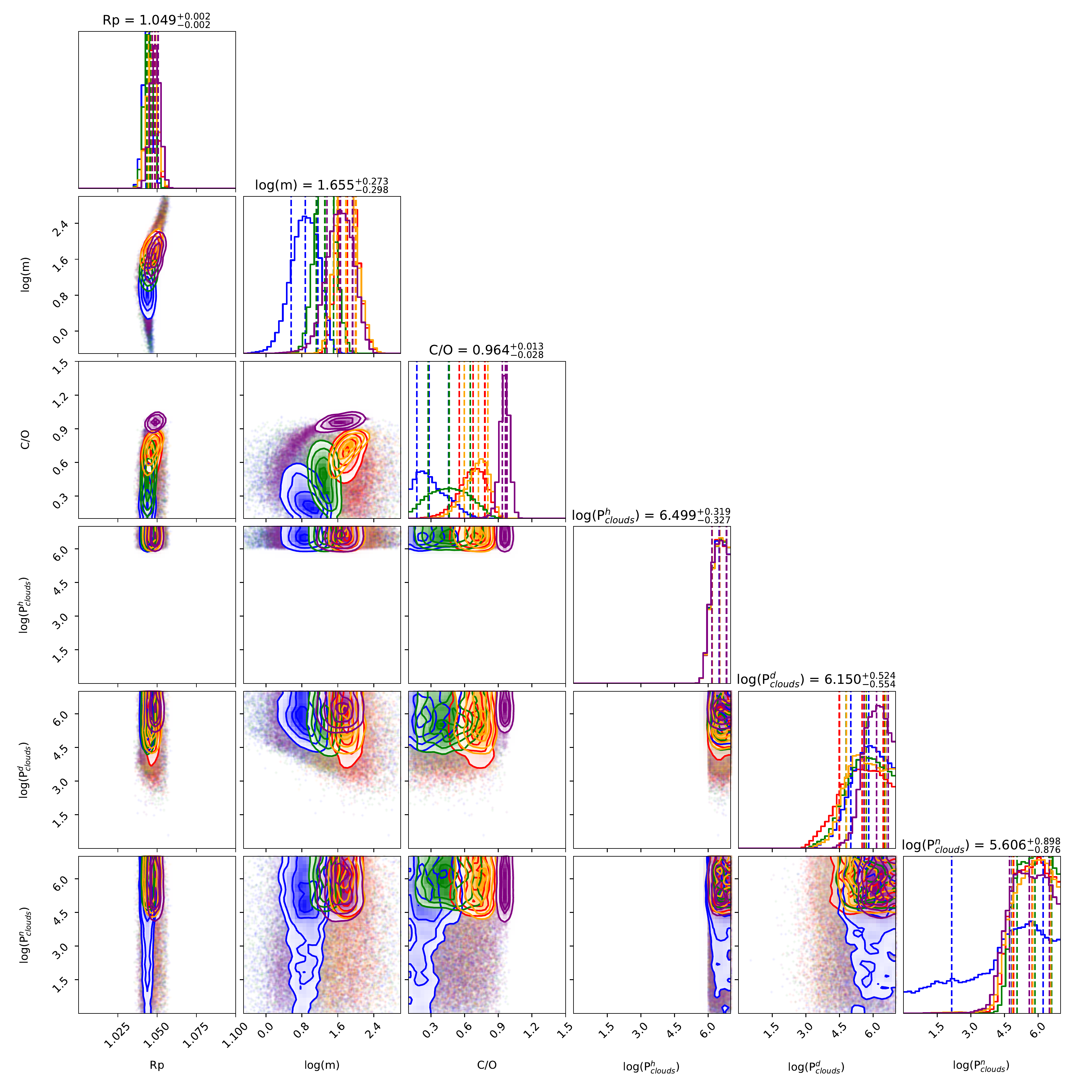}
    \caption{Posteriors from unified retrievals on the HST+Spitzer data when assuming different free temperature-pressure profiles. Blue: 3-point; Green: 4-point; Orange: 5-point (informed); Purple: 5-point (no inversion); Red: 7-point (fixed P)}
    \label{fig:post_SpzUnified_tp_explor}
\end{figure}

\begin{figure}
\centering
    \includegraphics[width = 0.4\textwidth]{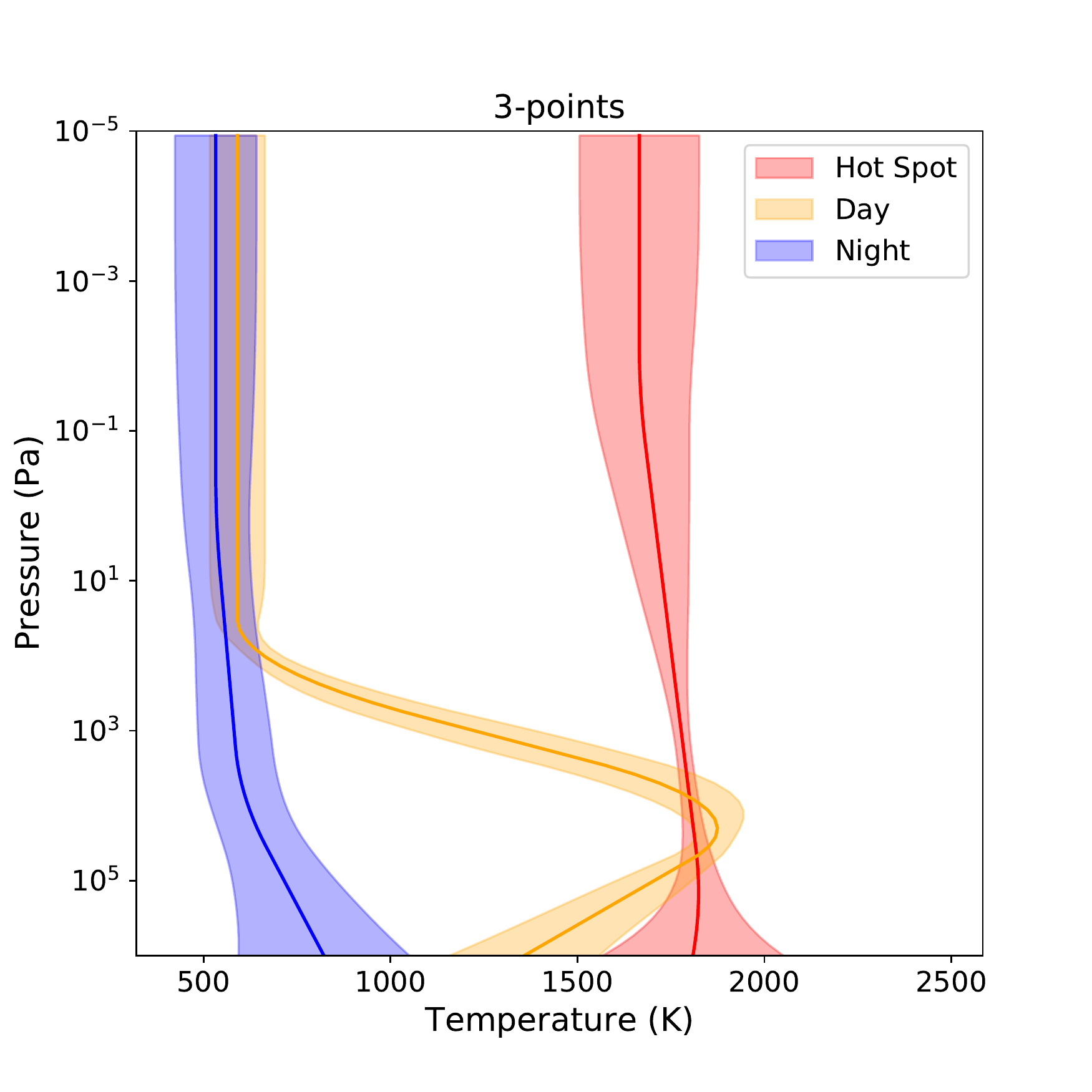}
    \includegraphics[width = 0.4\textwidth]{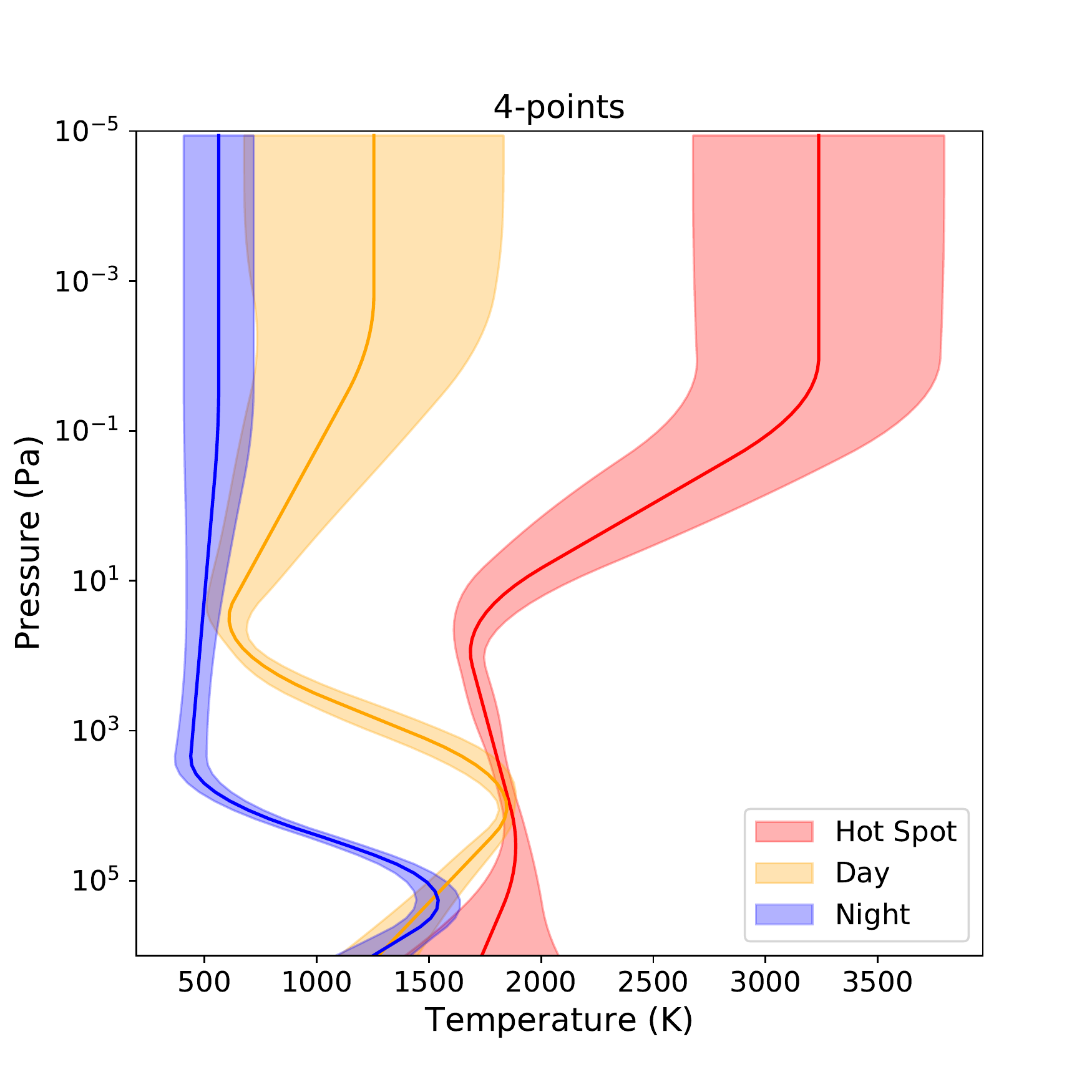}
    \includegraphics[width = 0.4\textwidth]{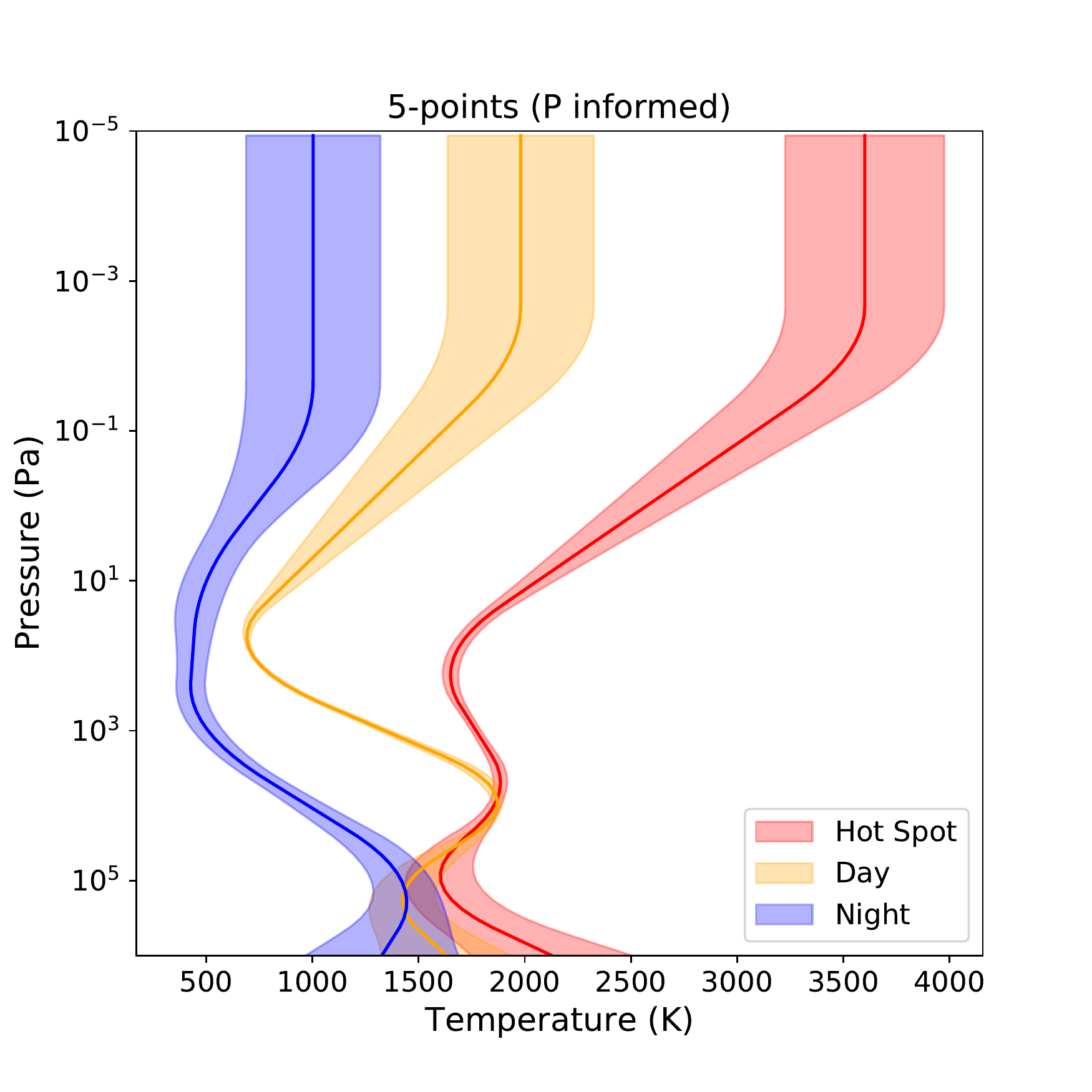}
    \includegraphics[width = 0.4\textwidth]{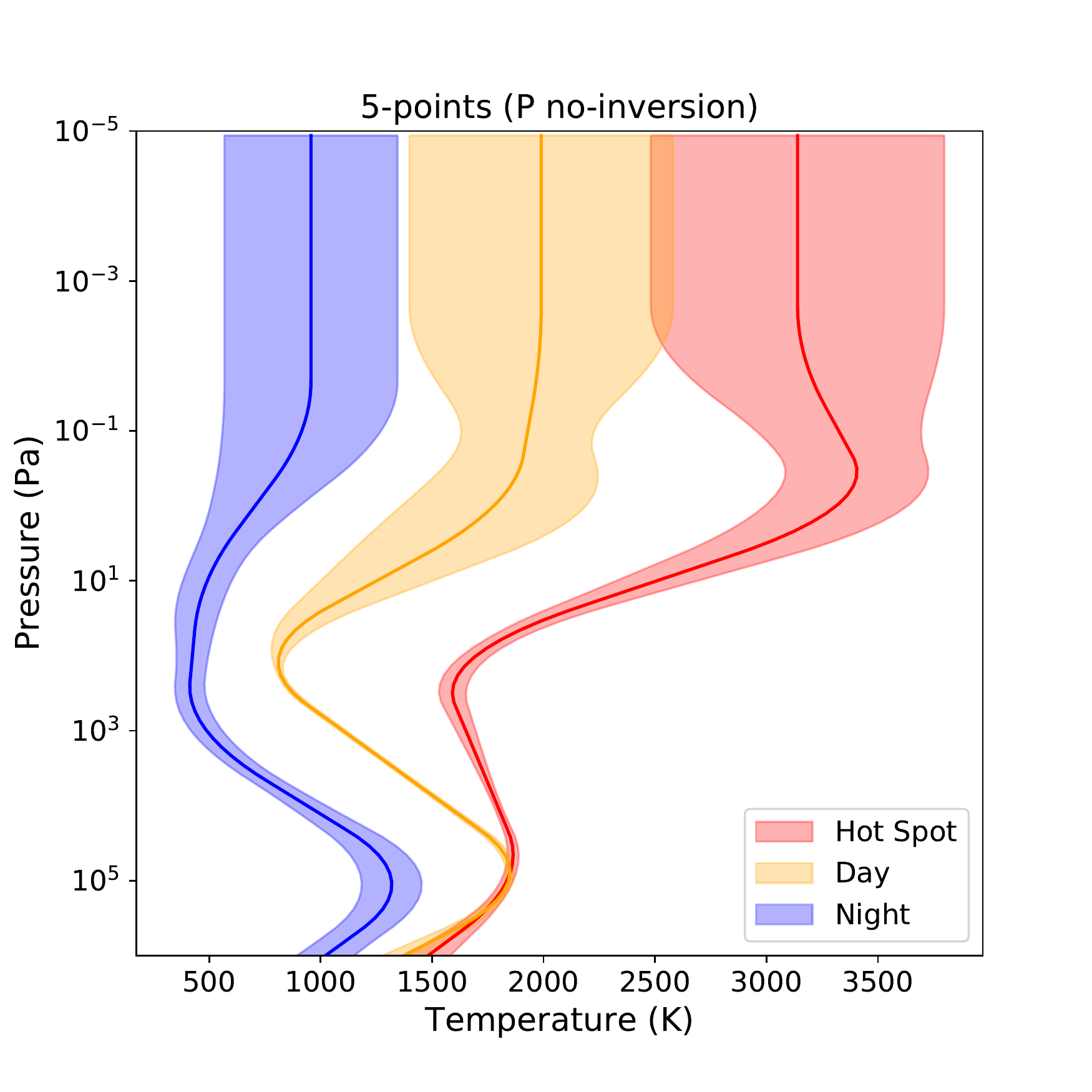}
    \includegraphics[width = 0.4\textwidth]{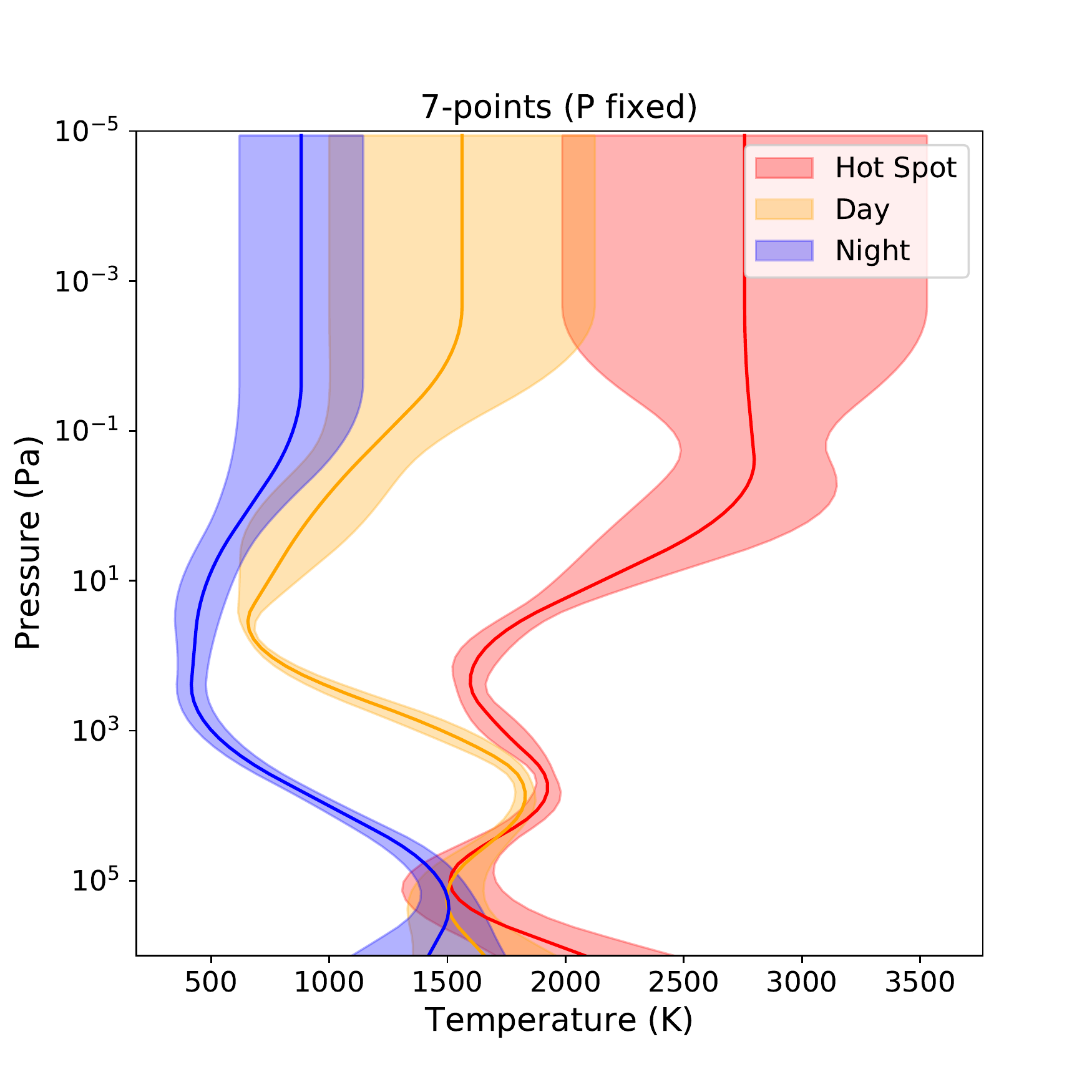}
    \caption{Retrieved temperature profiles from unified retrievals on the HST+Spitzer data when assuming different free temperature-pressure profiles. Note that the pressure of the temperature nodes is fixed for the 5-point and 7-point scenarios.}
    \label{fig:tp_SpzUnified_tp_explor}
\end{figure}

This time, we find that a more complex structure is required (see Figures \ref{fig:spectra_SpzUnified_tp_explor}, \ref{fig:post_SpzUnified_tp_explor} and \ref{fig:tp_SpzUnified_tp_explor}). The 3-point profile obtained log(E) = 2267.2, the 4-point profile obtained log(E) = 2275.0, the 5-point (informed) profile obtained log(E) = 2278.0, the 5-point (no inversion) obtained log(E) = 2272.6  and the 7-point profile with the fixed pressure node obtained log(E) = 2277.4. The significant difference in evidence between the simplest and the most complex models indicates that additional information is brought by the Spitzer data. In fact, similar thermal profiles shapes are found to the HST only runs, but with tighter contraints. For all runs, the metallicity is again, super-solar, while the C/O ratio is sub-solar to solar depending on the thermal profile considered. Now, having determined that a complex thermal profile is required to best extract the information contained in the WASP-43\,b phase curve data, we choose to use the 7-point profile as our baseline in this article. This is motivated by performance reasons (retrieving pressure has proven to require many more samples) and to avoid user dependant behaviour as shown with the 5-point profile. \\

In this exercise, we only explored scenarios of various complexities for the thermal profile. The model complexity of a model should match the information content in the data but, ensuring this is by no mean easy in the context of exoplanet atmospheres, since retrievals often include many modules dealing with different parts of the physics. For example, in case of a bad fit, should one increase the complexity of the thermal profile or the chemistry. This question is difficult to answer and will essentially always lead to user dependant choices. Here, our chemistry is kept relatively simple for practical reasons, but it is not inconceivable that a more complex chemistry or the addition of new absorbers would lead to different results.

\clearpage

\section{Retrieval with the free hot-spot offset.}\label{apx:free_hs_run}

\begin{figure}[H]
\centering
    \includegraphics[width = 0.86\textwidth]{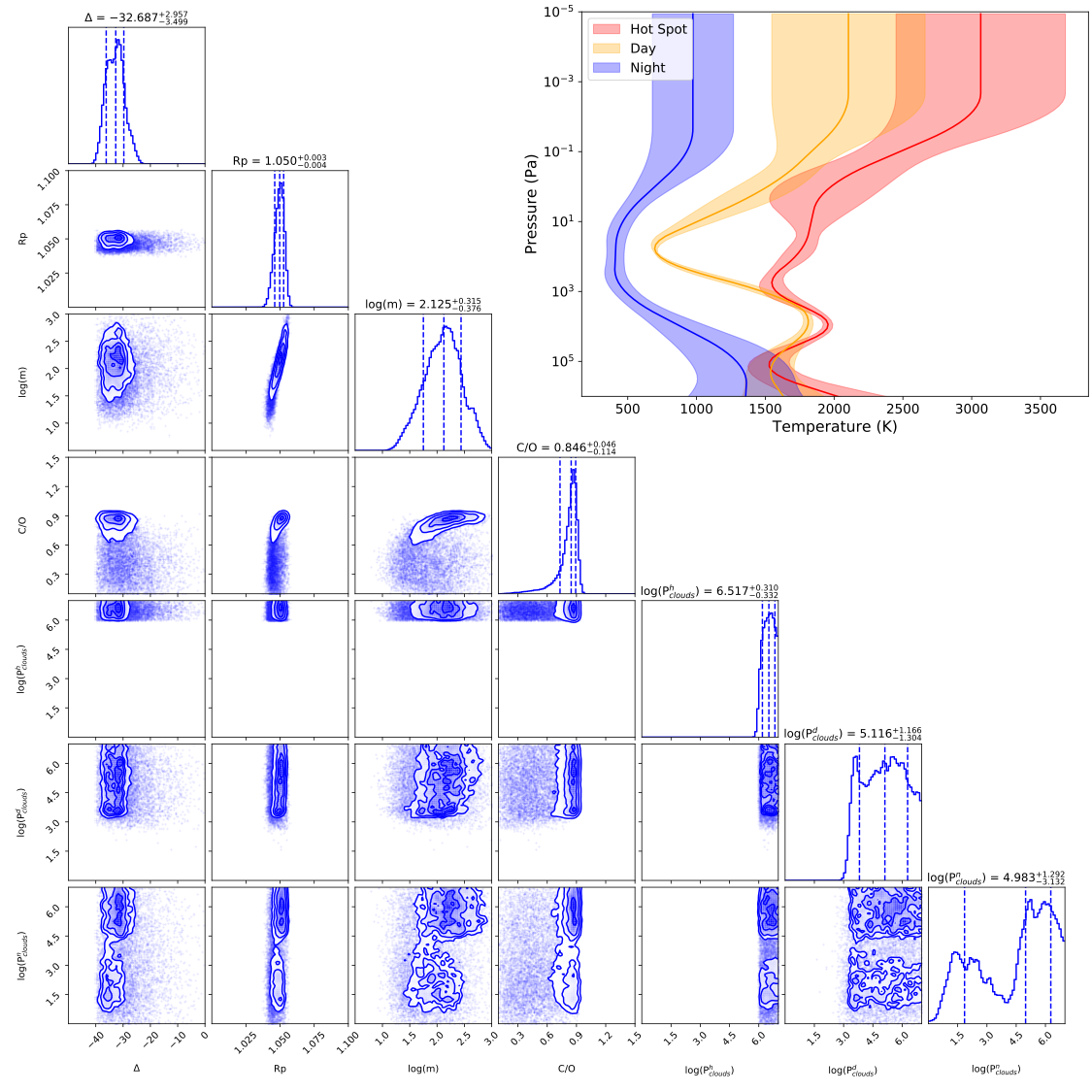}
    \caption{Posterior distribution and temperature structure (top right) of the full retrieval scenario with the free hot-spot offset parameter.}
    \label{fig:post_hot_spot_offset}
\end{figure}

In Figure \ref{fig:post_hot_spot_offset} we show the posterior distribution of the 'Full' retrieval run, leaving the hot-spot offset as a free parameter. The recovered parameter does not match the \cite{stevenson_w43_1} value of 12.2 degrees. In all the other runs of this paper, we fix the hot-spot value to 12.2 degrees, but we note that the conclusions from the free hot-spot shift run are the same as the full run.

\clearpage

\section{Best fit spectra of the 2-Faces free, the 2-Faces equilibrium and the Full retrievals.}\label{apx:spectra_compa}

\begin{figure}[H]
\centering
    \includegraphics[width = 0.86\textwidth]{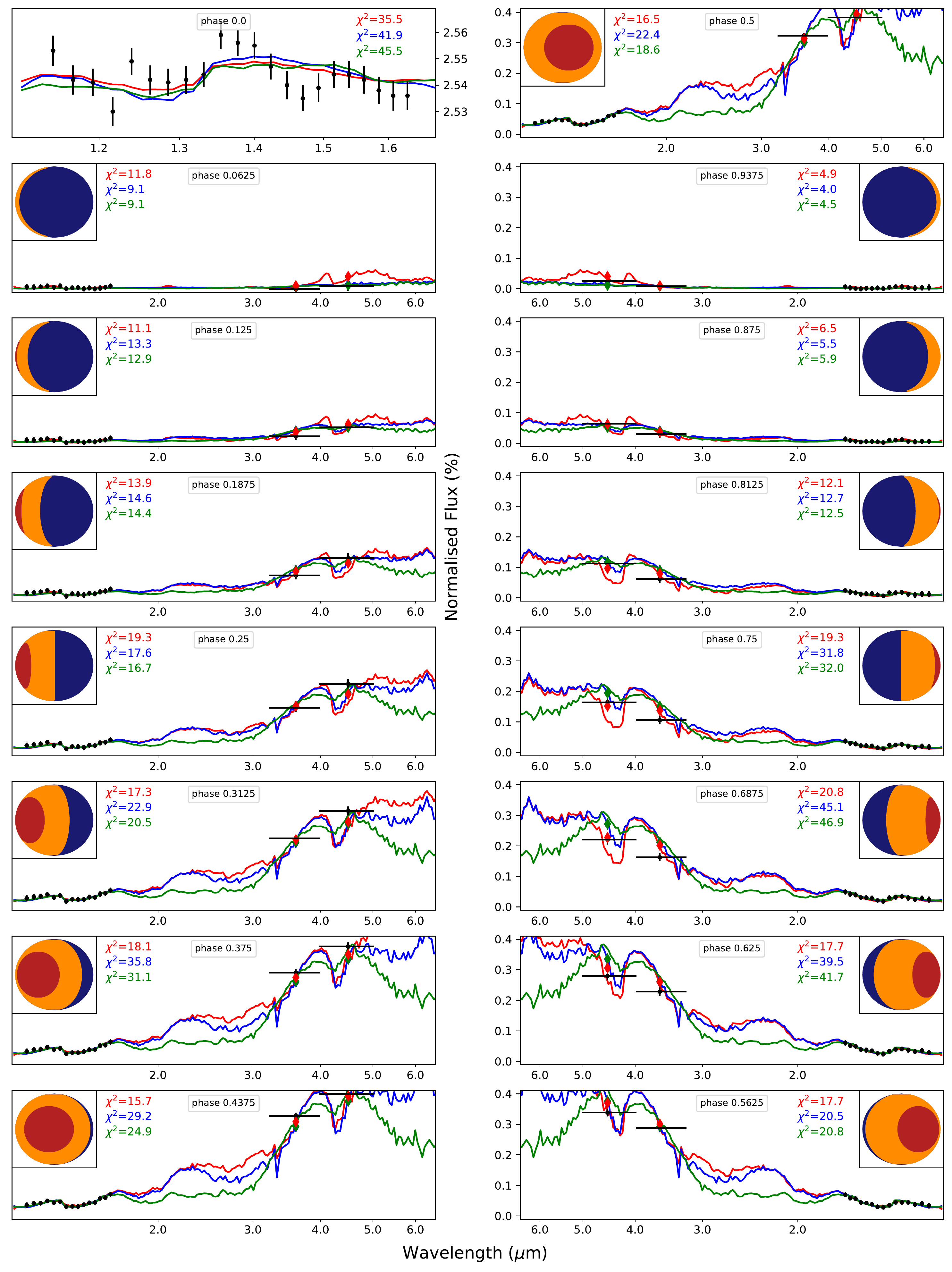}
    \caption{Best fit spectra and geometry of our WASP-43\,b phase-curve retrieval with the 2-Faces free chemistry model (green), the 2-Faces equilibrium chemistry model (blue) and the full model (red). The diamonds represent the the averaged spitzer bandpasses. The right panels have inverted wavelength axis.}
    \label{fig:spectra_full}
\end{figure}
\begin{figure}[H]
\centering
    \includegraphics[width = 0.86\textwidth]{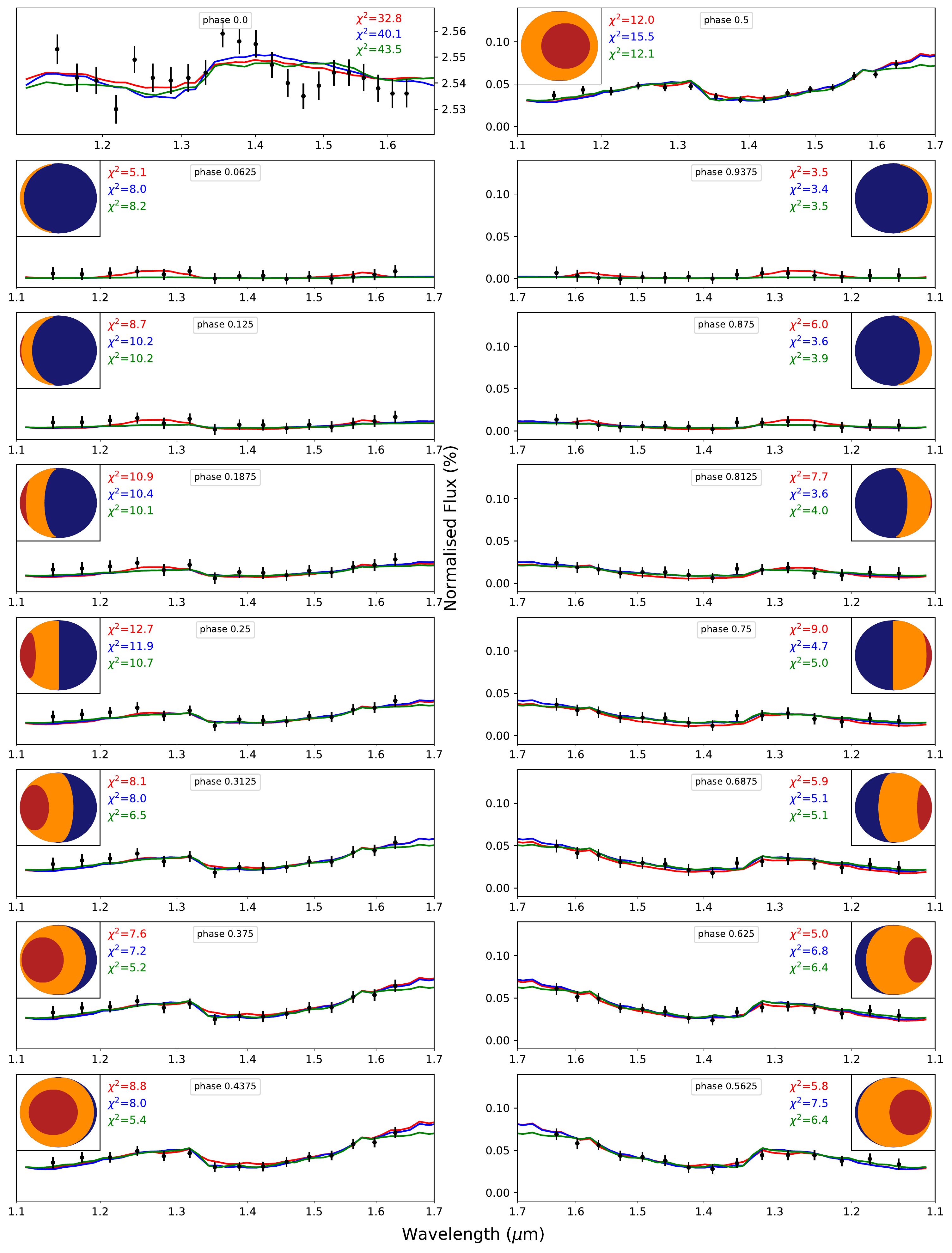}
    \caption{Best fit spectra and geometry of our WASP-43\,b phase-curve retrieval with the 2-Faces free chemistry model (green), the 2-Faces equilibrium chemistry model (blue) and the full model (red). It is the same as Figure \ref{fig:spectra_full}, but zoomed in the HST wavelength range. The right panels have inverted wavelength axis. The $\chi^2$ was calculated for the data between 1.1$\mu$m and 1.7$\mu$m only. }
    \label{fig:spectra_full_HST}
\end{figure}

\section{Complementary plots of the 2-Faces free chemistry retrieval.}\label{apx:free_scn}

\begin{figure}[H]
\centering
    \includegraphics[width = 0.86\textwidth]{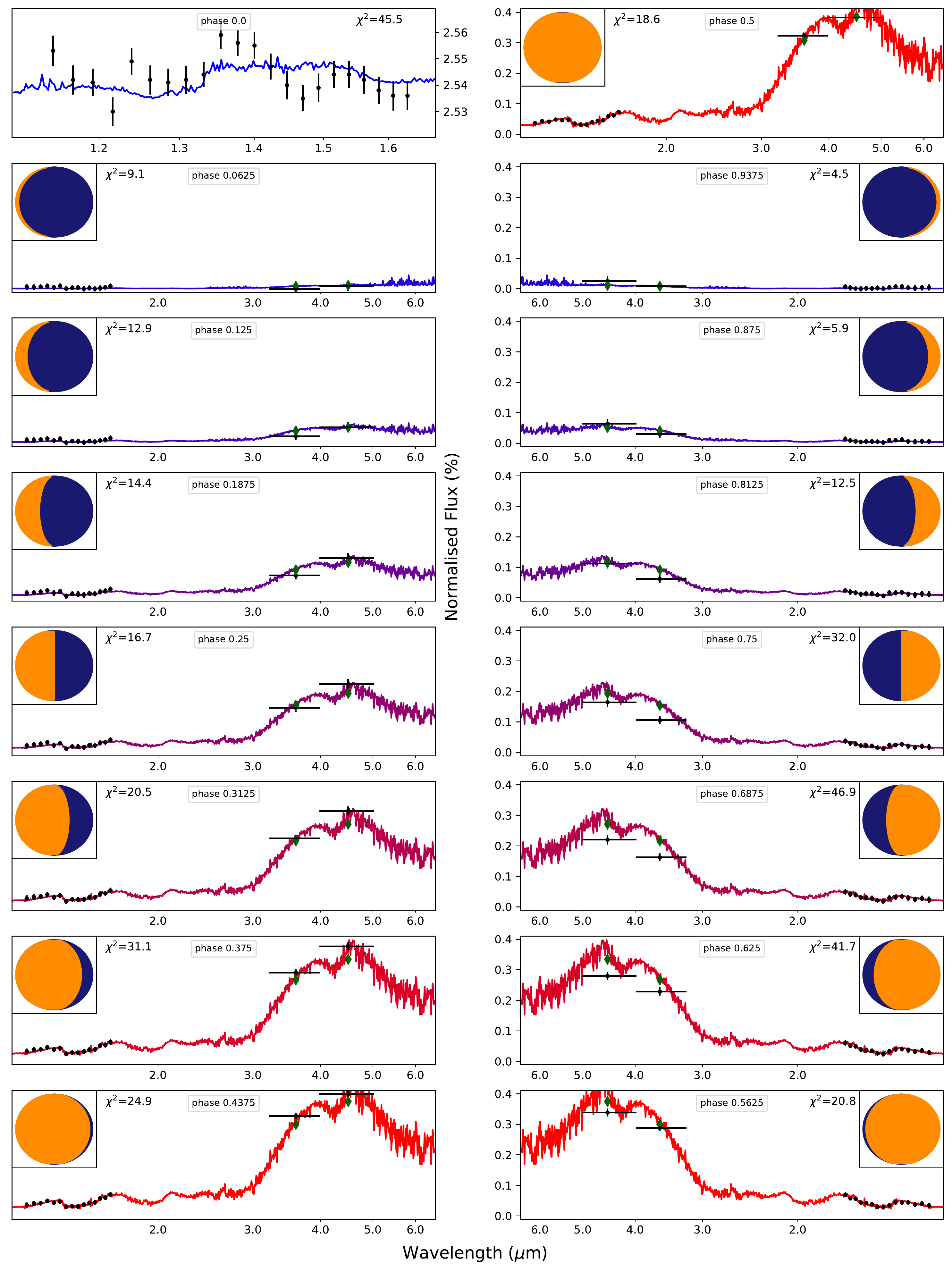}
    \caption{Best fit spectra and geometry of our WASP-43\,b phase-curve retrieval with the 2-Faces free chemistry model. The temperature is described by the N-point free profile and the chemistry is coupled between the day and night-sides. The diamonds represent the the averaged spitzer bandpasses. The right panels have inverted wavelength axis.}
    \label{fig:spectra_2F_free}
\end{figure}


\begin{figure}[H]
\centering
    \includegraphics[width = 0.8\textwidth]{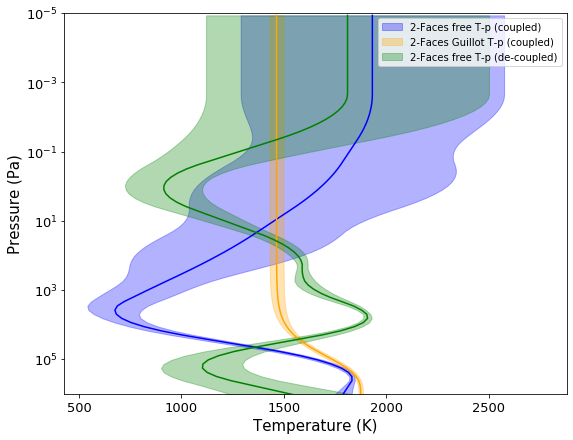}
    \includegraphics[width = 0.8\textwidth]{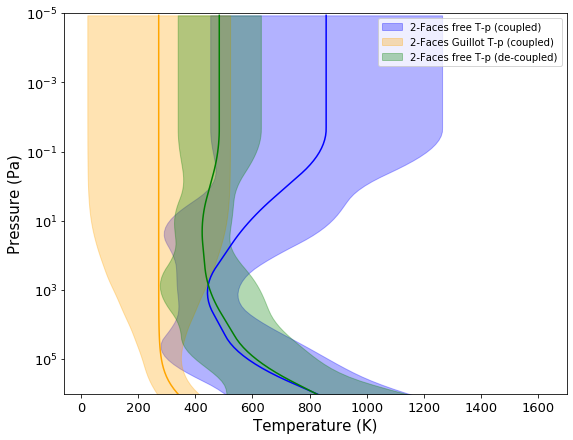}
    \caption{Temperature structure in the 2-Faces free scenarios for the day-side (top) and night-side (bottom). Blue: 2-Faces free chemistry with free T-p profile and coupled chemistry; Orange: 2-Faces free chemistry with \cite{Guillot_TP_model} profile and coupled chemistry; Green: 2-Faces free chemistry with free T-p profile and de-coupled chemistry.}    \label{fig:temp_2F_compa}
\end{figure}


\begin{figure}[H]
    \includegraphics[width = \textwidth]{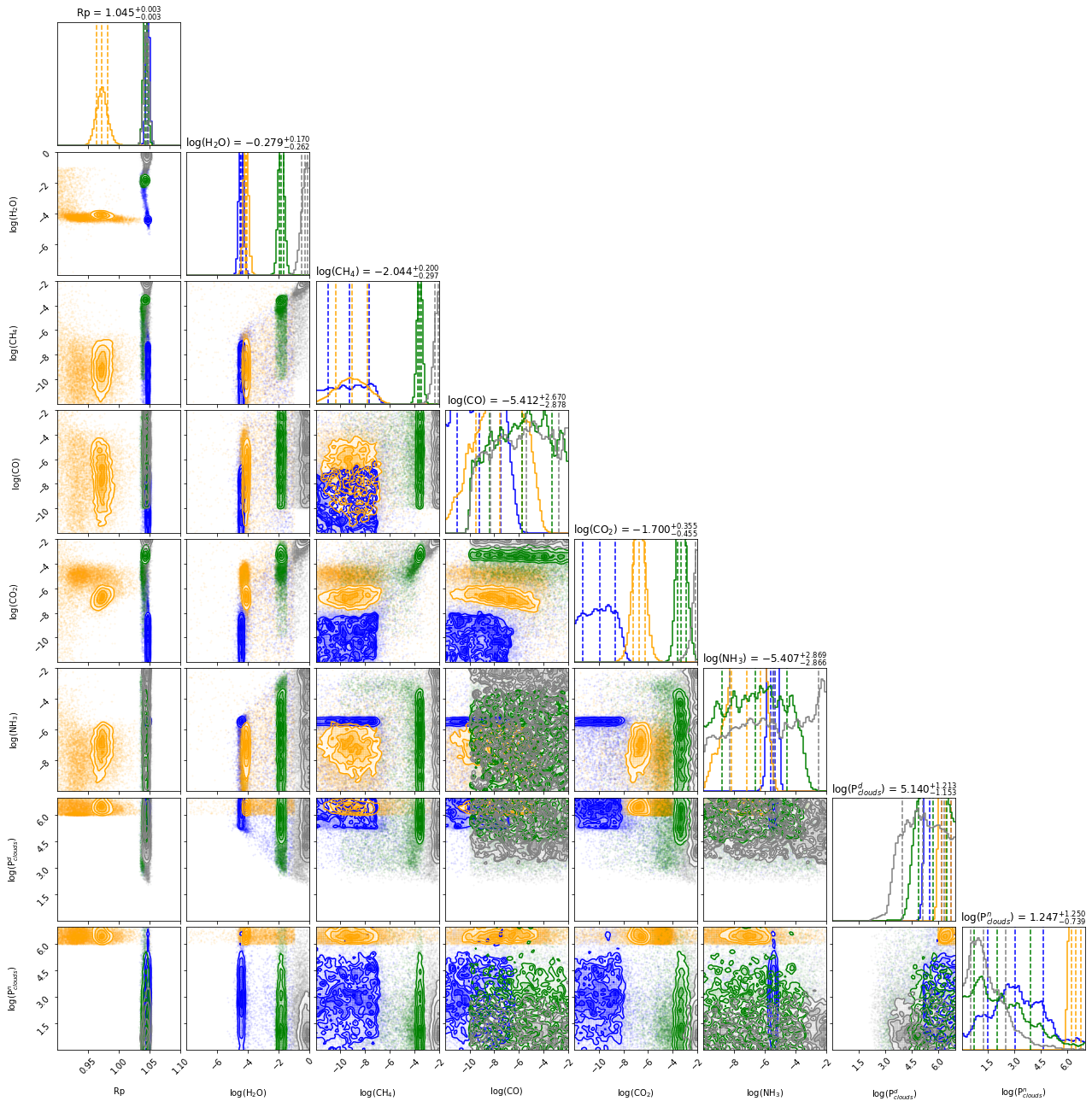}
    \caption{Posteriors distribution of the day-side chemistry and the clouds properties for the 2-Faces free runs. Blue: 2-Faces free chemistry with free T-p profile and coupled chemistry; Orange: 2-Faces free chemistry with \cite{Guillot_TP_model} profile and coupled chemistry; Green: 2-Faces free chemistry with free T-p profile and de-coupled chemistry (Main solution). The latter retrieval also converged to a high mean molecular weight solution (shown in Grey), which is discarded from our assumption of primary atmosphere for this planet.}
    \label{fig:post_2F_compa}
\end{figure}

\begin{figure}[H]
\centering
    \includegraphics[width = 0.6\textwidth]{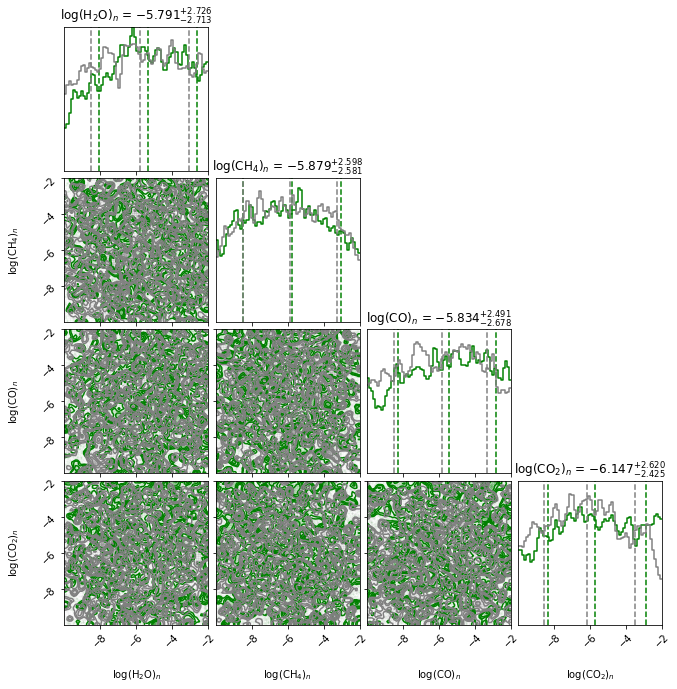}
    \caption{Posteriors distribution 2-Faces free chemistry with free T-p profile and de-coupled chemistry. Green: Main solution; Grey: high mean molecular weight solution. The night-side chemistry posteriors do not lead to the detection of molecules for this run.}
    \label{fig:post_2F_compa}
\end{figure}

\section{Complementary plots of the 2-Faces equilibrium chemistry retrieval.}\label{apx:2f_eq}

\begin{figure}[H]
\centering
    \includegraphics[width = 0.86\textwidth]{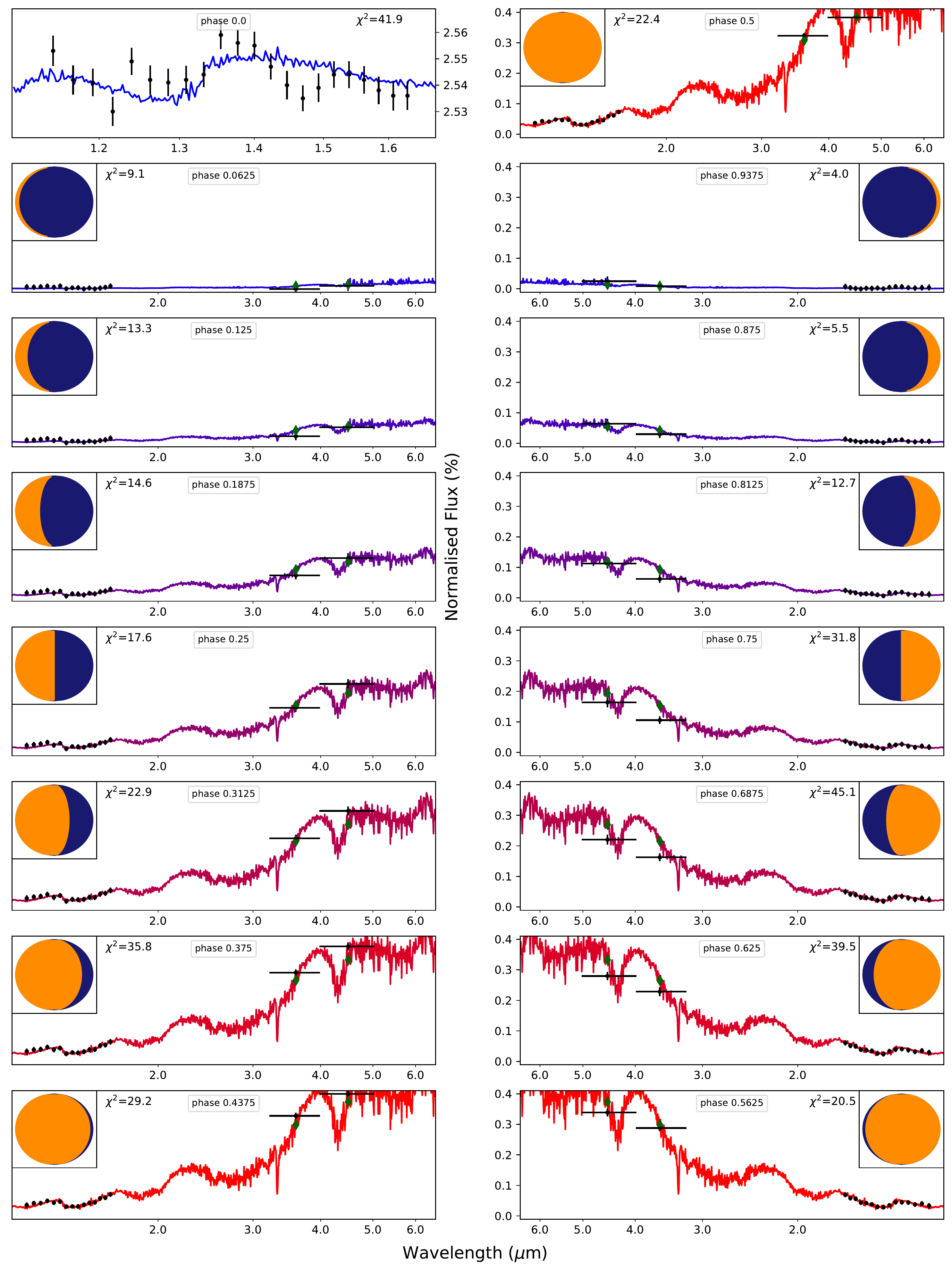}
    \caption{Best fit spectra and geometry of our WASP-43\,b phase-curve retrieval with the 2-Faces equilibrium chemistry model. The diamonds represent the the averaged spitzer bandpasses. The right panels have inverted wavelength axis.}
    \label{fig:spectra_2F_eq}
\end{figure}

\begin{figure*}
\begin{center}
\begin{minipage}{0.95\textwidth}
    \includegraphics[width = 0.49\textwidth]{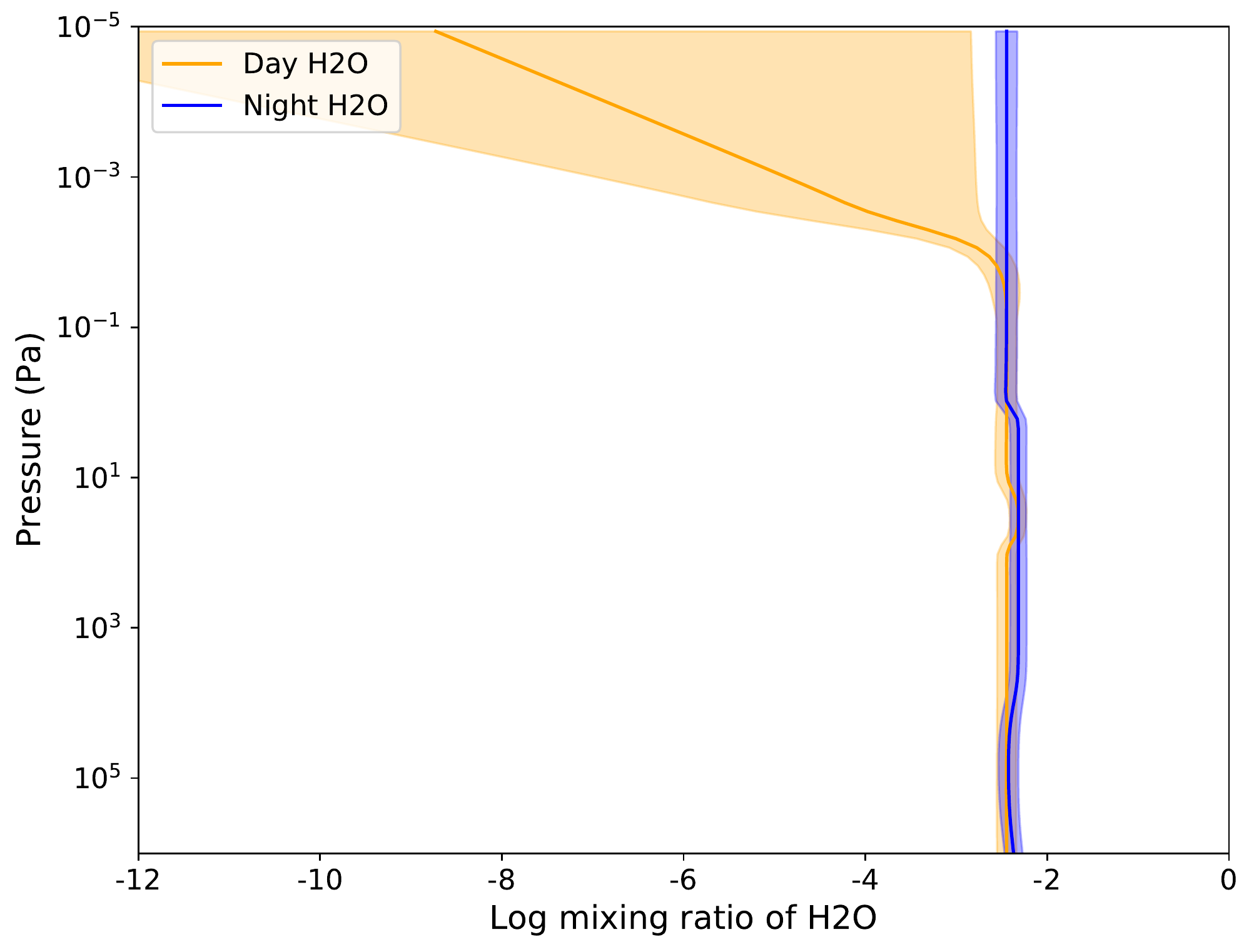}
    \includegraphics[width = 0.49\textwidth]{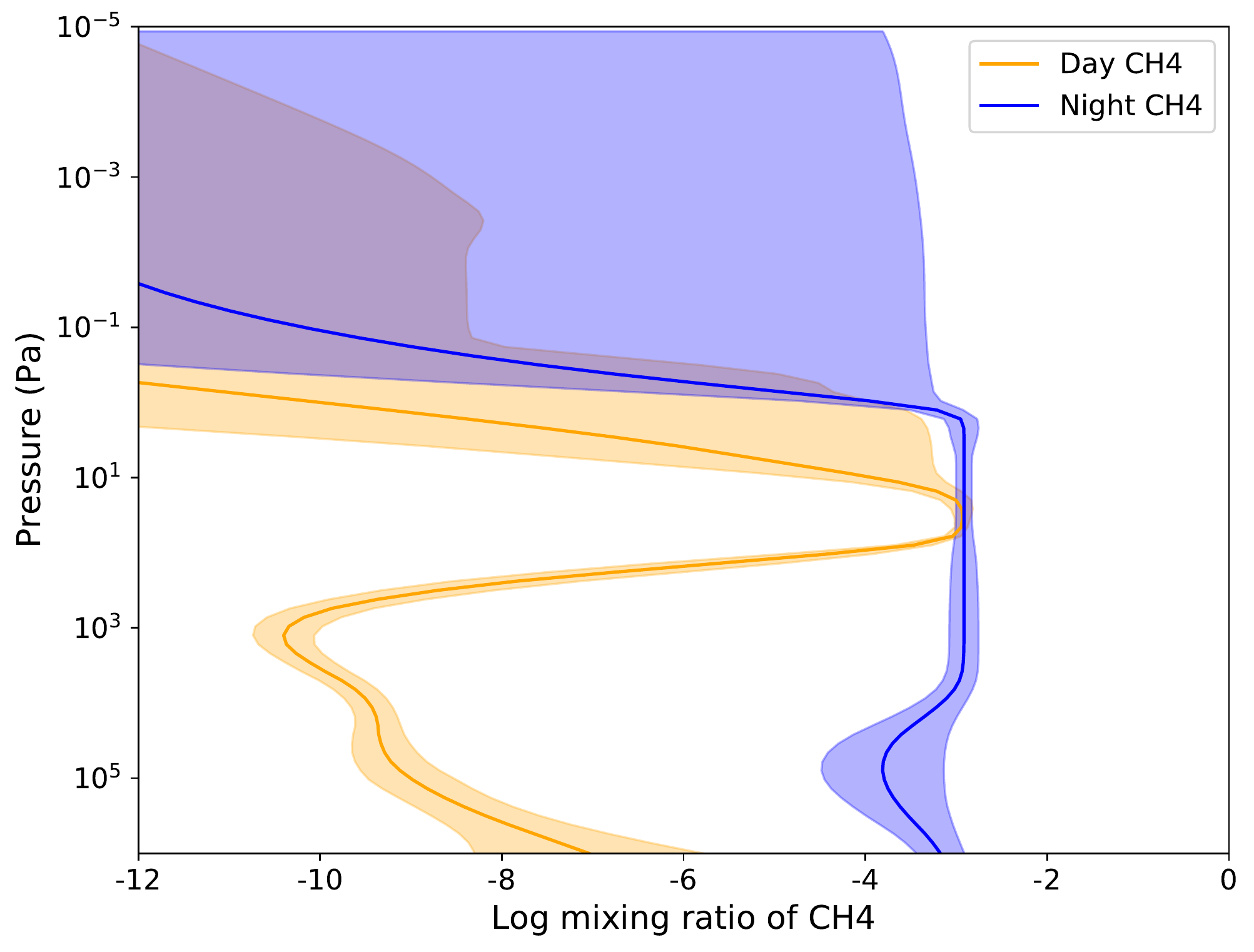} \\
    \includegraphics[width = 0.49\textwidth]{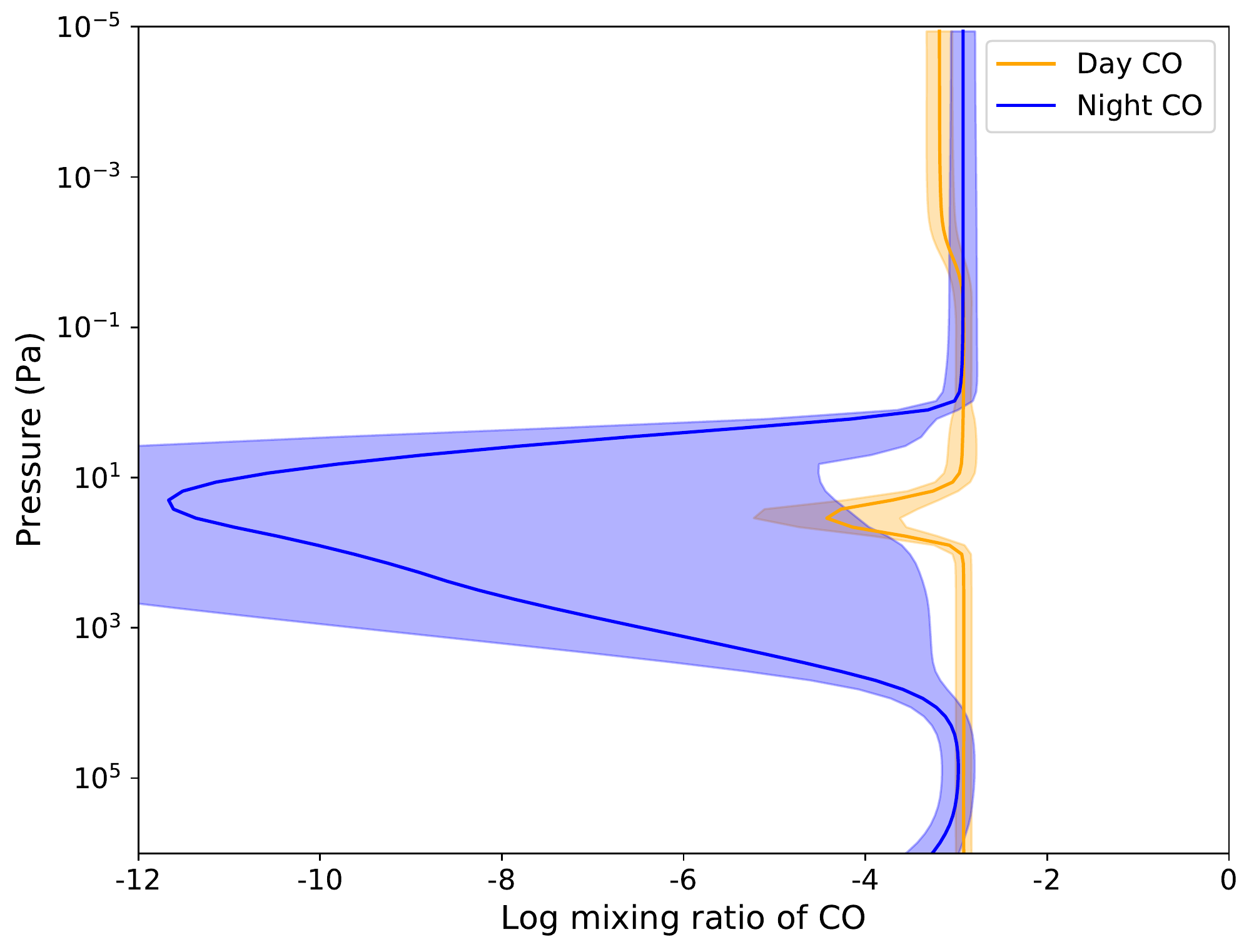}
    \includegraphics[width = 0.49\textwidth]{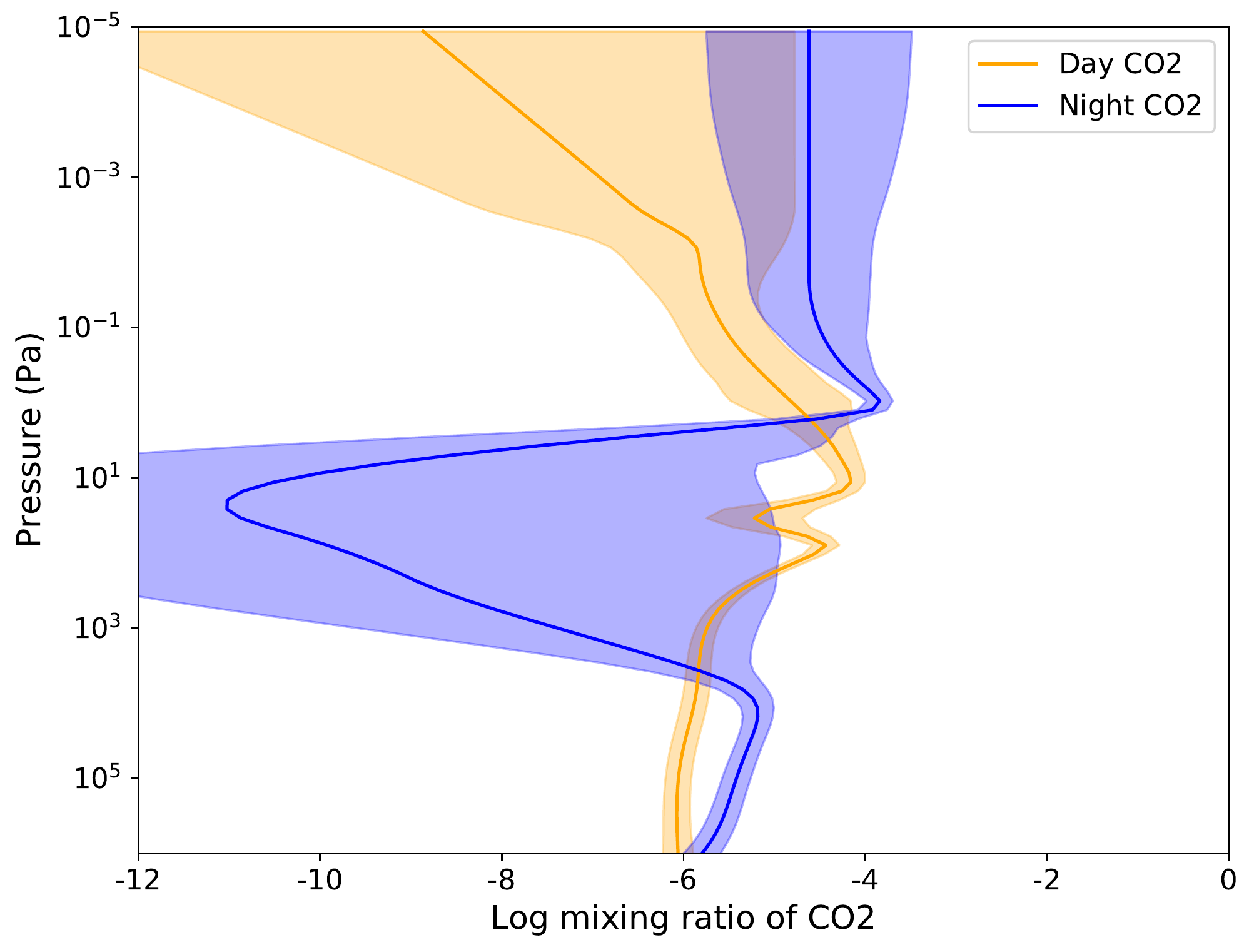} \\
    \includegraphics[width = 0.49\textwidth]{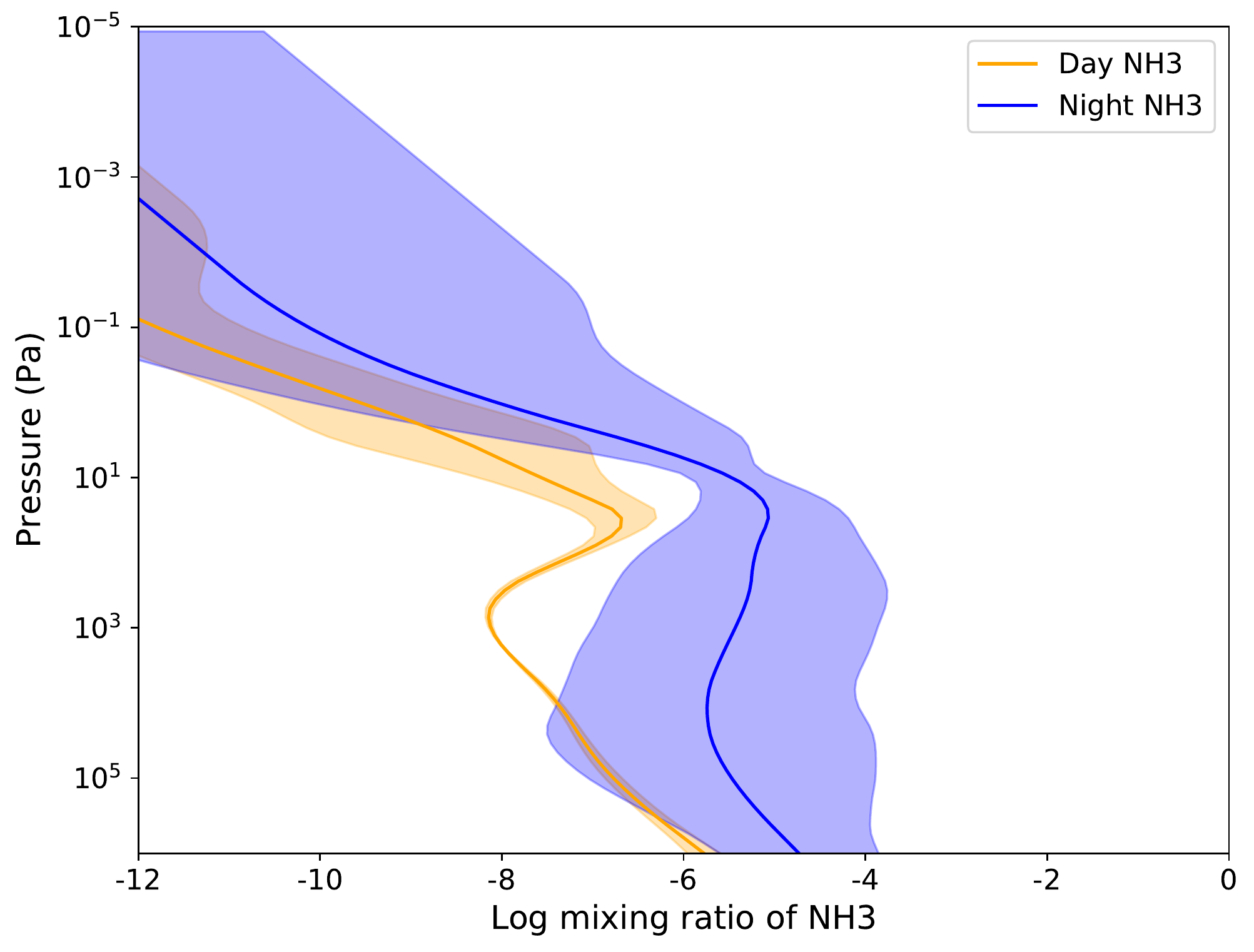}
    \includegraphics[width = 0.49\textwidth]{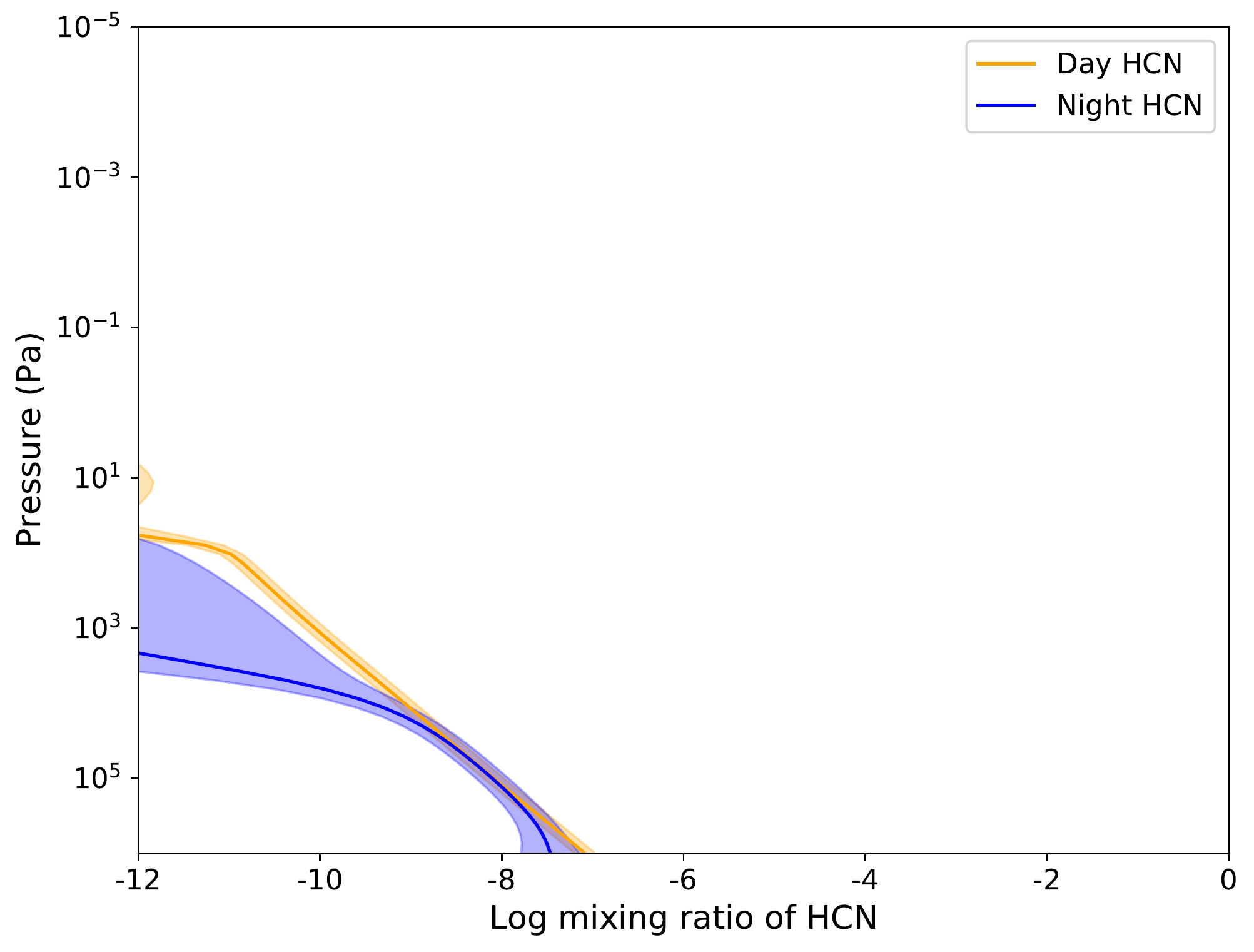}
    \caption{Molecular abundances in the different regions of the 2-Faces run, according to the chemical equilibrium scheme. Orange: day-side; Blue: night-side.}
    \label{fig:abundances_2F_eq}
\end{minipage}
\end{center}
\end{figure*}

\begin{figure}[H]
\centering
    \includegraphics[width = 0.92\textwidth]{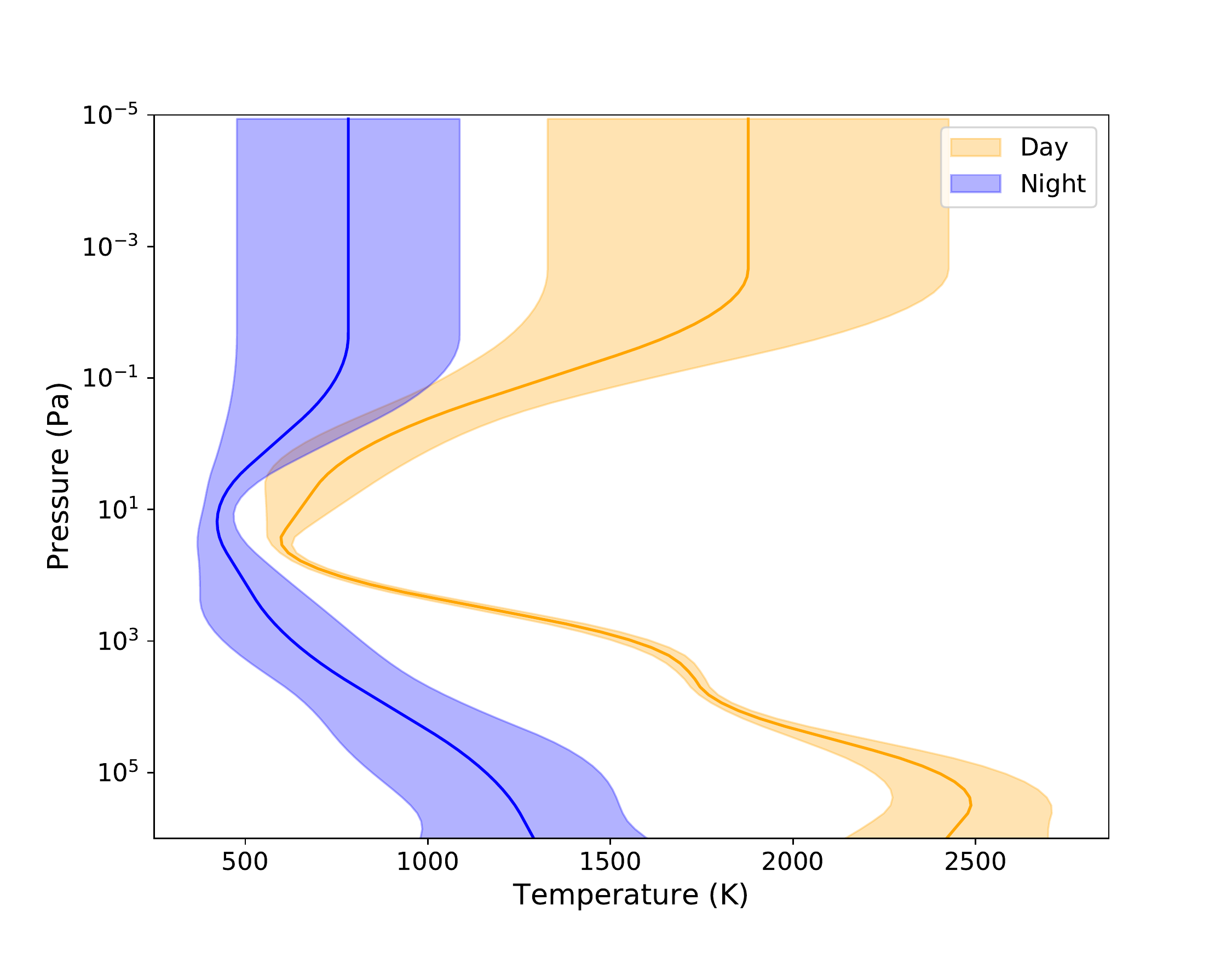}
    \caption{Temperature structure in the 2-Faces equilibrium scenario.}    \label{fig:temp_2F_eq}
\end{figure}

\begin{figure}[H]
    \includegraphics[width = \textwidth]{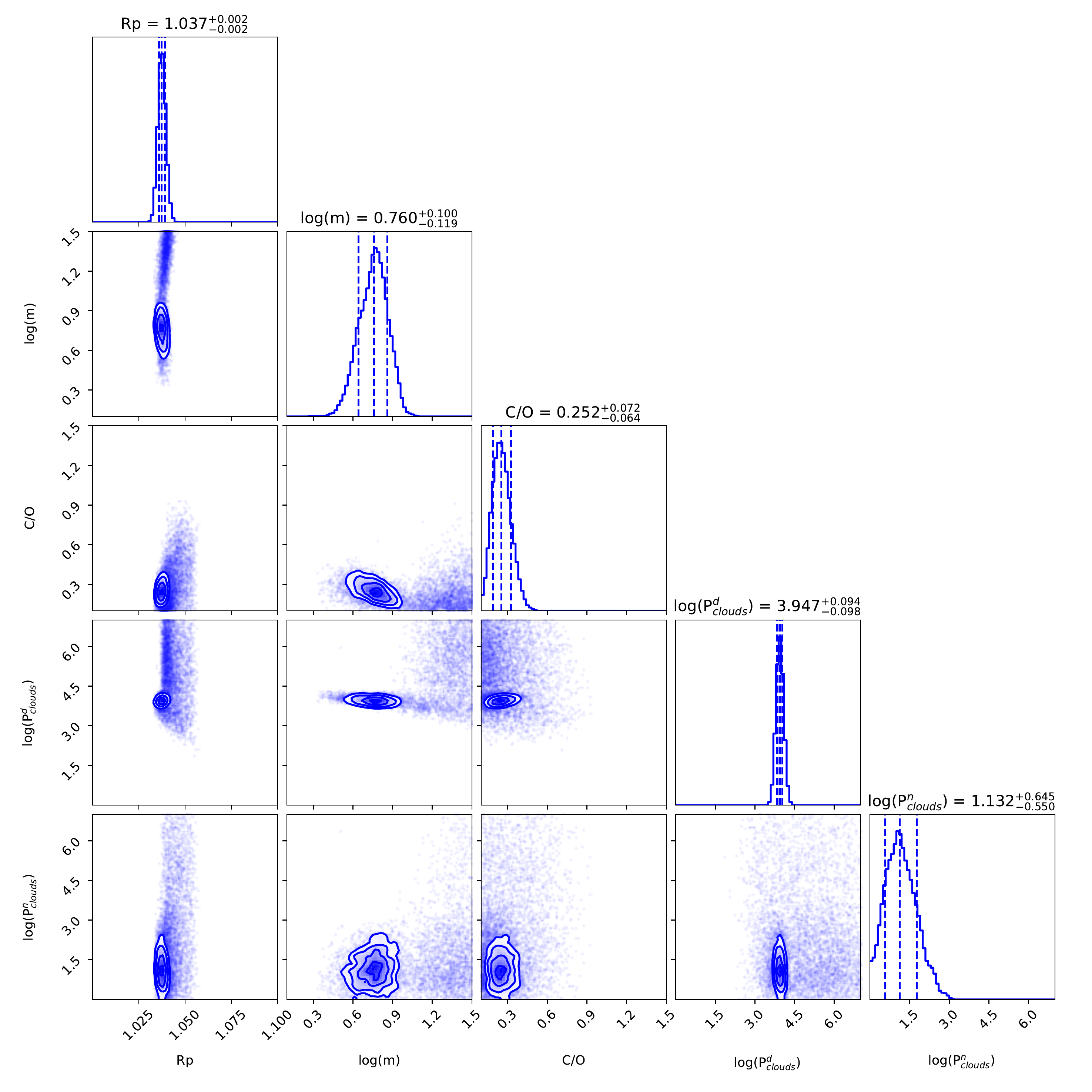}
    \caption{Posteriors distribution of the chemistry and the clouds properties for the 2-Faces equilibrium chemistry run.}
    \label{fig:post_2F_eq}
\end{figure}

\section{Complementary plots the full model retrieval (HST+Spitzer against HST only).}\label{apx:complem_full}

\begin{figure}[H]
\centering
    \includegraphics[width = 0.86\textwidth]{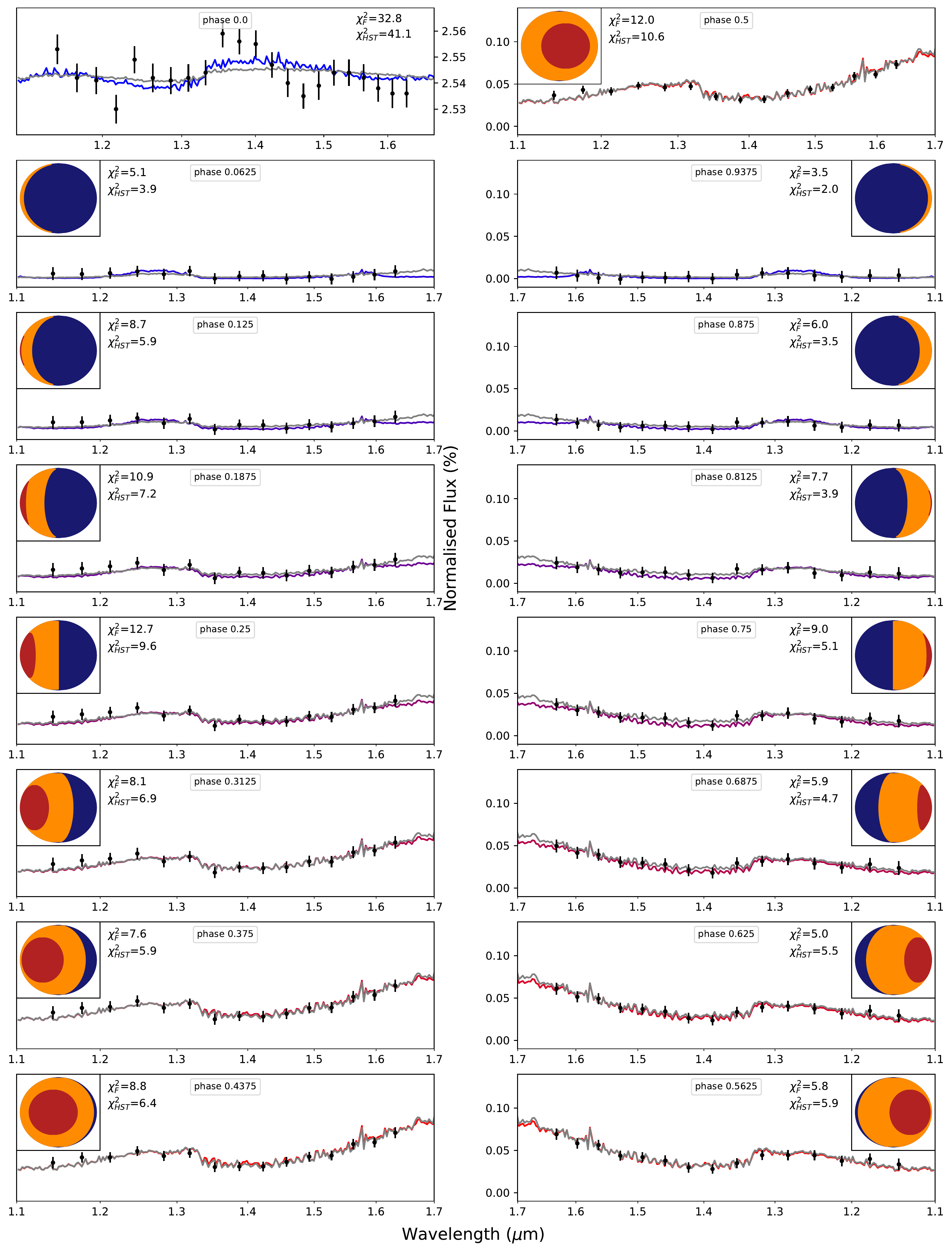}
    \caption{Best fit spectra and geometry of our WASP-43\,b phase-curve full retrieval. The colored spectrum is the same as Figure \ref{fig:spectra}, zoomed around the HST wavelengths. The grey one corresponds to the HST only retrieval. At the top, we show the transit (left) and the eclipse (right). The $\chi^2$ was calculated for the data between 1.1$\mu$m and 1.7$\mu$m only.}
    \label{fig:spectra_HST}
\end{figure}

\begin{figure}[H]
\centering
    \includegraphics[width = 0.86\textwidth]{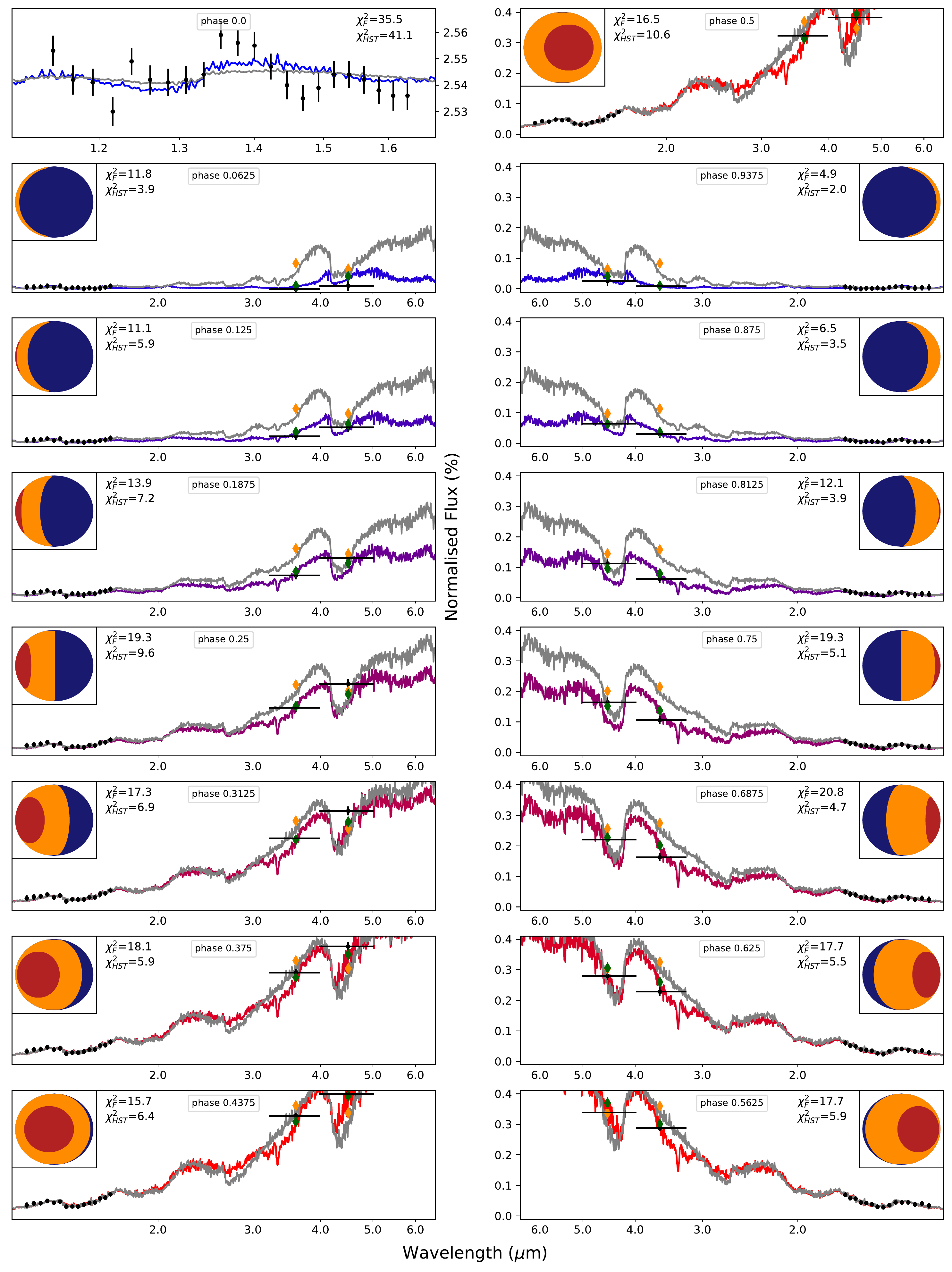}
    \caption{Best fit spectra and geometry of our WASP-43\,b phase-curve retrieval. The colored spectrum is the same as Figure \ref{fig:spectra}. The grey one corresponds to the HST only retrieval and shows the impact of adding the Spitzer data. At the top, we show the transit (left) and the eclipse (right). The diamonds represent the the averaged spitzer bandpasses (Orange: HST only; Green: HST+Spitzer). The right panels have inverted wavelength axis. The indicated $\chi_{HST}^2$ was computed from 1.1$\mu$m and 1.7$\mu$m only.}
    \label{fig:spectra_HST_Spz}
\end{figure}

\begin{figure}[H]
\centering
    \includegraphics[width = 0.92\textwidth]{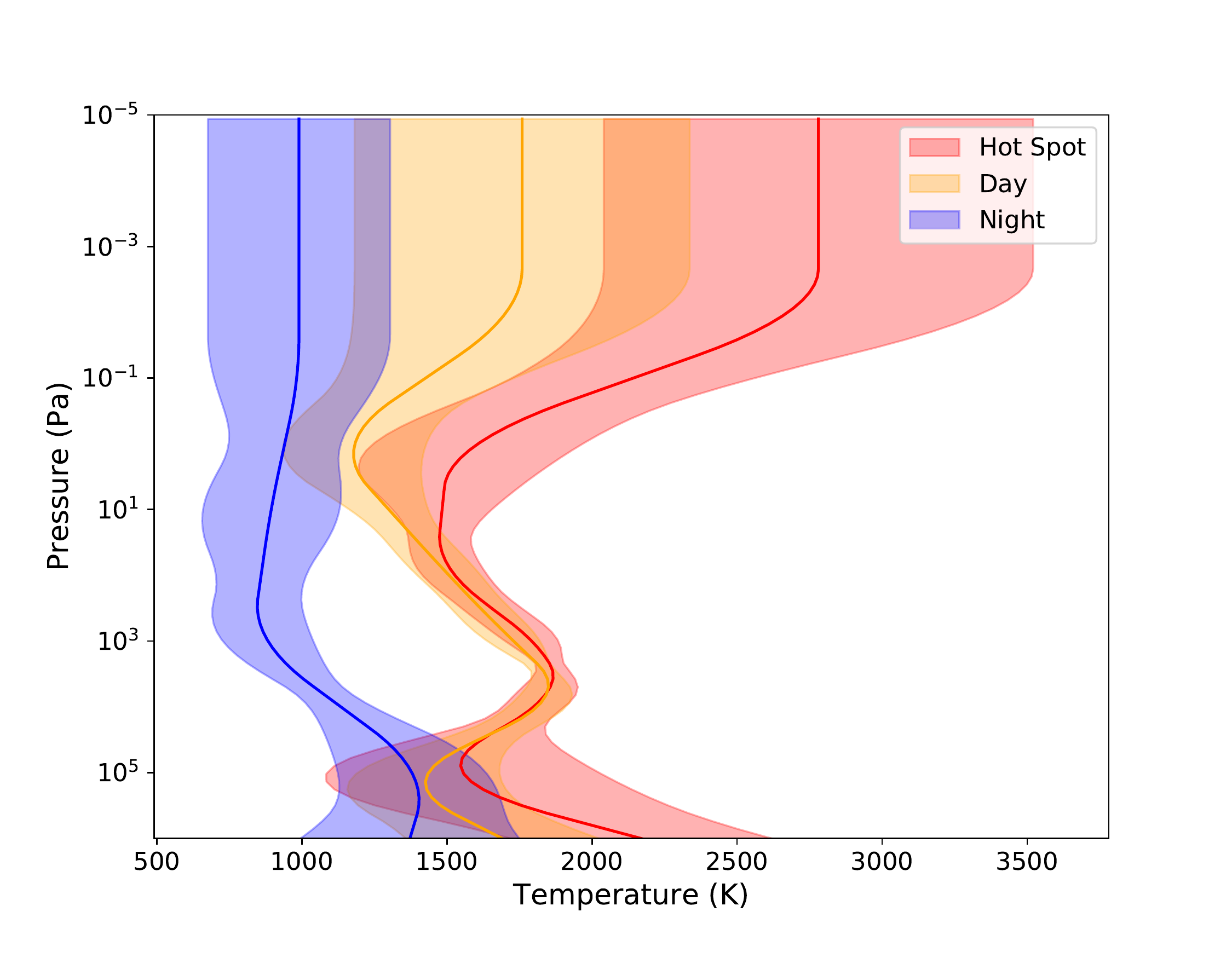}
    \caption{Temperature structure in the case of the HST only retrieval. The 1$\sigma$ error on the temperature profiles is much larger than in the HST+Spitzer case but the overall shape of those profiles is qualitatively consistent (see Figure \ref{fig:temperature}).}    \label{fig:temp_HST}
\end{figure}

\begin{figure}[H]
    \includegraphics[width = \textwidth]{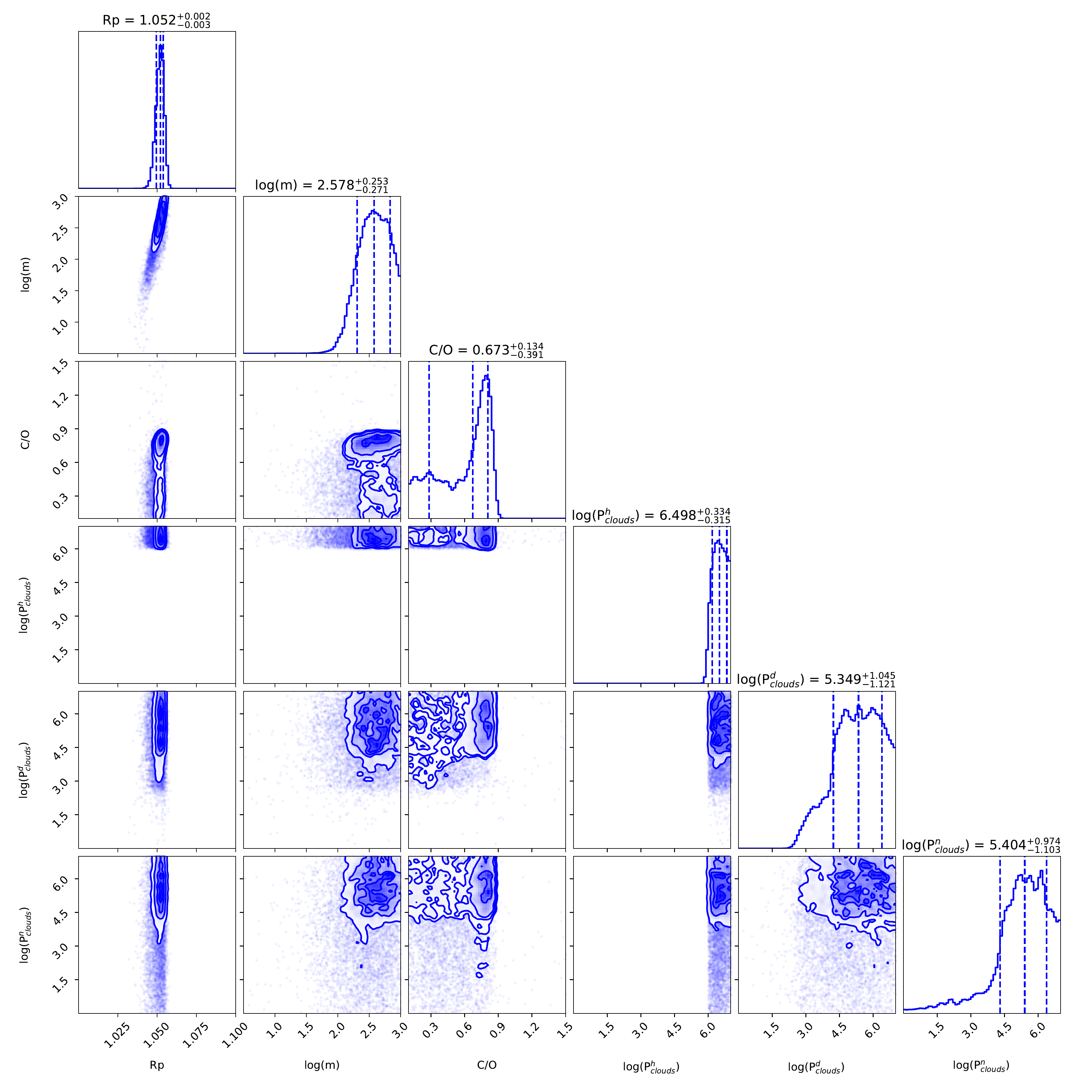}
    \caption{Posteriors distribution of the chemistry and the clouds properties for the full run on the HST only data. The retrieved metallicity reaches high values.}
    \label{fig:post_HST}
\end{figure}

\clearpage
\section{Contribution functions for the full scenario.}\label{apx:contrib_full}

\begin{figure}[H]
\centering
    \includegraphics[width = 0.89\textwidth]{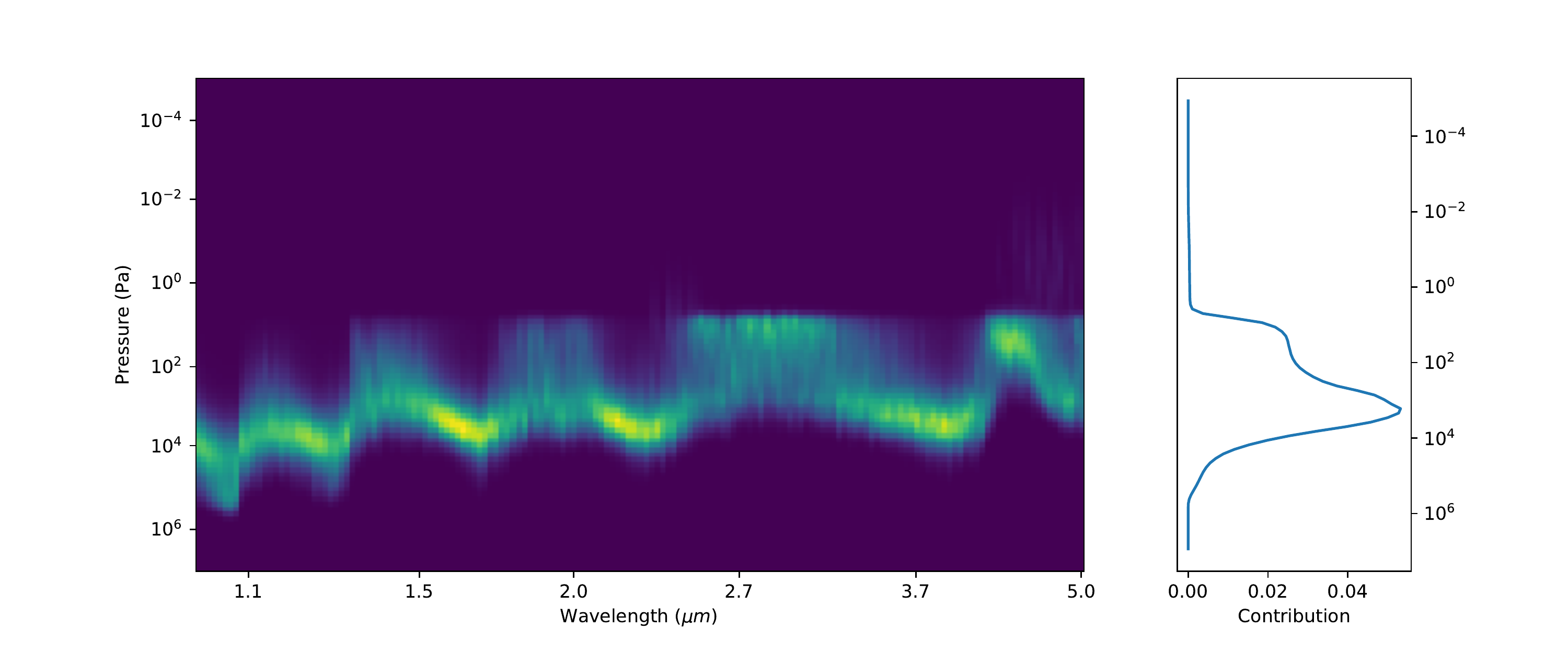}
    \includegraphics[width = 0.89\textwidth]{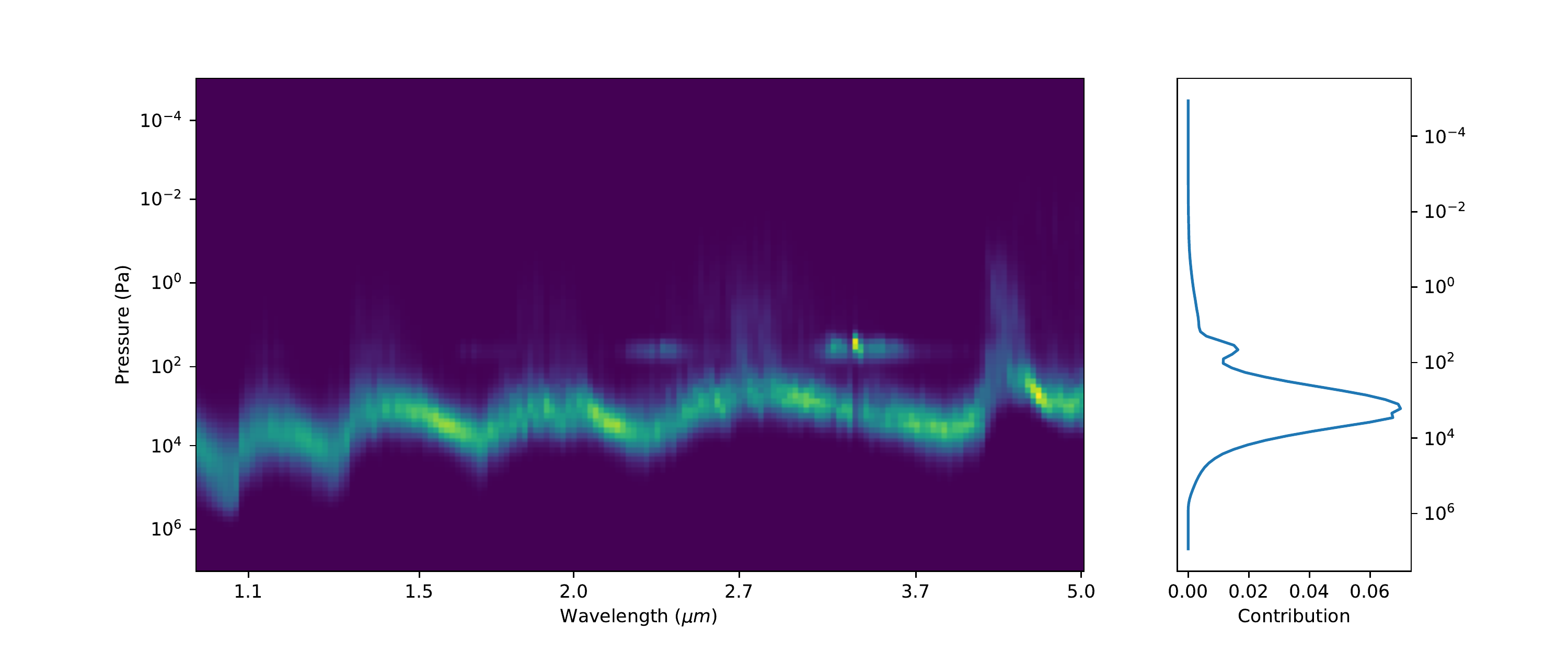}
    \includegraphics[width = 0.89\textwidth]{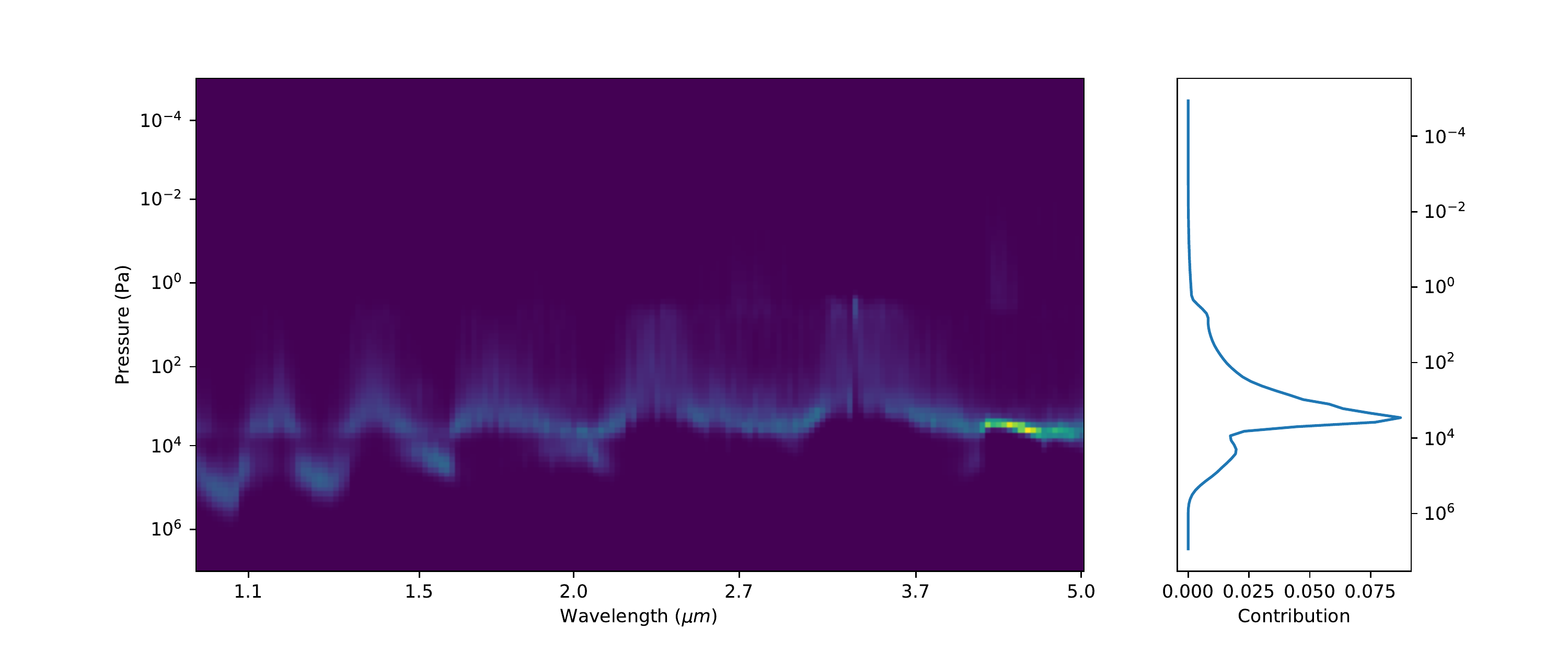}
    \caption{Contribution functions of the hot-spot (top), the day-side (middle) and the night-side (bottom) for the full run.}
    \label{fig:contrib_full_scenario}
\end{figure}

Figure \ref{fig:contrib_full_scenario} shows the contribution functions of each region for the full scenarios. The contribution function is defined as d$\mathcal{T}$/dP, where $\mathcal{T}$ is the transmittance.
\clearpage

\section{Posterior distributions of the full model for different hot-spot sizes.}\label{apx:post_full_alpha}

\begin{figure}[H]
    \includegraphics[width = \textwidth]{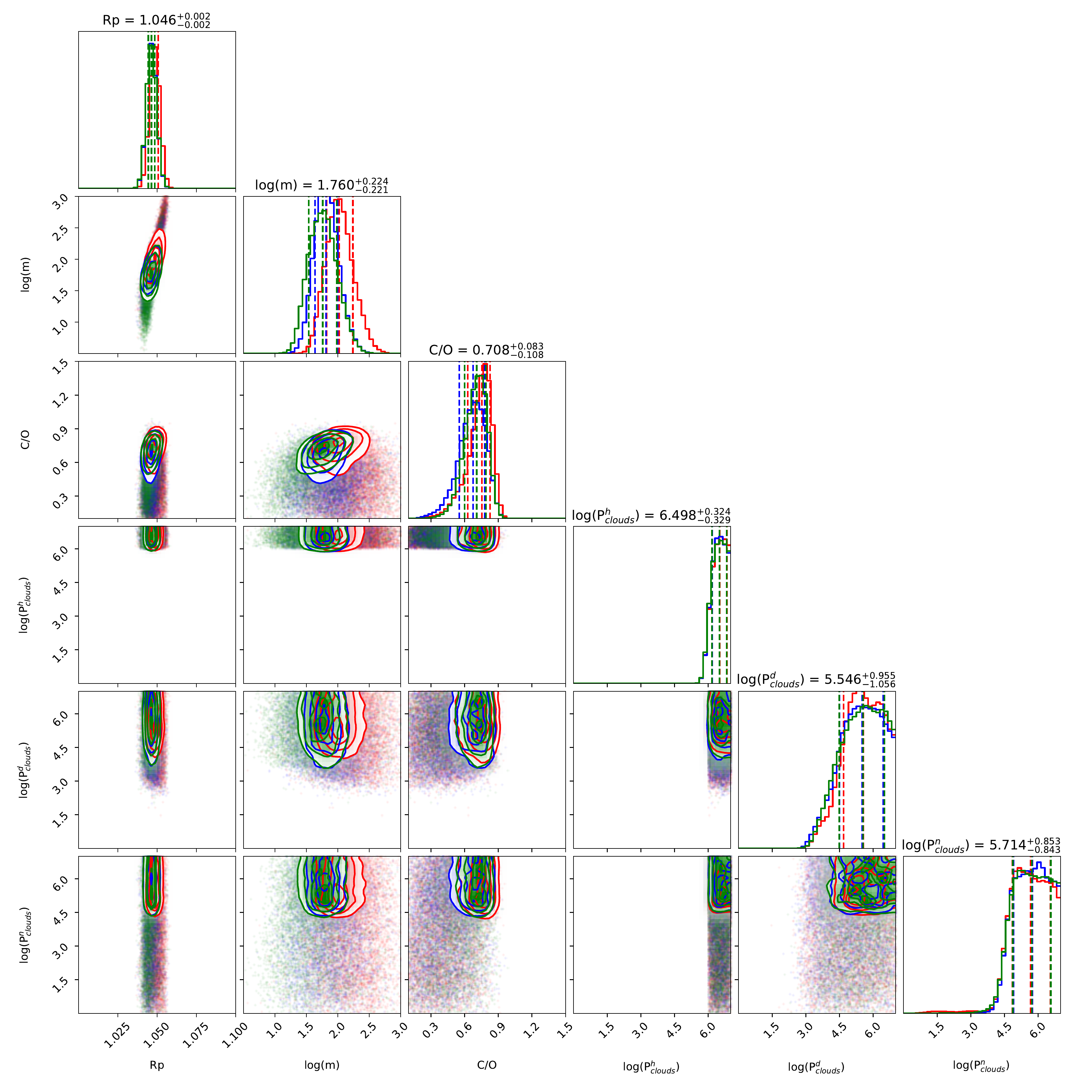}
    \caption{Posteriors distribution of the chemistry and the clouds properties for the run with $\alpha = 30$ degrees (green), $\alpha = 40$ degrees (blue) and $\alpha = 50$ degrees (red).}
    \label{fig:post_cumul1}
\end{figure}

\begin{figure}[H]
    \includegraphics[width = \textwidth]{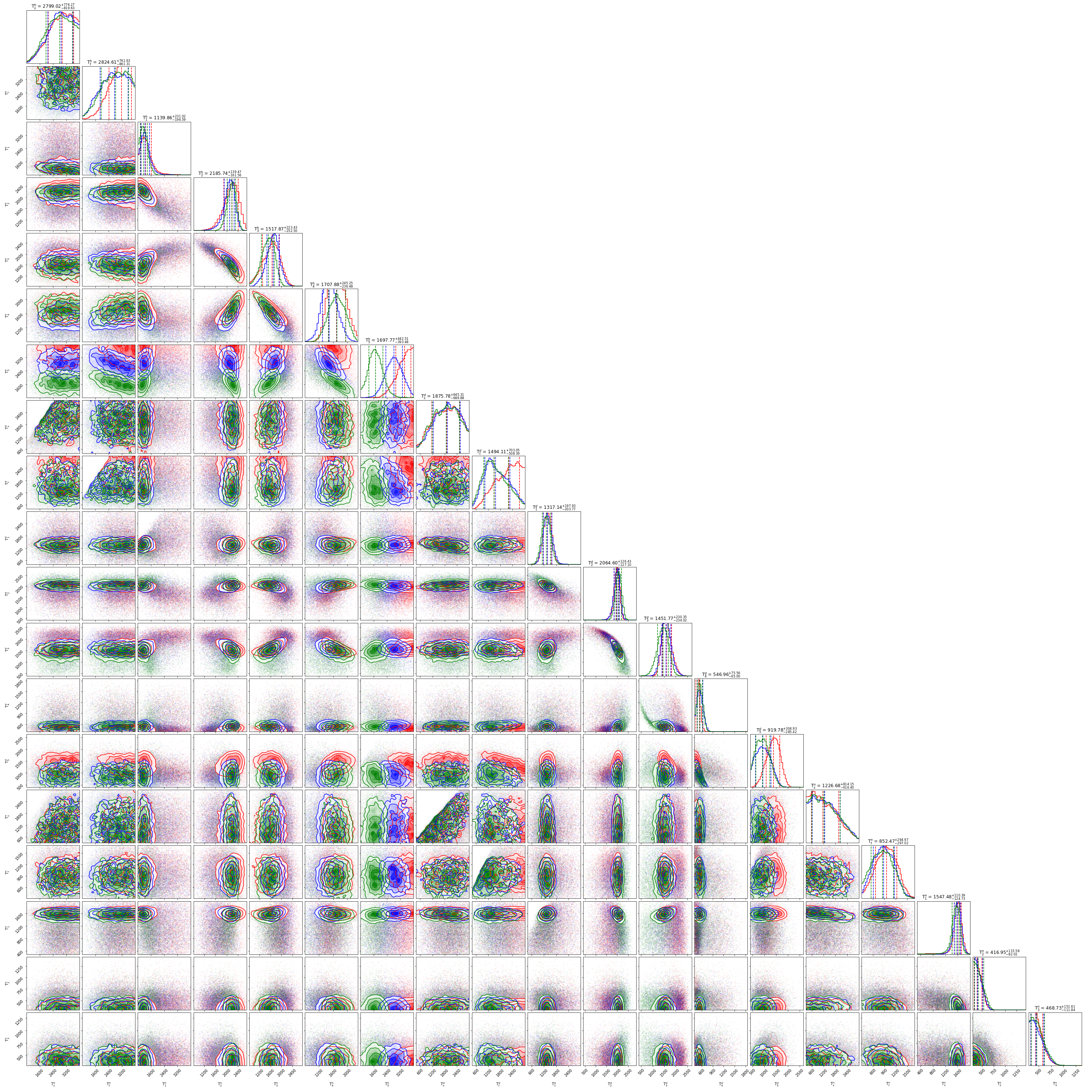}
    \caption{Posteriors distribution of the temperature points for the run with $\alpha = 30$ degrees (red), $\alpha = 40$ degrees (blue) and $\alpha = 50$ degrees (green). Since the T-p profile is smoothed, the actual profile does not intersect the retrieved pressure-temperature points.}
    \label{fig:post_cumul2}
\end{figure}

\section{Results of the full model with clouds allowed on the hot-spot.}\label{apx:full_cloudy}

\begin{figure}[H]
\centering
    \includegraphics[width = 0.86\textwidth]{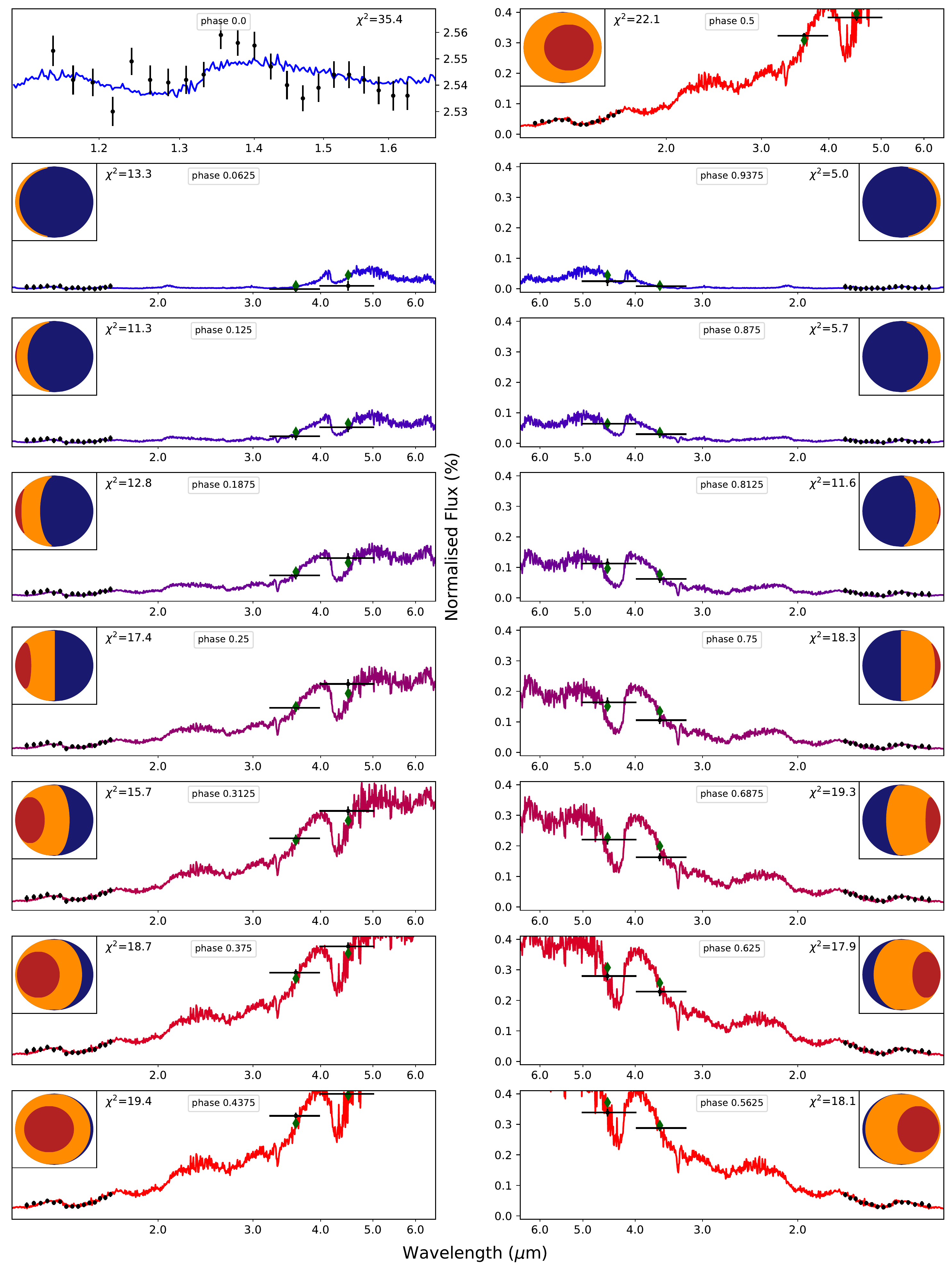}
    \caption{Best fit spectra and geometry of our WASP-43\,b phase-curve retrieval with clouds allowed on the hot-spot. At the top, we show the transit (left) and the eclipse (right). The diamonds represent the the averaged spitzer bandpasses. The right panels have inverted wavelength axis.}
    \label{fig:spectra_HST_Spz_clouds}
\end{figure}

\begin{figure}[H]
\centering
    \includegraphics[width = 0.92\textwidth]{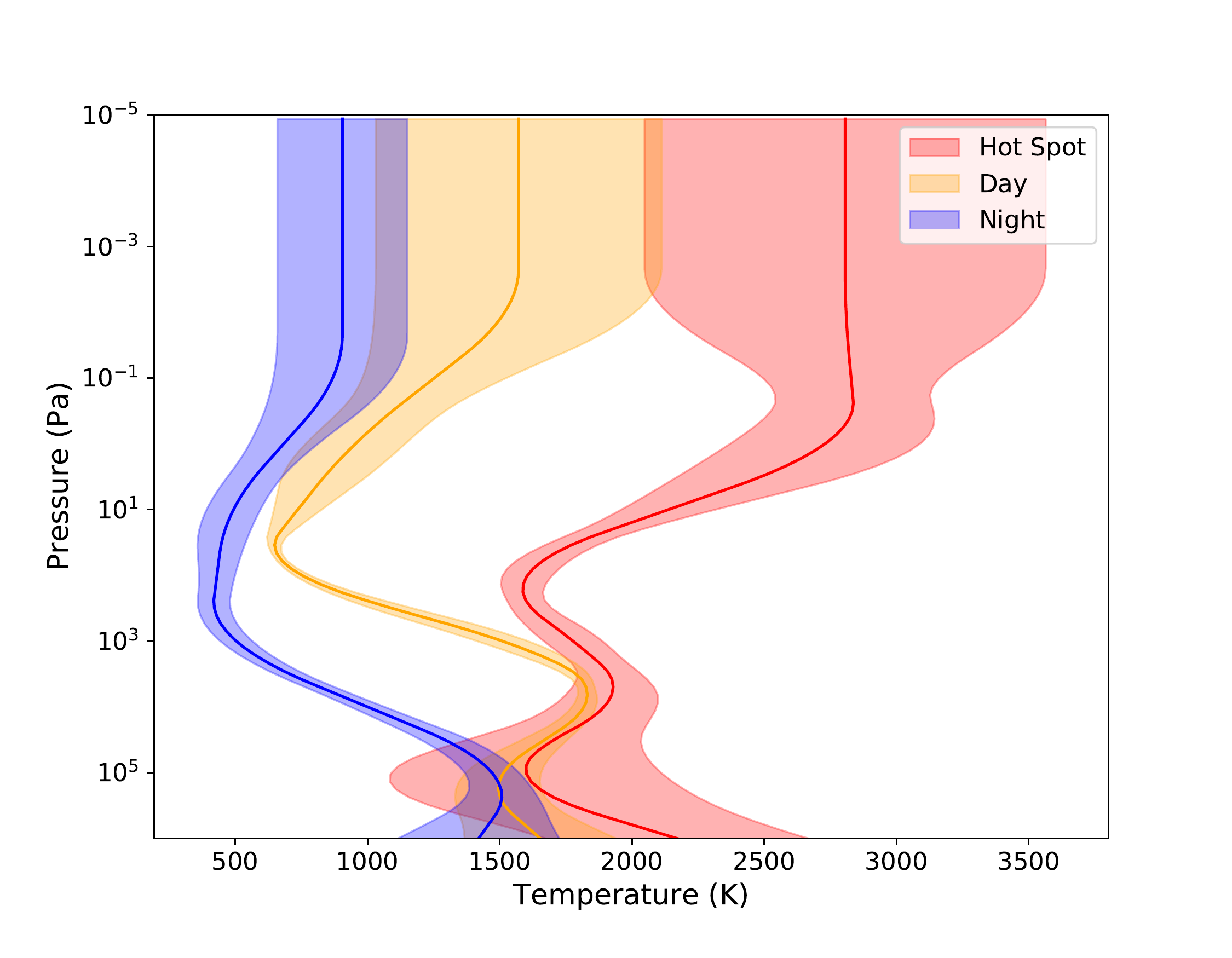}
    \caption{Temperature structure in the case of the HST+Spitzer retrieval with clouds allowed on the hot-spot. As compared with the clear hot-spot model (see Figure \ref{fig:temperature}), the uncertainties on the hot-spot temperature in the lower atmosphere are much larger.}
    \label{fig:temp_HST_Spz_clouds}
\end{figure}

\begin{figure}[H]
    \includegraphics[width = \textwidth]{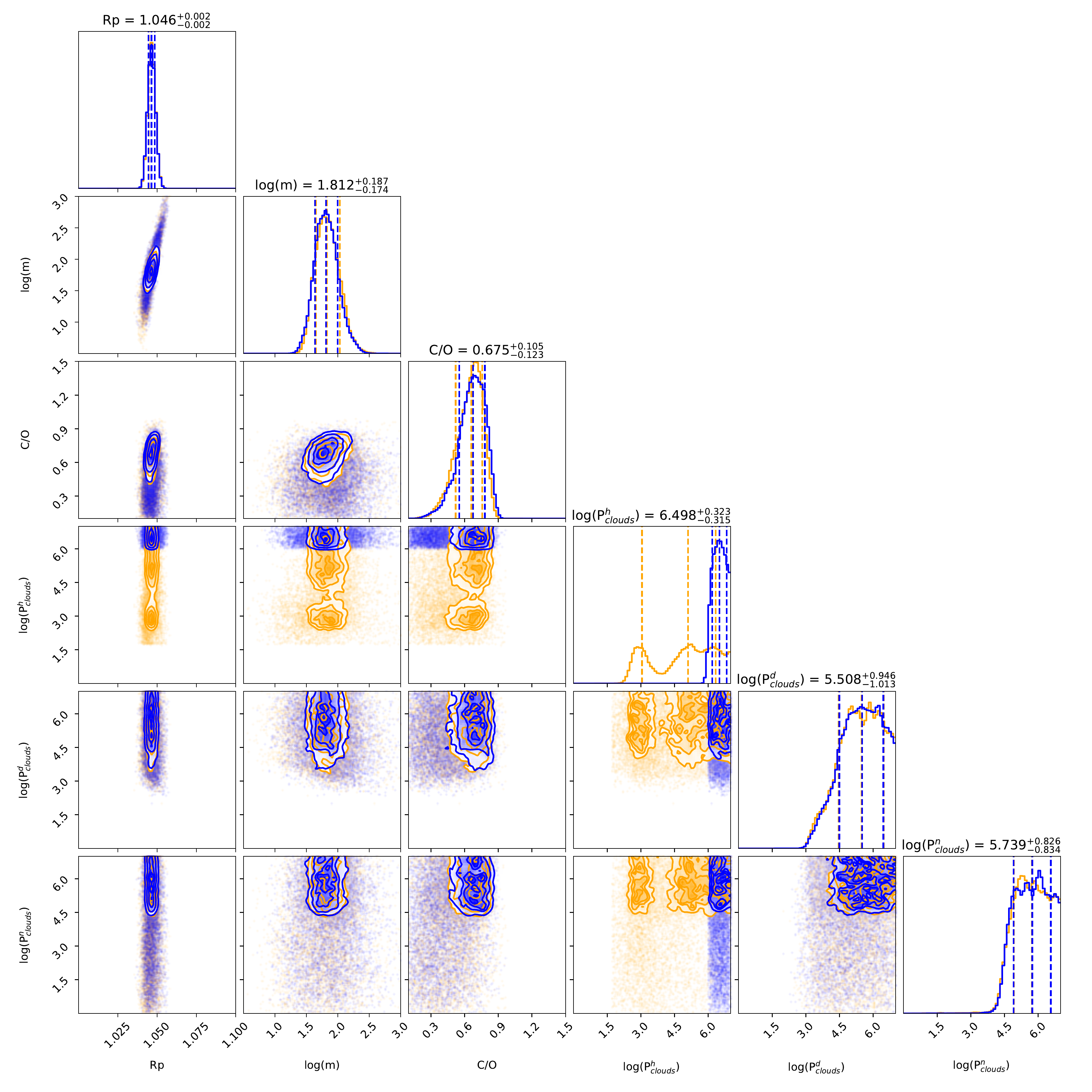}
    \caption{Posteriors for the chemistry and the cloud parameters in our runs with a fixed hot-spot size at 40 degrees. In blue: clear hot-spot (minimum cloud pressure forced at 10$^6$ Pa); In orange: cloud allowed on the hot-spot.}
    \label{fig:posteriors}
\end{figure}

\section{Results of the full model with 2500 live points.} \label{apx:full_2500}

\begin{figure}[H]
\centering
    \includegraphics[width = 0.86\textwidth]{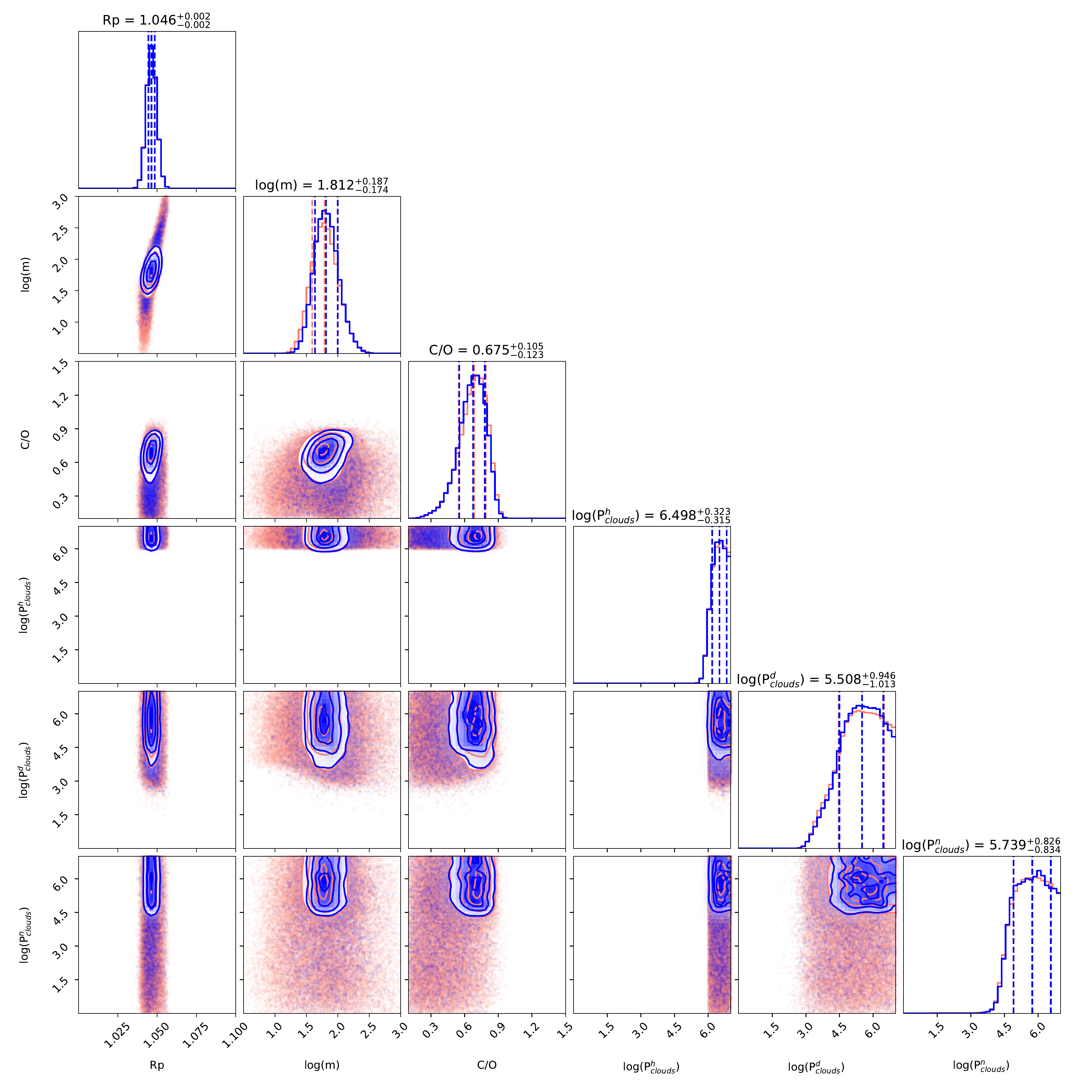}
    \caption{Posteriors for the chemistry and the cloud parameters in our full scenario with 2500 live points (Pink), compared with the standard run (Blue). While not shown, the temperature structure was also unchanged.}
    \label{fig:post_full_2500}
\end{figure}







\renewcommand{\floatpagefraction}{.9}%




\end{document}